\documentclass{jfm}

\usepackage{amsmath,amsfonts,amssymb}
\usepackage{graphicx}
\usepackage{epstopdf, epsfig}
\usepackage[english]{babel}
\usepackage[utf8]{inputenc}
\usepackage{xcolor}
\usepackage{float}

\usepackage{bbold}
\usepackage{bbm}
\usepackage{float}
\usepackage{caption}
\usepackage{comment}
\usepackage{booktabs}
\usepackage{verbatim}
\usepackage{mathdots}
\usepackage{array}
\usepackage[colorlinks=true, allcolors=blue]{hyperref}
\usepackage{tikz} 
\usetikzlibrary{calc}
\usepackage{pgfplots}
\usepackage{pgfplotstable}
\usepackage{wrapfig}
\usepackage{ifthen}
\usepackage{sidecap}

\usepackage{multirow}

\providecommand\xx{\mathbf{x}}

\providecommand\uu{\mathbf{u}}
\providecommand\phuu{\left\langle\mathbf{u}\right\rangle}
\providecommand\phu{\left\langle u \right\rangle}
\providecommand\phv{\left\langle v \right\rangle}
\providecommand\php{\left\langle p \right\rangle}

\providecommand\qq{\mathbf{q}}
\providecommand\phqq{\left\langle\mathbf{q}\right\rangle}
\providecommand\oqq{\overline{\mathbf{q}}}

\providecommand\ff{\mathbf{f}}

\providecommand\php{\left \langle p \right \rangle}

\providecommand\QQ{\mathbf{Q}}

\shorttitle{Identification and reconstruction of high-frequency broadband fluctuations}
\shortauthor{L. Franceschini, D. Sipp, O. Marquet, J. Moulin and J. Dandois}

\title{Identification and reconstruction of high-frequency broadband fluctuations in a turbulent flow exhibiting low-frequency deterministic motions}

\author{Lucas Franceschini\aff{1},
 Denis Sipp\aff{1}, Olivier Marquet\aff{1} \\
    Johann Moulin\aff{1}, \and Julien Dandois\aff{1}}

\affiliation{\aff{1}DAAA, ONERA, Université Paris-Saclay, 92120, France}

\begin{document}

\maketitle

\begin{abstract}
The turbulent flow around a squared-section cylinder at $Re=22000$ exhibits both a low-frequency Vortex-Shedding (VS), a common feature of all bluff-body flows, and high-frequency Kelvin-Helmholtz (KH) structures evolving on the shear layers formed at the top and bottom side of the cylinder.
The VS motion triggers a periodic movement of these shear layers, inducing oscillations of the strength of their velocity gradients. The high-frequency KH structures respond to these low-frequency oscillations of the base-flow gradients in a quasi-steady manner. 
In this paper we will propose a general framework to capture and reconstruct stochastic high-frequency fluctuations (the KH structures) evolving on top of a low-frequency deterministic motion (the periodic VS). Based on a generic triple decomposition, that allows separation of the stochastic component from the deterministic one, we propose a Phase-Conditioned Localized Spectral Proper Orthogonal Decomposition (PCL-SPOD), which isolates high-frequency spatio-temporal structures within the stochastic component at different phases of the VS motion, with a short-time Fourier transform. Finally, we compare these structures to those obtained with a Phase-Conditioned Localized Resolvent formalism (PCL-Resolvent), which consists in a Resolvent analysis around instantaneous snapshots of the periodic VS.
\end{abstract}

\begin{keywords}
Spectral Proper Orthogonal Decomposition, Resolvent Analysis, Triple Decomposition
\end{keywords}

\section{Introduction}

The flow around a squared-section cylinder is a benchmark case, showing numerous interesting well documented behaviors at several Reynolds numbers. At low Reynolds numbers, it consists in a typical example of supercritical Hopf bifurcation where, for Reynolds numbers higher than $\approx 50$ (see \citet{sohankar1998low}), the Vortex-Shedding phenomenon arises, persisting up to very high Reynolds numbers (see, for example, \cite{roshko1961experiments} for circular cylinders). This phenomenon has been extensively studied, either with data-based techniques through, for example, Proper Orthogonal Decomposition (POD) or Dynamical Mode Decomposition (DMD) \citep{chen2012variants,noack2016recursive} or model-based techniques, such as linear stability analyses, among other techniques. It was shown  \citep{pier2002frequency,barkley2006linear,sipp2007global}) in low Reynolds numbers regimes that there exists a marginal eigenmode of the Navier-Stokes equations linearized around the mean-flow whose spatial structure and frequency are very close to those of the dominant Fourier mode of the flow. 

For higher Reynolds numbers (typically higher than $O(10^3)$),
the flow still exhibits the VS motion with characteristics that can be captured with a mean-flow linear stability analysis \citep{meliga2012sensitivity,mettot2014computation}.
 Yet, an additional phenomenon steps in.
For example, flow around a squared section cylinder also exhibits small-scale, high-frequency Kelvin-Helmholtz (KH) structures developing on the top and bottom shear layers of the cylinder. Due to the periodic VS motion, those shear-layers move periodically, which induces an oscillation of both the spatial location of the KH structures and their strength. This was clearly shown by \citet{brun2008coherent} through Large-Eddy Simulations (LES) and experimental data: probes located on the upper/lower vicinity of the cylinder showed an intermittent behavior, with KH structures present at certain phases of the VS cycle, but not at others. Typically, the KH structures develop over a broad range of frequencies. In this context, one of the most successful techniques to capture such modes is the Spectral Proper Orthogonal Decomposition (SPOD) that unveils, for a given frequency, all the modes underlying the turbulent spectral cross-correlation tensor \citep{sieber2016spectral,towne2018spectral,pickering2020optimal}. In cases where the flowfield exhibits a broadband-frequency content, such as jets, or boundary layers, it was shown that a Resolvent analysis \citep{mckeon2010critical,cossu2009optimal,beneddine2016conditions} was capable to accurately reconstruct those SPOD modes. Such an analysis is based on the input / output properties of the linearized Navier-Stokes equations around the mean-flow and extracts the structures that are most amplified in the frequency domain. Yet, those techniques are not able to describe the
impact of a low-frequency deterministic modulating motion (such as the VS cycle) on the spatio-temporal features of the broadband-frequency fluctuation field.

The present paper aims at extending the classical SPOD and Resolvent analyses in order to capture and model the dependency of high-frequency stochastic fluctuations with respect to a low-frequency deterministic motion. To that aim, we first need to introduce a decomposition able to separate those two components. The triple decomposition, based on the phase-average introduced by \citet{reynolds1972mechanics,hussain1983coherent}, suggests a first method to identify the deterministic component in the case of a periodic external forcing. Yet,
we stress that more complex deterministic components could be encountered, for example, quasi-periodic ones in statistically stationary flows, for which harmonic averages can be employed \citep{mezic2013analysis,arbabi2017study}, or even general externally-driven flows where arbitrary deterministic forces are imposed in a repeated (but not necessarily periodic) manner. In the latter case, a decomposition based on the expectancy over several realizations of the same experiment can be used to define the deterministic component.
Then, we advocate for the use of a Phase-Conditioned Localized SPOD where the spectral correlation tensor is defined as a function of the phase of the deterministic motion. The hypothesis allowing us to construct such a tensor is the large separation in time-scales between the deterministic and the stochastic motion, allowing the latter to experience the former in a quasi-steady manner. The construction of this tensor will be based on the short-time Fourier Transform of the stochastic component, which will be evaluated over several realizations of the flow, each time around a given phase of the deterministic motion. This analysis can be viewed as a spectral version of the conditional space-time POD proposed by \citet{schmidt2019conditional}, which aims at capturing rare events. Then, we introduce a Phase-Conditioned Localized resolvent analysis, capable to model the high-frequency stochastic component. It is also based on the hypothesis that this component evolves quasi-steadily, leading to the usual resolvent analysis, but based on an instantaneous snapshot of the deterministic field at a given phase rather than then mean-flow. 

\begin{figure}
	\centering
	\raisebox{1.2in}{(a)}\includegraphics[trim={0.5cm 0.5cm 0.5cm 0.5cm},clip,height=120px]{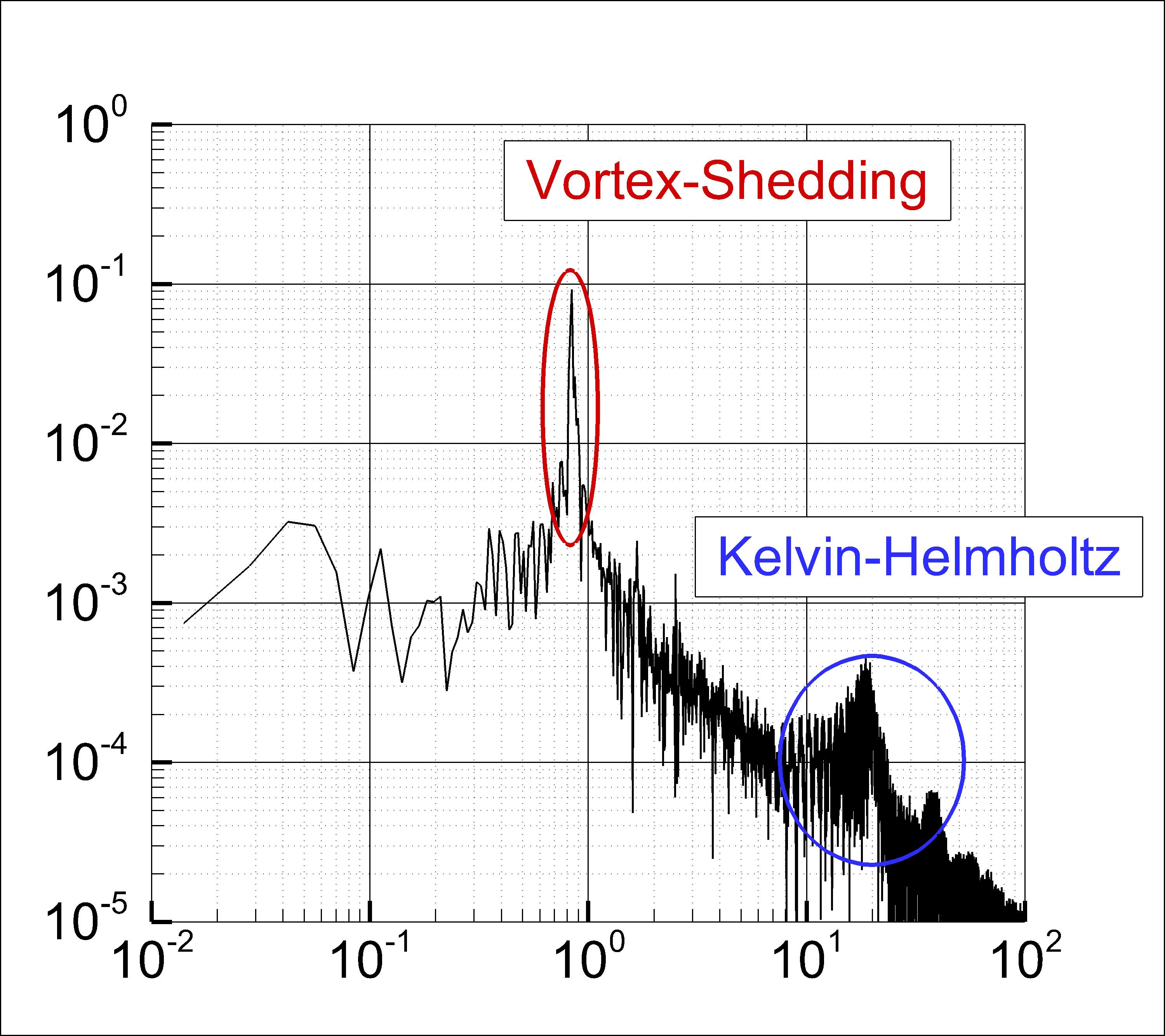}\raisebox{0.2in}{$\omega$}
	
    
    \raisebox{0.5in}{(b)}\includegraphics[trim={1cm 10cm 30cm 10cm},clip,height=80px]{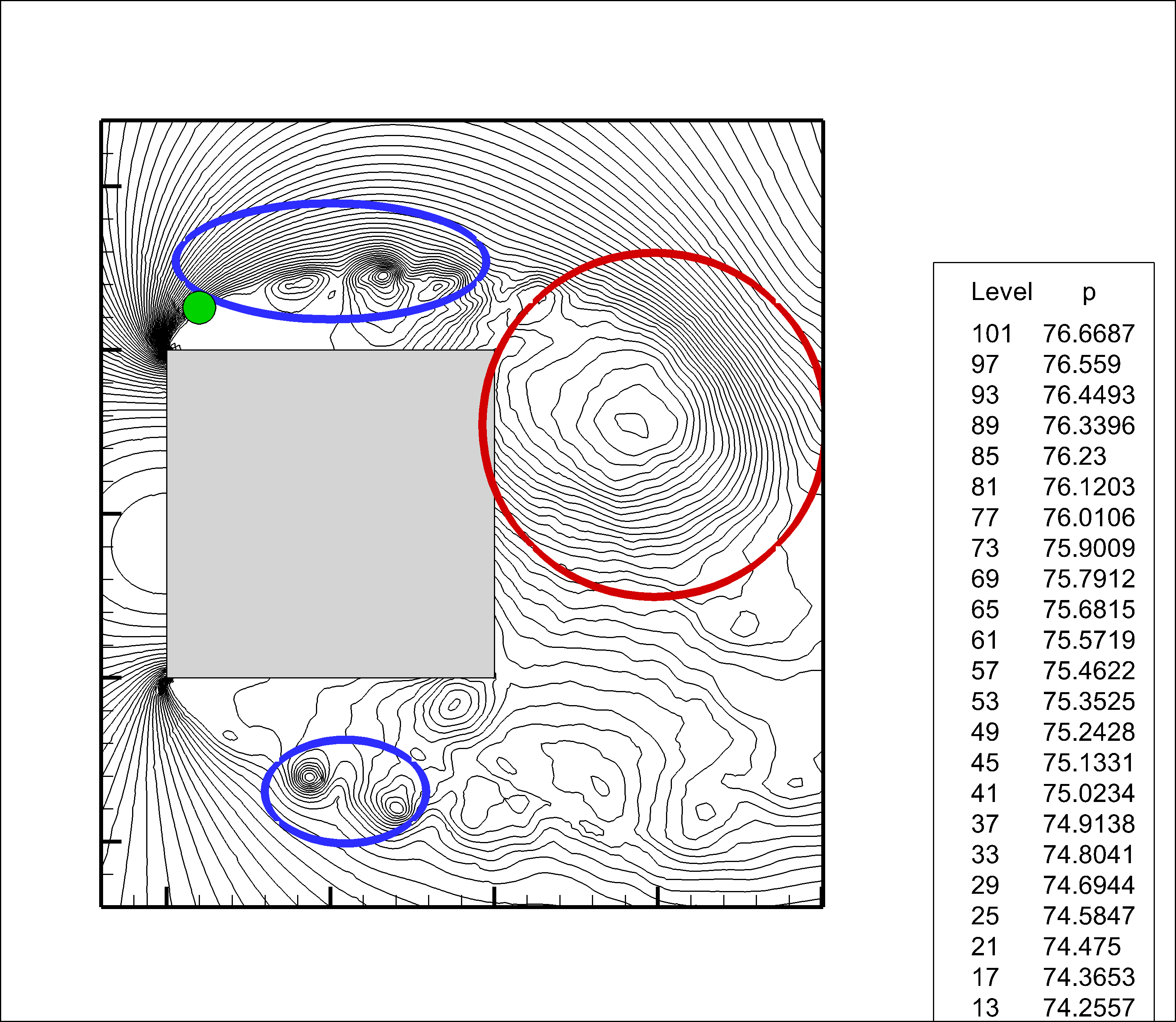}
    \includegraphics[trim={8cm 10cm 30cm 10cm},clip,height=80px]{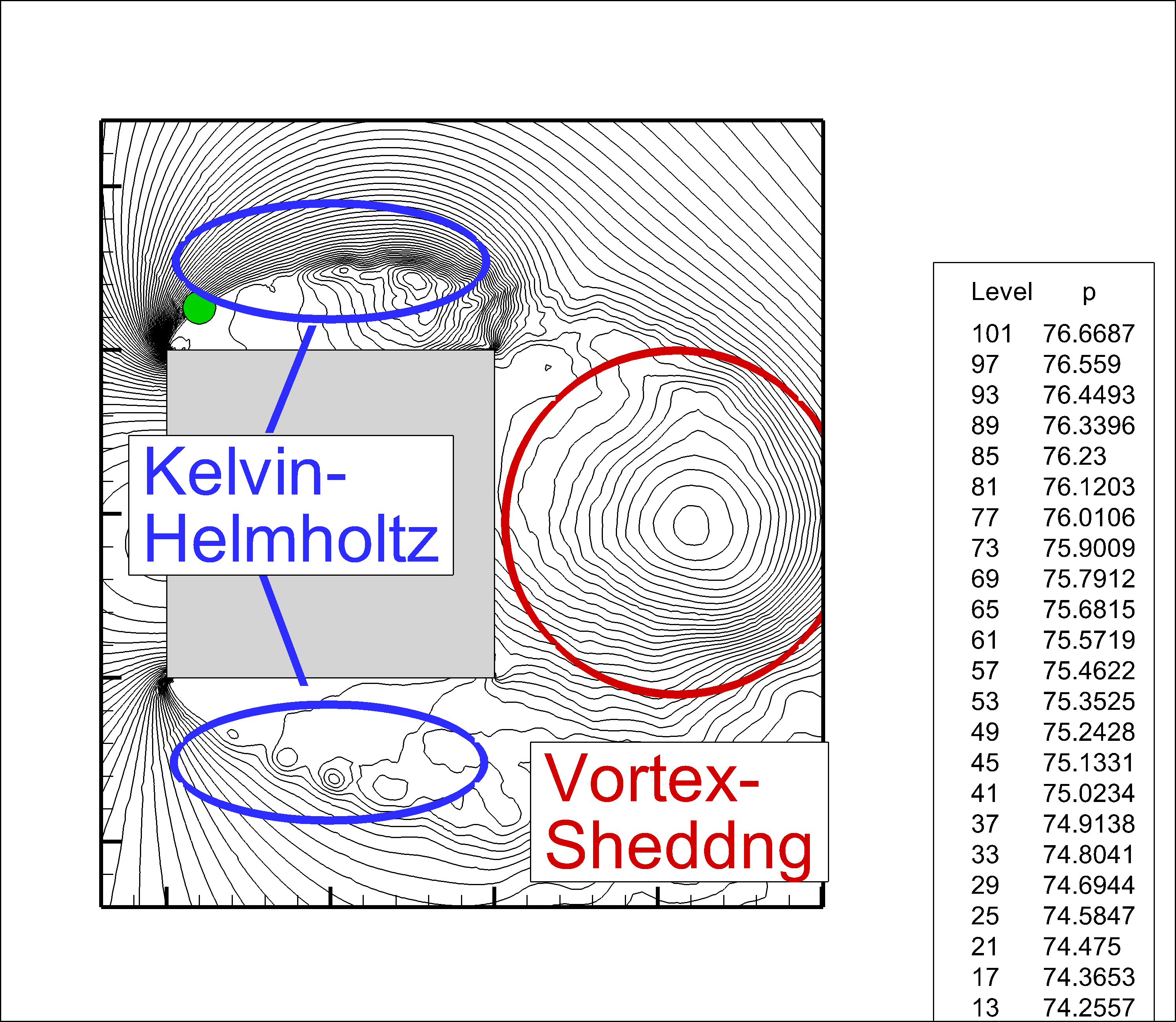}
    \includegraphics[trim={8cm 10cm 30cm 10cm},clip,height=80px]{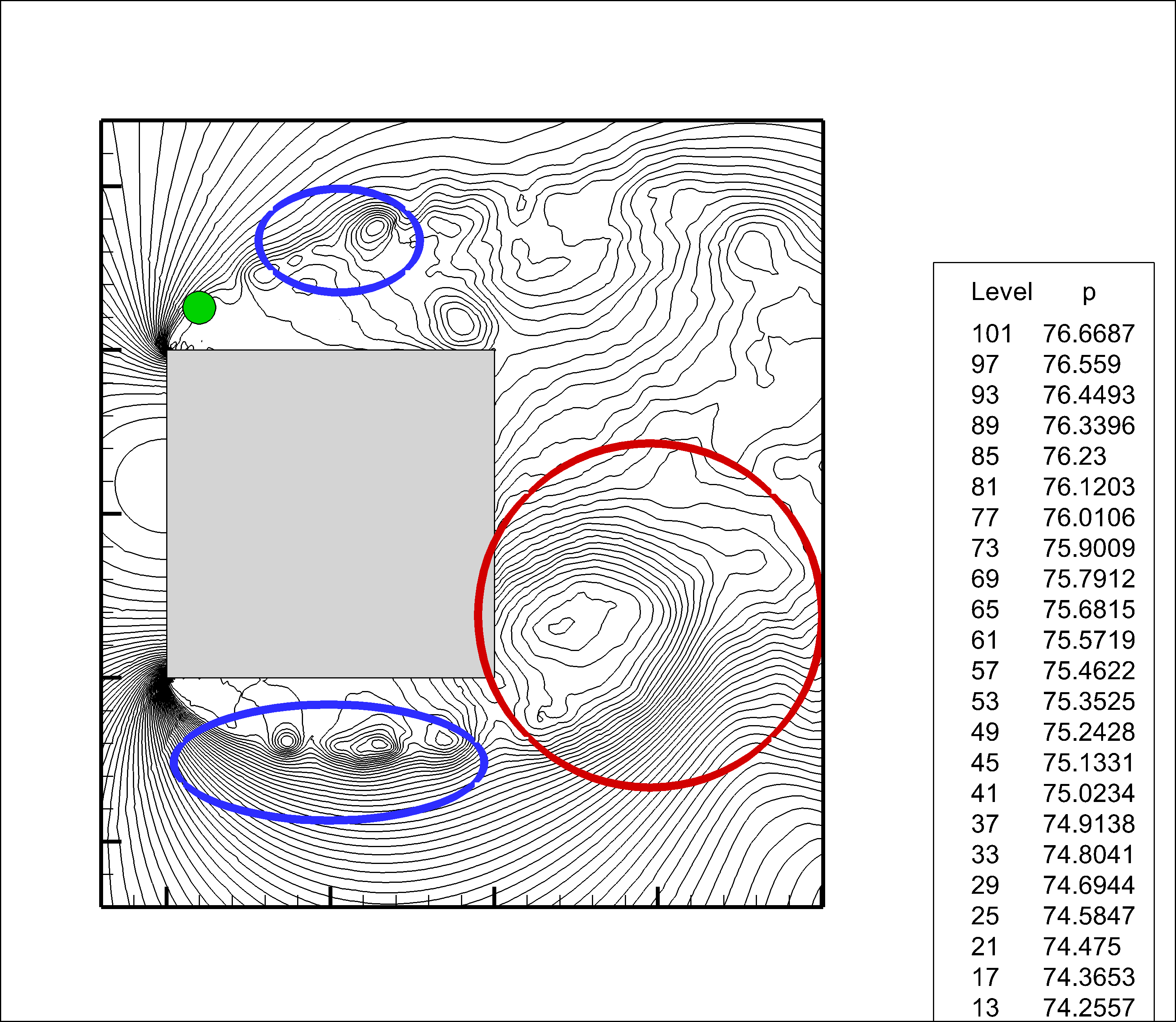}
    \includegraphics[trim={8cm 10cm 30cm 10cm},clip,height=80px]{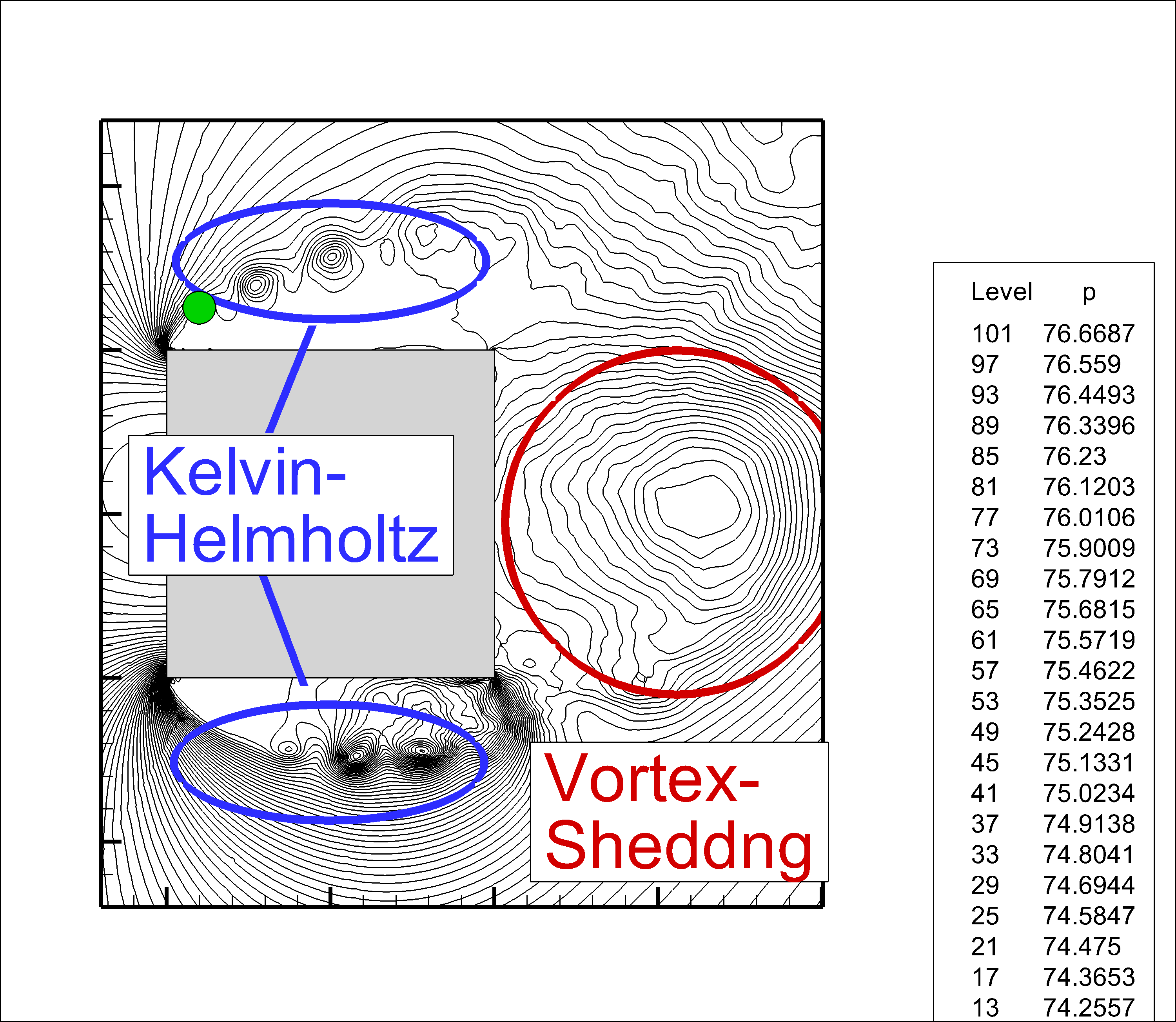}

    \caption{Typical signal produced by a DNS run on a squared-section cylinder at $Re=22000$. (a) Power-Spectral Density of the point $(-0.4,0.63)$ (from the center of the cylinder) on the shear-layer region. (b) Snapshots of the (span-wise-averaged) pressure field at four different phases, exhibiting the high-frequency Kelvin-Helmholtz phenomenon (blue circles) and its dependency on the lower-frequency vortex-shedding (red circles). The green point represents the probing location.}\label{fig:Sketch}
\end{figure}

We will illustrate these new techniques on the case of the flow around a squared-section cylinder at $Re=22000$. More precisely we aim at identifying and reconstructing the high-frequency KH fluctuations evolving on top of the low-frequency periodic VS one. Since the deterministic motion is periodic, by ergodicity, we may employ the Welch algorithm to compute the spectral correlation tensor. Indeed, each period of the VS phenomenon, may be taken as an independent (and phased) realization, allowing us to perform the PCL-SPOD and PCL-Resolvent analyses based on data from a single DNS run.
Some results of the DNS simulation that has been specifically run for this work, have been displayed in figure \ref{fig:Sketch}, showing both phenomena KH/VS at four different phases of the VS phenomenon (b), together with a Power-Spectral Density (PSD), in which both the low-frequency periodic VS motion and the high-frequency broadband KH fluctuations can be seen (a).

The article is structured as follows. In section \ref{Theory}, we introduce the triple decomposition, the PCL-SPOD and PCL-Resolvent analysis. Then, in section \ref{Results}, we illustrate these techniques on the squared-section cylinder flow at $ Re=22000$. A last section (\S \ref{Conclusion}) concludes the article.

\section{Triple Decomposition, PCL-SPOD and PCL-Resolvent analyses}\label{Theory}

In this section, we introduce the theoretical basis for the analysis of the turbulent flow around the squared-section cylinder at high Reynolds number $Re=22000$. The first step of the analysis (\S \ref{Theory:GeneralExternalSignal}), will be the decomposition of the flow variables (here grouped in the state vector $\qq$) into a deterministic motion (denoted $\phqq$), that later will be particularized to represent the Vortex-Shedding (VS), and a stochastic one (denoted $\qq'$), that will represent the Kelvin-Helmholtz structures (KH). The hypothesis that $\phqq$ is a deterministic field is important to allow us to phase several realizations. Indeed, if we are capable to identify a deterministic $\phqq(t)$ from an ensemble average of realisations, we will be able to study the dynamics of fluctuations $\qq'(t)$ on top of $\phqq(t)$ at all times $t$ with a statistical viewpoint (average over realisations). 
We propose an SPOD and a resolvent analysis capable of analysing the dynamics of the signal $\qq'$ conditioned to the deterministic one $\phqq$. An important hypothesis in the following is that a separation of time-scales holds between the component $\phqq$ and the fluctuation field $\qq'$, allowing us to use a quasi-steady hypothesis. With this hypothesis, we will be able to analyse the high-frequency content of the signal according to the phase of the deterministic motion using a Short-Time Fourier Transform (STFT). With the aid of this tool, we introduce (\S \ref{Theory:POD}) a Phase-Conditioned Localized (PCL) SPOD technique. Using a similar reasoning, a PCL-resolvent analysis is proposed (\S \ref{sec:PCLresolvent}).

In section \S \ref{Theory:Periodic}, we then particularize the analysis to the case of a periodic deterministic component. In such a case, all realizations can be phased with respect to the phase of the periodic cycle. Moreover, if we regard each period as one realization (ergodicity), we can define, in this context, the phase-average (see \citet{reynolds1972mechanics,hussain1983coherent}), allowing us to obtain $\phqq$ (\S \ref{Theory:ExtractionPeriodic}) and statistics (\S \ref{Theory:PeriodicImplementation}) from a single run by averaging over periodic cycles (instead of realizations).
We provide also details about the implementation of the PCL-SPOD technique in such a periodic case. Finally (\ref{sec:Meandering}), we recall that turbulent flows exhibiting low-frequency motion generally exhibit the meandering phenomenon, which makes the low-frequency motion being weakly non-periodic. We will show how such flows may still be analysed in a periodic framework by defining statistics for $\qq'$ based on a redefinition of $\phqq$ within each period.

\subsection{General case}\label{Theory:GeneralExternalSignal}


Turbulent flows, in general, can be viewed as stochastic processes where several realizations produce different outcomes. Those realizations though exhibiting common statistical features are indeed initialized differently. Each realization then exhibits common features with respect to the others, that are gathered within the deterministic component.

If an external conditional signal is available, then all realizations can be phased, and this component can be determined by:
\begin{equation} \label{eqn:average}
     \phqq(\mathbf{x},t) = E[\qq(\mathbf{x},t)],
\end{equation}
where $E[\cdot]$ is the expectancy operator over an ensemble of realizations.

If such an external conditional signal is not available, 
we can build such a signal with harmonic averages in the case of
statistically stationary flows.
According to \cite{mezic2013analysis,arbabi2017study}, there exists a countable set of frequencies $\{ \omega_j \}$ such that the harmonic averages, defined as
\begin{equation}\label{eqn:Harmonic}
    \hat{\qq}(\omega_j) = \lim_{T_f \rightarrow+\infty} \frac{1}{T_f} \int_{0}^{T_f} e^{- \mathrm{i} \omega_j t} \qq(t) \; dt,
\end{equation}
converge to a finite non-zero quantity as the final time of the realization $T_f$ becomes large. 
The resulting modes $\hat{\qq}(\omega_j)$ are called Koopman modes.
The quasi-periodic component of a given realization is then reconstructed thanks to the Koopman Mode Decomposition (KMD):
\begin{equation}\label{eqn:Koopman}
    \phqq(\mathbf{x},t) \equiv \oqq (\mathbf{x}) + \sum_{\omega_j} e^{\mathrm{i} \omega_j t} \hat{\qq}(\omega_j) + c.c.,
\end{equation}
where $\phqq$ groups together all non-zero harmonic averages of $\qq$ described by equations \eqref{eqn:Harmonic}. All realizations will generate the same KMD (up to a dephasing) and the different realizations may therefore be phased.
We have therefore intentionally used, for this new quantity, the same notation $\phqq$ as in \eqref{eqn:average}, since it will also correspond to the deterministic component of any other realizations after phasing and averaging. We remark that the mean-flow is also a Koopman mode associated with the zero frequency, i.e. $\oqq=\hat{\qq}(\omega=0)$. A conditioning signal may then be defined by considering the value of a component of $\phqq$ at some given point in the domain. Such a framework may describe for example complex fluid systems where several instability mechanisms are present (such as the lid-driven cavity, see \citet{arbabi2017study} or the shear-driven cavity, see \citet{leclercq2019linear,bengana2019bifurcation}). In the context of fluid-structure interaction, it may also be adapted to the description of quasi-periodic flow-induced vibrations observed for a single spring-mounted cylinder \citep{prasanth2008vortex} or airfoil \citep{menon2019flow} and for a double spring-mounted plate  \citep{moulin2020bifurcation}.


The total flowfield $\uu$ may then be decomposed as:
\begin{equation} \label{eqn:PhaseDecomposition}
	\qq(\mathbf{x},t) = \phqq(\mathbf{x},t) + \qq'(\mathbf{x},t),
\end{equation}
where $\qq'$ is the stochastic component. If $\phqq$ was obtained with Koopman decomposition, $\qq'$ holds a broadband distribution in frequency, as the corresponding harmonic averages of $\hat{\qq}' (\omega)$ converge to zero for any frequency $\omega$ different from the Koopman modes.
In the following, we describe (\S \ref{Theory:POD}) a data-driven approach (in the spirit of the SPOD analysis) that manages to extract, at given time instants of the conditioning signal, coherent structures from the flow snapshots. Then (\S \ref{sec:PCLresolvent}) a resolvent analysis is presented that allows reconstruction of these structures just from knowledge of $ \phqq $ at the considered time instants.

\subsubsection{A Phase Conditioned Localized Spectral Proper Orthogonal Decomposition}\label{Theory:POD}

{In the following, we will look at structures $ \qq'$ that exhibit high frequencies with respect to those of the field $\phqq(t)$. Due to this different time-scales, we will say in the following that $\tau$ is a slow time-scale. We may thus write $\phqq(\tau)$. The structures in $\qq'$ are therefore characterized by short life-cycles of the order of several periods $T = 2\pi/\omega $ and   may be parametrized by the slow-time $\tau $.
}
In order to study the dependency of $\qq'$ on $\phqq$, we consider in the following the quantity:
\begin{equation}
    \qq'_{\tau}(t) = w(t-\tau) \qq'(t),
\end{equation}
which is the fluctuation $\qq'$ multiplied by a time-window function $w(\zeta=t-\tau)$ introduced to focus the analysis of the signal $\qq'(t)$ around  $\tau$. To that aim, this window should be null for $| \zeta | = |t-\tau| > T / 2$, selecting this way only high-frequencies larger than $2 \pi/ T$. In other words, this window is centered around $t \approx \tau $ and its support is the fast period $T$. By varying the time $\tau$, we thus select the high-frequency fluctuation field at different times along the slow time-scale. 

To extract the dynamics around the phase $\tau$, we will perform a space/time POD (see \citet{towne2018spectral}) on the quantity $\qq_{\tau}'$. This is done by looking for $ \phi_{\tau}(\xx,t) $ that maximizes the gain \citep{schmidt2019conditional}:
\begin{equation} \label{eqn:POD}
    \lambda_{\tau}^2 = \frac{E [ | \left \langle \qq'_{\tau}(\xx,t) , \phi_{\tau}(\xx,t) \right \rangle_{\Omega,t} |^2 ] }{ \left \langle  \phi_{\tau}(\xx,t) , \phi_{\tau}(\xx,t) \right \rangle_{\Omega,t} }.
\end{equation}
Here, $\phi_{\tau}(\xx,t)$ and $\lambda_{\tau}^2$ designate the dominant POD mode and its energy, while 
the inner product is defined as $\left \langle \phi , \psi \right \rangle_{\Omega,t} = \int_{-\infty}^{+\infty} \int_{\Omega} \phi \cdot \psi^* dt d\xx$ {on a domain $\Omega$ that will be specified later.}
With those definitions, the maximization of $\lambda_{\tau}^2$ leads to the eigenvalue problem:
\begin{equation} \label{eqn:PODEigen}
    \int_{\Omega} \int_{-\infty}^{+\infty} \mathbf{C}_{\tau}(\xx,\xx',t,t') \phi_{\tau}(\xx',t') \; dt' d\xx' = \lambda_{\tau}^2 \, \phi_{\tau}(\xx,t),
\end{equation}
where $\mathbf{C}_{\tau}(\xx,\xx',t,t') = E [ \qq'_{\tau}(\xx,t) \qq'_{\tau}(\xx',t') ]$ is the cross-correlation tensor, here parametrized by the time $\tau$ and such that $\mathbf{C}_{\tau}(\xx,\xx',t,t') = 0 $ for $ |t-t'| > T$.
Again, the averaging is performed over the realizations using the expectation operator $ E[\cdot]$.
In this work, we intend to further specify the frequency content of this tensor. 
Before doing so, we need to consider an approximation that allows for a decoupling of space and time,
 namely that the correlation tensor is only a function of the time difference:
\begin{equation}
    \mathbf{C}_{\tau}(\xx,\xx',t,t') \rightarrow \mathbf{C}_{\tau}(\xx,\xx',t-t')
\end{equation}
This approximation is expected to be valid when $T$ is much lower than the characteristic time-scale of $ \tau $, that is to say when the broadband fluctuation $\qq'$ experiences the component $\phqq$ as a slowly-varying mean-flow. The usual SPOD (see \citet{towne2018spectral}) corresponds to the case where $\phqq(t) = \oqq$ is a constant, and the above property automatically holds for all frequencies. 

Then, expressing the tensor $\mathbf{C}_{\tau}(\xx,\xx',t-t')$ in Fourier space, we have:
\begin{equation} \label{eqn:spectralCorr}
    \mathbf{C}_{\tau}(\xx,\xx',t-t') = \frac{1}{2\pi} \int_{-\infty}^{+\infty} \mathbf{S}_{\tau}(\xx,\xx',\omega) e^{ \mathrm{i} \omega (t-t')} d\omega,
\end{equation}
where $\mathbf{S}_{\tau}(\xx,\xx',\omega)$ denotes the Fourier transform of the tensor $\mathbf{C}_{\tau}$. Taking the Fourier transform of 
\eqref{eqn:PODEigen} and simplifying by use of the Fourier transform definition, we obtain the following eigenvalue problem:
\begin{equation} \label{eqn:SPODEigen}
    \int_{\Omega} \mathbf{S}_{\tau}(\xx,\xx',\omega) \hat{\phi}_{\tau}(\xx',\omega) \, d\xx' = \lambda_{\tau}^2 \hat{\phi}_{\tau}(\xx,\omega),
\end{equation}
where $\hat{\phi}_{\tau}(\xx,\omega) = \int_{-\infty}^{+\infty} \phi_{\tau} e^{- \mathrm{i} \omega t} dt$.
The tensor $\mathbf{S}_{\tau}(\xx,\xx',\omega)$, being defined as the Fourier transform of $\mathbf{C}_{\tau}(\mathbf{x},\mathbf{x}',t-t')$, can be written as:
\begin{equation} \label{eqn:StauDef}
    \mathbf{S}_{\tau}(\xx,\xx',\omega) = E \left[ \hat{\qq}_{\tau}(\mathbf{x},\omega) \hat{\qq}_{\tau}^* (\mathbf{x}',\omega) \right], 
\end{equation}
where the quantity $\hat{\qq}_{\tau}$ is the Fourier transform of $\qq'_{\tau}$ and can alternatively be seen as the short-time Fourier transform of $\uu'$ (see \citet{griffin1984signal}):
\begin{equation} \label{eqn:ShortTime}
    \hat{\qq}_{\tau}(\omega) \equiv \mathcal{F}_{w}(\qq')(\tau,\omega) = \int_{-\infty}^{\infty} \qq'_{\tau}(t) e^{- \mathrm{i} \omega t} dt = \int_{-\infty}^{\infty} w(t-\tau) \qq'(t) e^{- \mathrm{i} \omega t} dt .
\end{equation}

The present approach is from now on referred to as the Phase Conditioned Localized Spectral Proper Orthogonal Decomposition (or PCL-SPOD). 


\subsubsection{A Phase Conditioned Localized Resolvent Analysis}\label{sec:PCLresolvent}

In this section, we introduce the Resolvent analysis, capable to reconstruct the fluctuation $\qq'(t)$ and hence the PCL-SPOD modes $\hat{\phi}_{\tau}$. To do so, we suppose that the flowfield $\qq$, composed by the velocity and pressure fields $\qq=(\uu,p)$, is given by the incompressible (non-dimensional) Navier-Stokes equations:
\begin{equation} \label{eqn:NS}
	\partial_t \uu + \uu \cdot \bnabla \uu + \bnabla p - \bnabla \cdot ( Re^{-1}(\bnabla \uu + \bnabla \uu^T) ) = \mathbf{0}, \;\;\;\; \bnabla \cdot \uu = 0,
\end{equation}
Applying the decomposition introduced in \eqref{eqn:PhaseDecomposition}, we can derive the following equation for ${\mathbf q}'=(\uu',p')$ (see \citet{reynolds1972mechanics}) 
\begin{equation} \label{eqn:BroadbandNS}
    Q \partial_t \qq' + L(\tau) \qq' = P \ff',
\end{equation}
where $Q$ and $P$ are two linear operators such that $Q{\mathbf q}'=P{\mathbf u}'=({\mathbf u}',0)^T $,
$L(\tau)$
is the linearized Navier-Stokes operator around the deterministic flow $\phuu (\tau)$, varying with the slow time-scale $\tau$, and $\ff' = \bnabla \cdot \left( \left \langle \uu' \otimes \uu' \right \rangle - \uu' \otimes \uu' \right)$ is the nonlinear forcing term stemming from the turbulent fluctuations $\uu'$. In order to analyse the dynamics in a small window around a phase $\tau$, similarly to what was done before, we consider the short-time Fourier transform of equation \eqref{eqn:BroadbandNS}. First, by applying \eqref{eqn:ShortTime} to the time derivative $\partial_t {\mathbf q}'$ (we omit the operator $Q$ for clarity), we have:
\begin{subequations}
    \begin{align}
        \mathcal{F}_{w}(\partial_t {\mathbf q}') (\tau,\omega) & =
        \int_{-\infty}^{\infty} w(t-\tau) \partial_t {\mathbf q}' e^{- \mathrm{i} \omega t} dt = - \int_{-\infty}^{\infty} {\mathbf q}' \partial_t \left( w(t-\tau)  e^{- \mathrm{i} \omega t} \right) dt \\
        & = \int_{-\infty}^{\infty} {\mathbf q}' \left( \partial_{\tau} w(t-\tau) + \mathrm{i} \omega w(t-\tau) \right) e^{- \mathrm{i} \omega t} dt = \partial_{\tau} \hat{\mathbf q}_{\tau} + \mathrm{i} \omega  \hat{\mathbf q}_{\tau}
    \end{align}
\end{subequations}
This term, together with the remaining ones in equation \eqref{eqn:BroadbandNS}, leads to:
\begin{equation} \label{eqn:BroadbandNSfreq}
    Q \partial_{\tau} \hat{\mathbf q}_{\tau} + \mathrm{i} \omega Q \hat{\mathbf q}_{\tau} + L(\tau) \hat{\qq}_{\tau} = P \hat{\ff}_{\tau}.
\end{equation}
Although this equation can define a linear transfer function between the forcing and the fluctuation, its resolution can be costly since one has to integrate it in time. Moreover, questions related to the initial/final conditions of $\hat{\mathbf q}_{\tau}$ (due to the compact support of the window $w(\zeta)$) complexify the analysis of it. However, similarly as for the PCL-SPOD, we use the hypothesis that the frequency for this quantity is much larger than the one from the slow time-scale $\tau$. For this reason, it is reasonable to drop out the term $Q \partial_{\tau} \hat{\mathbf q}_{\tau}$ in comparison with $\mathrm{i} \omega Q \hat{\mathbf q}_{\tau}$, leading to the following linear input/output relationship:
\begin{equation} \label{eqn:BroadbandNSfreq}
    \left( \mathrm{i} \omega Q + L(\tau) \right) \hat{\mathbf q}_{\tau} = P \hat{\ff}_{\tau},
\end{equation}
or, in a more compact form:
\begin{equation} \label{eqn:BroadbandNSfreq}
    \hat{\mathbf u}_{\tau} = R_\tau \hat{\ff}_{\tau},
\end{equation}
where $R_{\tau} = P^T (\mathrm{i} \omega Q + L(\tau))^{-1} P$ is the resolvent operator (
the quantity $ (\cdot)^{-1}$ refers to the solution of the linear problem and 
$P^T$ is the restriction operator such that $P^T\hat{\mathbf q}_{\tau}=\hat{\mathbf u}_{\tau}$).
The properties of this input/output relationship can be studied by identifying (\citet{mckeon2010critical,beneddine2016conditions}) its most energetic input-output modes. We thus want to maximize the following gain:
\begin{equation}
    G_{\tau} = \frac{ \int_{\Omega} || \hat{\uu}_{\tau} ||^2 d\xx }{ \int_{\Omega} ||\hat{\ff}_{\tau} ||^2 d\xx } = \frac{\hat{\uu}_{\tau}^* W_\Omega \hat{\uu}_{\tau} }{ \hat{\ff}_{\tau}^*W_\Omega \hat{\ff}_{\tau}}.
\end{equation}
In the last equality, we have introduced as well the spatial discretization (where, for clarity the same notation for the spatial functions and discrete vectors are maintained), for which the integration over $\Omega$ is given by the weight-matrix $W_\Omega$. This leads to the final following eigenvalue problem:
\begin{equation} \label{eqn:SVDResolvent}
R^{*}_{\tau} W_\Omega R_{\tau}  \hat{\mathbf{g}}_{\tau} = \mu_{\tau}^2 W_\Omega \hat{\mathbf{g}}_{\tau},
\end{equation}
where $R^{*}_{\tau}$ is the transconjugate of the matrix $R_{\tau}$.
 The optimal forcing $\hat{\mathbf{g}}_{\tau}$ and optimal gain $ \mu_{\tau}^2 $ are linked to the optimal response mode $\hat{\psi}_{\tau}$ through:
$ \hat{\psi}_{\tau} = \mu_{\tau}^{-1}R_{\tau} W_\Omega \hat{\mathbf{g}}_{\tau}$.

\subsection{Periodic external conditional signal}\label{Theory:Periodic}

In the case of a periodic conditional signal of period $ T_0$, by ergodicity, all expected values over realizations can be replaced by averages over periodic segments of the conditional signal.
Hence, the phase $ \tau $ is now within $ (0,T_0)$ and a single realization is now sufficient to define 
$\phqq (\mathbf{x},\tau)$ and the correlation tensors $\mathbf{C}_{\tau}$, $\mathbf{S}_{\tau}$.

\subsubsection{Extraction of periodic component $\phqq (\mathbf{x},t)$} \label{Theory:ExtractionPeriodic}

If an external {\it periodic} conditional signal of period $ T_0$ is available, the signal $\uu(t)$ of a given realization may be divided in $N_b$ bins of temporal length $T_0$. In each bin $k$, the time referring to the phase $ \tau \in (0,T_0) $ will be denoted $ \tau_k=\tau+kT_0$. Then we may define the phase-average (\citet{reynolds1972mechanics,hussain1983coherent}):
\begin{equation}
  \phqq (\mathbf{x},\tau)=\frac{1}{N_b} \sum_{k=1}^{N_b} \qq(\mathbf{x},\tau_k).
\end{equation}
We may think here, for example, of rotating blades, where the period $ T_0 $ is given by the rotation speed of the blades. 

In the case of statistically stationary flows, we resort to the Koopman decomposition \eqref{eqn:Koopman} for the definition of $ \phqq$ and assume in the following that this field is periodic of period $ T_0 $.

\subsubsection{Analysis of high-frequency fluctuations with PCL-SPOD}\label{Theory:PeriodicImplementation}

\begin{figure}
	\centering
	\Large
	$$\vdots$$ \\
	\includegraphics[width=0.9\textwidth]{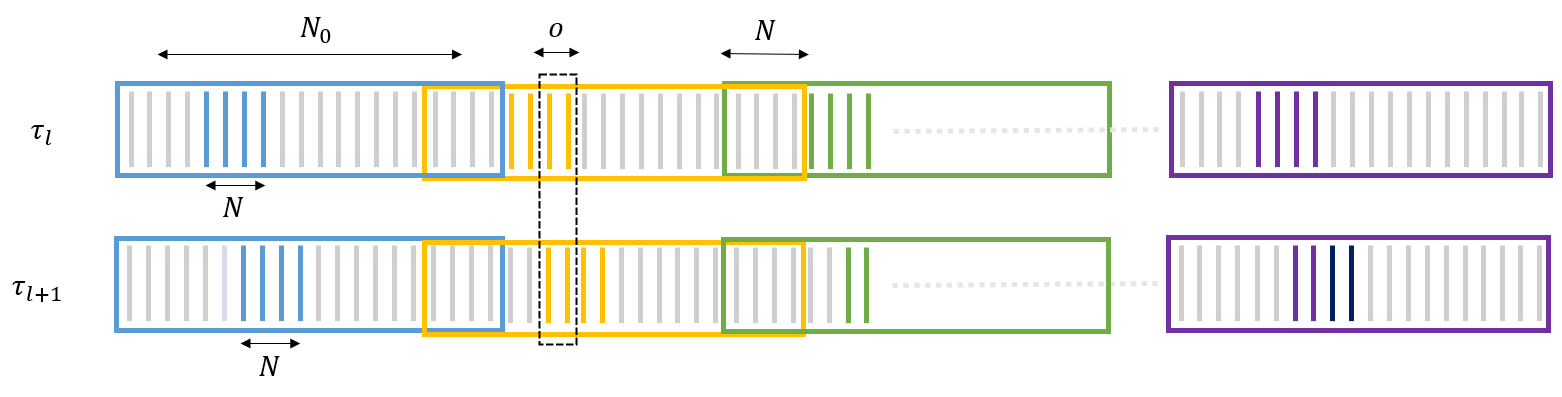} \\
    \Large
	$\vdots$
    \caption{Principle sketch of the Phase Conditioned Localized-SPOD using Welch algorithm. The total amount of snapshots are ranged in $N_b$ bins (indicated by different colors) of size $N_0+N$ snapshots (vertical bars) separated by $\Delta t$, with $T_0 \approx N_0 \Delta t$ the slow period and $T \approx N \Delta t$ the support of the window function. The SPOD algorithm is then performed around each phase $\tau_{l}$, represented by the bars in colors. $o$ is the number of overlapping snapshots for the definition of the time instants $\tau_l$.}\label{fig:PCL-SPOD}
\end{figure}

In this section, we describe how to perform the PCL-SPOD for periodic flows $\phqq$ (of period $T_0$), as it is the case of the configuration of interest.
The approximation of the Spectral Correlation tensor $\mathbf{S}_{\tau}$ defined in \eqref{eqn:StauDef}, necessary for the PCL-SPOD, relies on the Welch method (see \citet{welch1967use,bendat2011random}),
which allows to converge the statistics with data collected from a single DNS run. The signal of interest ($\qq'$ or $\qq'_{\tau}$) is divided in $N_b$ overlapping bins of temporal length $T_0+T$, as sketched in Figure \ref{fig:PCL-SPOD}. Half of the period $T/2$ is added to each side of the bin to accommodate for the half time-window required to handle the phases $ 0 \leq \tau < T/2 $ and $ T_0-T/2 < \tau < T_0$. 

With that in mind, the spectral tensor $\mathbf{S}_{\tau}$ defined in \eqref{eqn:StauDef}
can be approximated as 
\begin{equation}
    \mathbf{S}_{\tau}(\xx,\xx',\omega) \approx \frac{1}{N_b} \sum_{k=1}^{N_b} \hat{\qq}_{\tau_k}(\omega) \hat{\qq}^{*}_{\tau_k}(\omega) \equiv \hat{\QQ}_{\tau} \hat{\QQ}_{\tau}^*,
\end{equation}
where $\hat{\qq}_{\tau_k}(\omega)$ is, now, the discrete Short-Time Fourier Transform (STFT), evaluated for each bin $k$ and computed using the standard Hann window for $w(\zeta)$. {The short-time (discrete) Fourier Transform is computed using the signal treatment library from scipy. The inputs are the temporal series $\qq'(t_{l=0,1,\cdots})$ of the $N_b$ bins, the size $N$ of the window $T/\Delta t$ and the (discrete) overlap $ o $ that we allow between the windows $[\tau-T/2,\tau+T/2]$. This last parameter is necessary for us to obtain a richer phase-discretization $ \tau $ (finer than having only $T_0/T$ phases, when no overlap is employed). The output of this analysis is the STFT, $\hat{\uu}_{\tau_i}(\omega_j)$, evaluated on a given set of phases $0 \leq \tau_i < T_0 $ and on a frequency grid $\omega_{j=0,1,\cdots}=j2\pi/T$.} 
This data can be rearranged, for simplicity, in the matrix $\hat{\QQ}_{\tau}$, containing a collection of those quantities, for every bin, stacked in columns and normalized by $1/\sqrt{N_b}$. The PCL-SPOD can then be obtained by solving the following matrix eigenproblem:
\begin{equation} \label{eqn:SPOD}
    \hat{\QQ}_{\tau} \hat{\QQ}_{\tau}^* W_{\Omega} \hat{\phi}_{\tau}(\omega) = \lambda_{\tau}^2 \hat{\phi}_{\tau}(\omega),
\end{equation}
where, as before, $W_{\Omega}$ is related to spatial integration over $\Omega$. It is worth-mentioning that this problem involves finding the eigenvalues/eigenvectors of a large dense matrix $\hat{\QQ}_{\tau} \hat{\QQ}_{\tau}^*$. Instead, we solve the following eigenvalue/eigenvector problem:
\begin{equation} \label{eqn:SPOD2}
    \hat{\QQ}_{\tau}^* W_{\Omega} \hat{\QQ}_{\tau} \hat{\mathbf{y}}_{\tau}(\omega) = \lambda_{\tau}^2 \hat{\mathbf{y}}_{\tau}(\omega),
\end{equation}
involving a much smaller matrix $\hat{\QQ}_{\tau}^* W_{\Omega} \hat{\QQ}_{\tau}$, whose dimension is the number of bins. The PCL-SPOD mode can then be recovered with $\hat{\phi}_{\tau} = \lambda_{\tau}^{-1} \hat{\mathbf{Q}}_{\tau} \hat{\mathbf{y}}_{\tau}$.


\subsection{Taking into account low-frequency meandering}\label{sec:Meandering}

In this paragraph we briefly address the meandering phenomenon, a known feature in cylinder flow (see \citet{lehmkuhl2013low}), whose characteristic frequency is much lower than the VS one. This phenomenon, is present in the spectrum shown in figure \ref{fig:Sketch}(a) where a small "bump" occurs at $\omega<10^{-1}$. The VS motion, $\phqq$, is thus modulated by this frequency, making it hard to be captured with harmonic averages \eqref{eqn:Harmonic}, which converge only on very large time spans, typically $T_f = O(10^3 T_0)$. However, if we were to compute the harmonic averages with $T_f=T_0$, we would obtain a non-converged quantity $\phuu_k$ that would be slightly different from one bin to the other. This means that, to obtain a meaningful fluctuation field $\qq_k'$ for a given bin $k$ (which mitigates the effect of this low frequency motion), it is desirable to use the periodic field computed with snapshots restricted to that bin, namely $\qq_k' = \qq_k - \phqq_k$. This procedure, preferred here, has shown to be more effective than using $\phqq$, computed with the totality of our signal (which corresponds to only $40T_0$ since the focus of this work is the capture of high-frequency KH).

\section{Results for the squared-section cylinder}\label{Results}

The configuration corresponds to a squared-section cylinder at Reynolds number $Re=U_{\infty} D /\nu = 22000$, where $ D $ is the cylinder's diameter and $U_{\infty}$ the incoming uniform velocity. These two reference scales are used to non-dimensionalize all quantities in the following. 

The DNS code solves the three-dimensional compressible Navier-Stokes equations, with a standard Sutherland law for the dynamic viscosity. We use the FastS code, developed by ONERA, which is a highly optimised solver for high performance computing clusters. For further details, see \citet{dandois2018large} and references herein. The code is run at a low inflow Mach number, $M=0.1$, to be close to an incompressible flow regime. The spatial discretization used in the solver corresponds to a second-order accurate finite-volume method based on a modification of the AUSM+(P) scheme \citep{mary2002large}. The time-integration is handled with a second-order accurate backward scheme of Gear, with a time step of $ 3.3\times 10^{-4}$. The size of the simulated time-window (after an initial transitory phase was convected away) was of around 300 time units, corresponding to approximately 40 vortex shedding periods. The spatial domain for the DNS consists of a circle of diameter $100$. This domain is discretized with a mesh built by extruding a 2D mesh, of around $255 \times 10^3$ cells, clustered around the cylinder, along 4 diameters in the span, discretized with equally-spaced 960 cells.

A typical signal generated by this simulation was already depicted in figure \ref{fig:Sketch}, presenting the Vortex-shedding (VS) frequency, whose peak occurs at $\omega_0 \approx 0.837$ (Strouhal number of $St=0.133$, in accordance with \citet{trias2015turbulent}), and also the Kelvin-Helmholtz (KH) structures at higher frequencies. The spanwise-averaged quantities (velocity and pressure) were stored on disk every $\Delta t = 0.0209 $, which corresponds to a sampling frequency discretizing the frequency $\omega=30$ with 10 points. 

\begin{figure}
	\centering
	\raisebox{0.5in}{(a)}\includegraphics[trim={0.5cm 0.5cm 0.5cm 0.5cm},clip,height=80px]{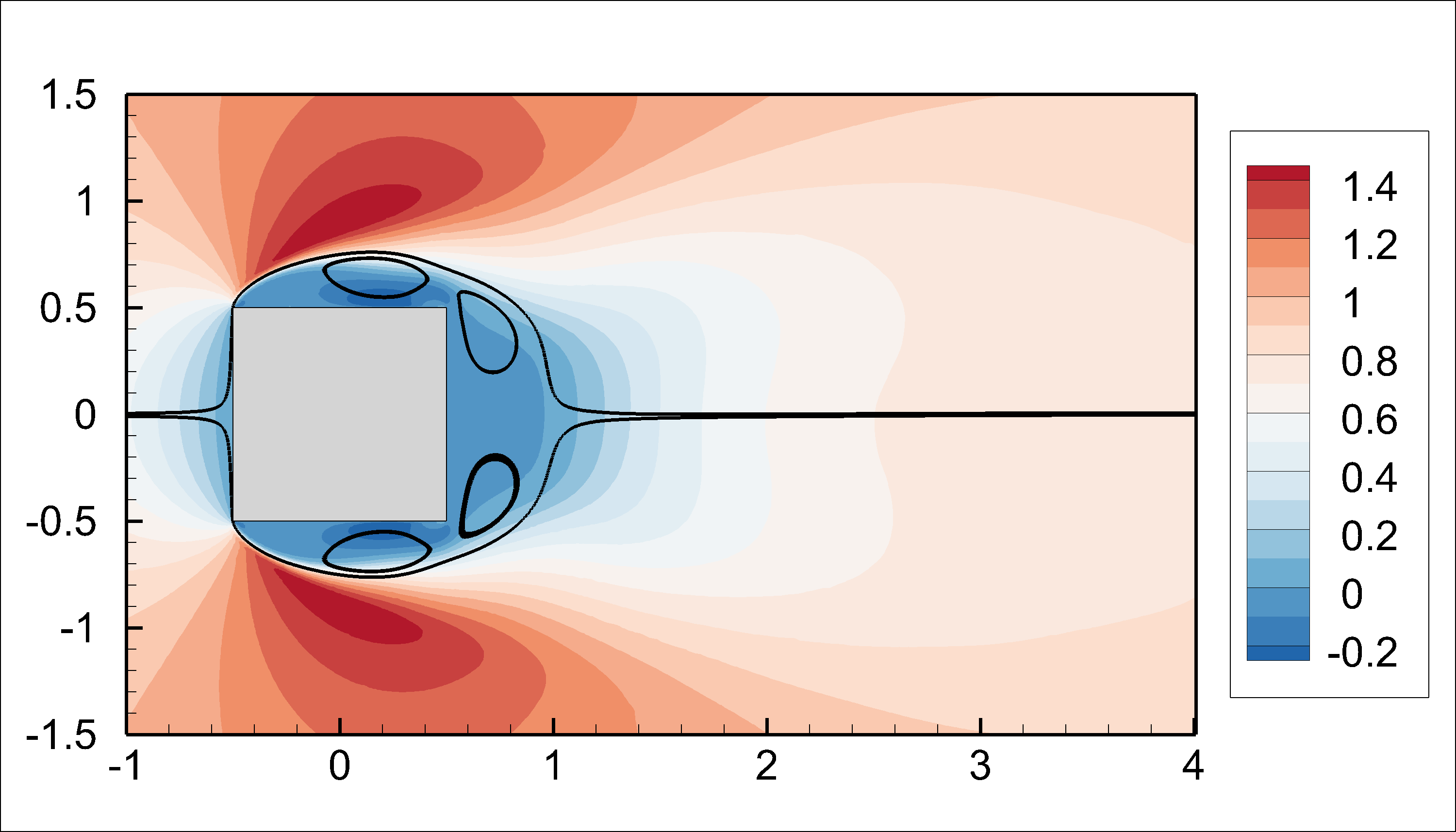}
	\raisebox{0.5in}{(b)}\includegraphics[trim={0.5cm 0.5cm 0.5cm 0.5cm},clip,height=80px]{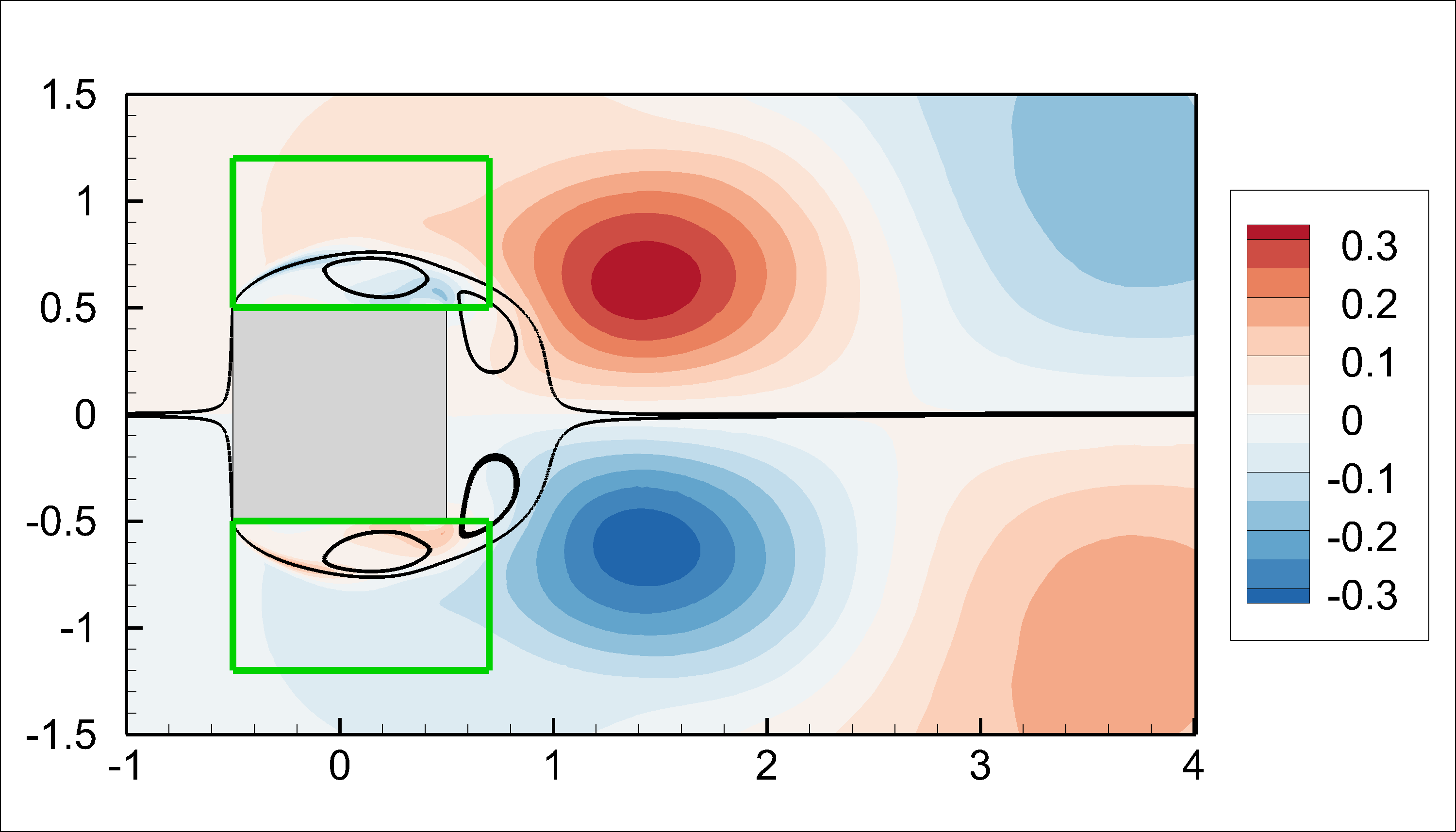}
	\caption{(a) Mean-flow and (b) first harmonic of periodic component (VS). Both were computed from a time series of around 40 periods. The first harmonic was computed by the harmonic-averaging procedure on the frequency $\omega_0=0.837$, for which the PSD shown in Fig. \ref{fig:Sketch}(a) was maximum. The green window represents the integration region $\Omega$ (defining $W_\Omega$) used for both the PCL SPOD and Resolvent analysis.}\label{fig:Harmo}
\end{figure}

We have computed the mean-flow and first 4 harmonics (representative of the VS mode) using harmonic-averages on the totality of the DNS run, $T_f=40 T_0$. 
The streamwise component of the mean-flow and the first harmonic
have been represented in Figure \ref{fig:Harmo}. 
The periodic component presents a space/time symmetry on $\phqq$, namely $(\phu,\phv,\php)(x,y,t)=(\phu,-\phv,\php)(x,-y,t+T_0/2)$ (see for instance \cite{jallas2017}). It states that, at a time $t$ , it corresponds to the (quasi) mirror-image (with respect to the symmetry axis $y=0$) of the flow half-a-period later, at time $t+T_0/2$.

In the next section, we discuss the PCL-SPOD procedure implemented to unveil the high-frequency dynamics, followed by the PCL-resolvent analysis, to model them.
 
\subsection{PCL-SPOD results}\label{sec:SPODres}

\begin{figure}
    \centering
    \begin{tabular}{llll}
          &  &  & \\
    \raisebox{0.5in}{(a)}
    \includegraphics[trim={1cm 10cm 30cm 10cm},clip,height=80px]{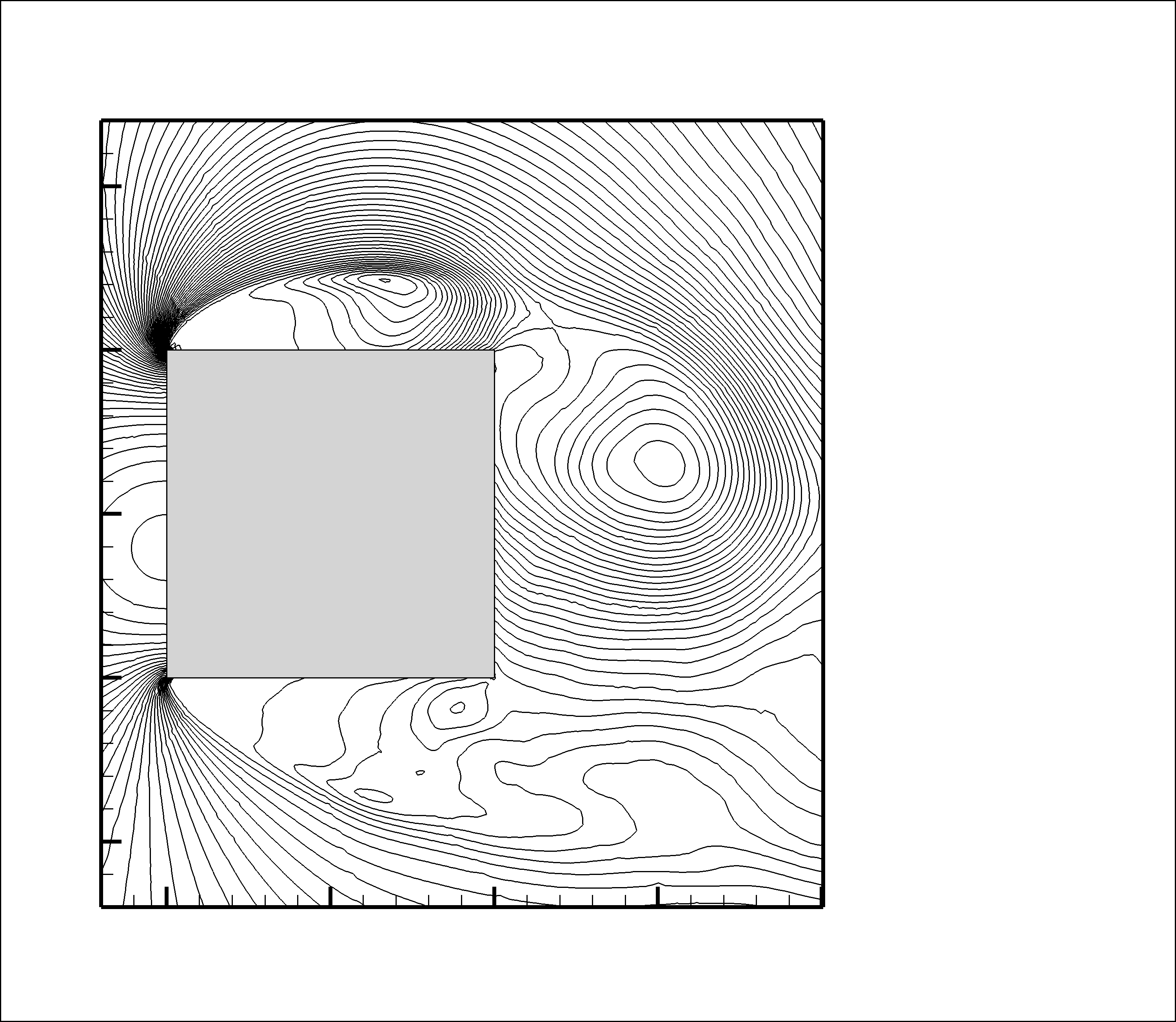}&
    \includegraphics[trim={8cm 10cm 30cm 10cm},clip,height=80px]{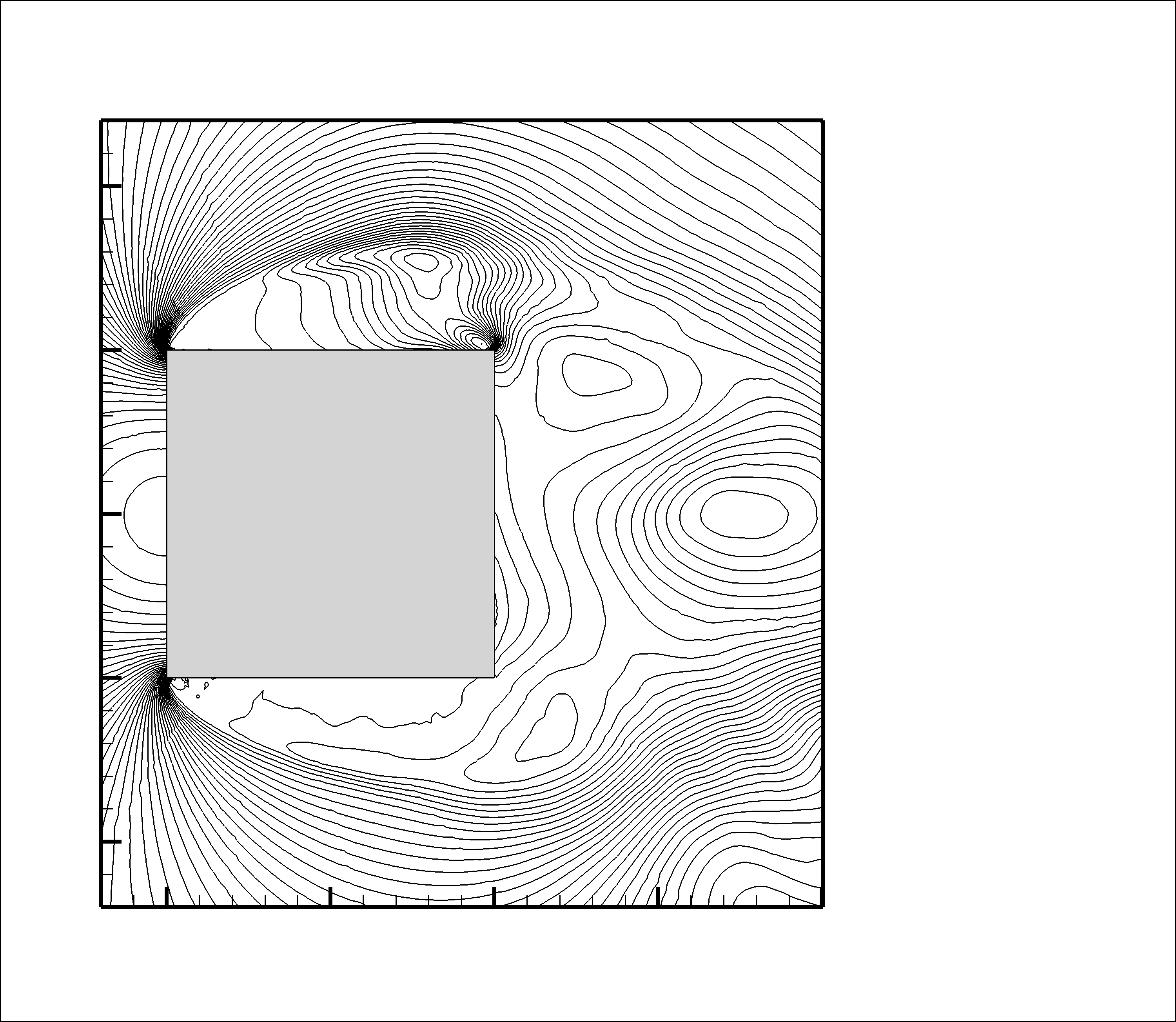}&
    \includegraphics[trim={8cm 10cm 30cm 10cm},clip,height=80px]{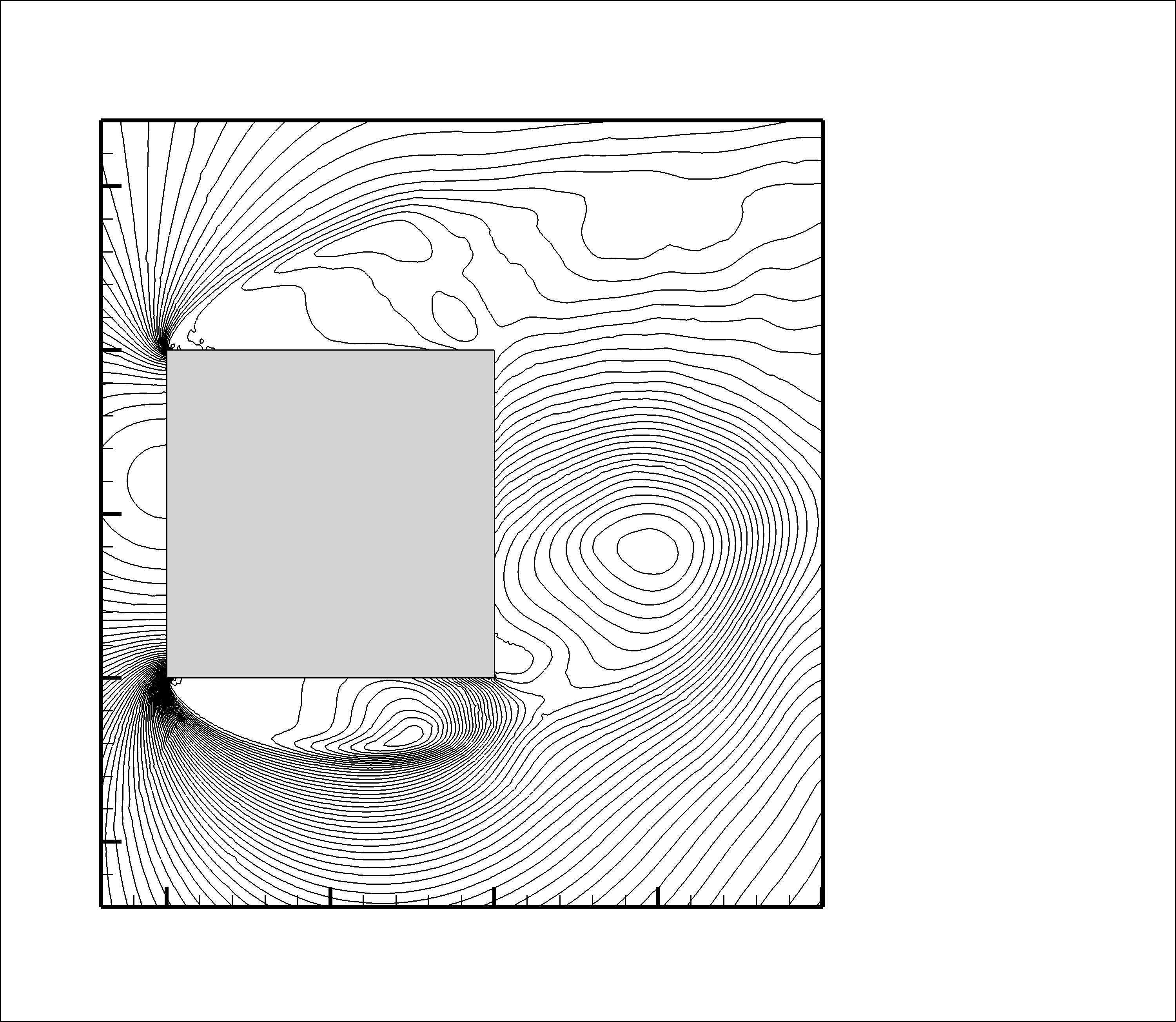}&
    \includegraphics[trim={8cm 10cm 30cm 10cm},clip,height=80px]{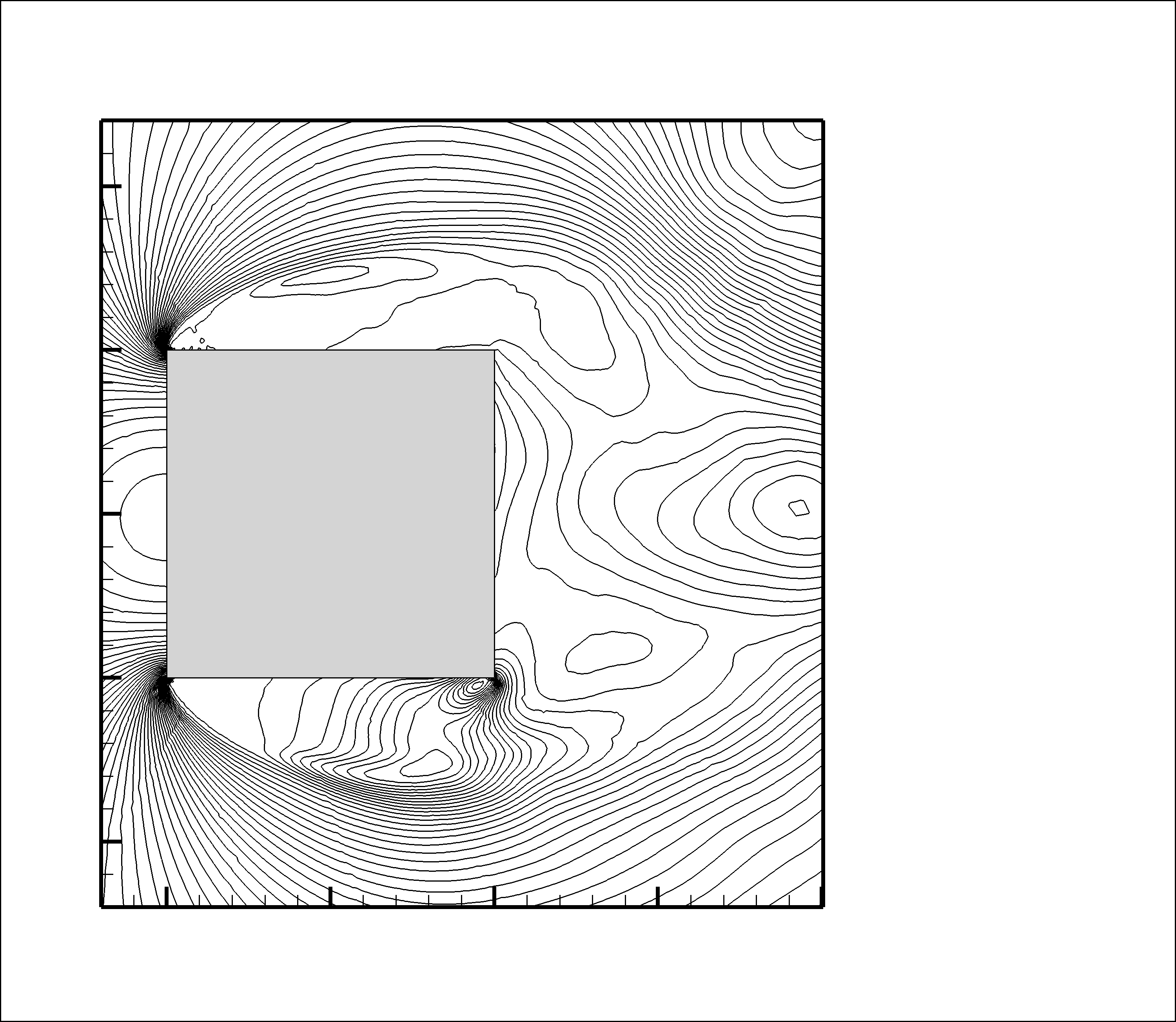} \\
    \raisebox{0.5in}{(b)}
    \includegraphics[trim={1cm 10cm 30cm 10cm},clip,height=80px]{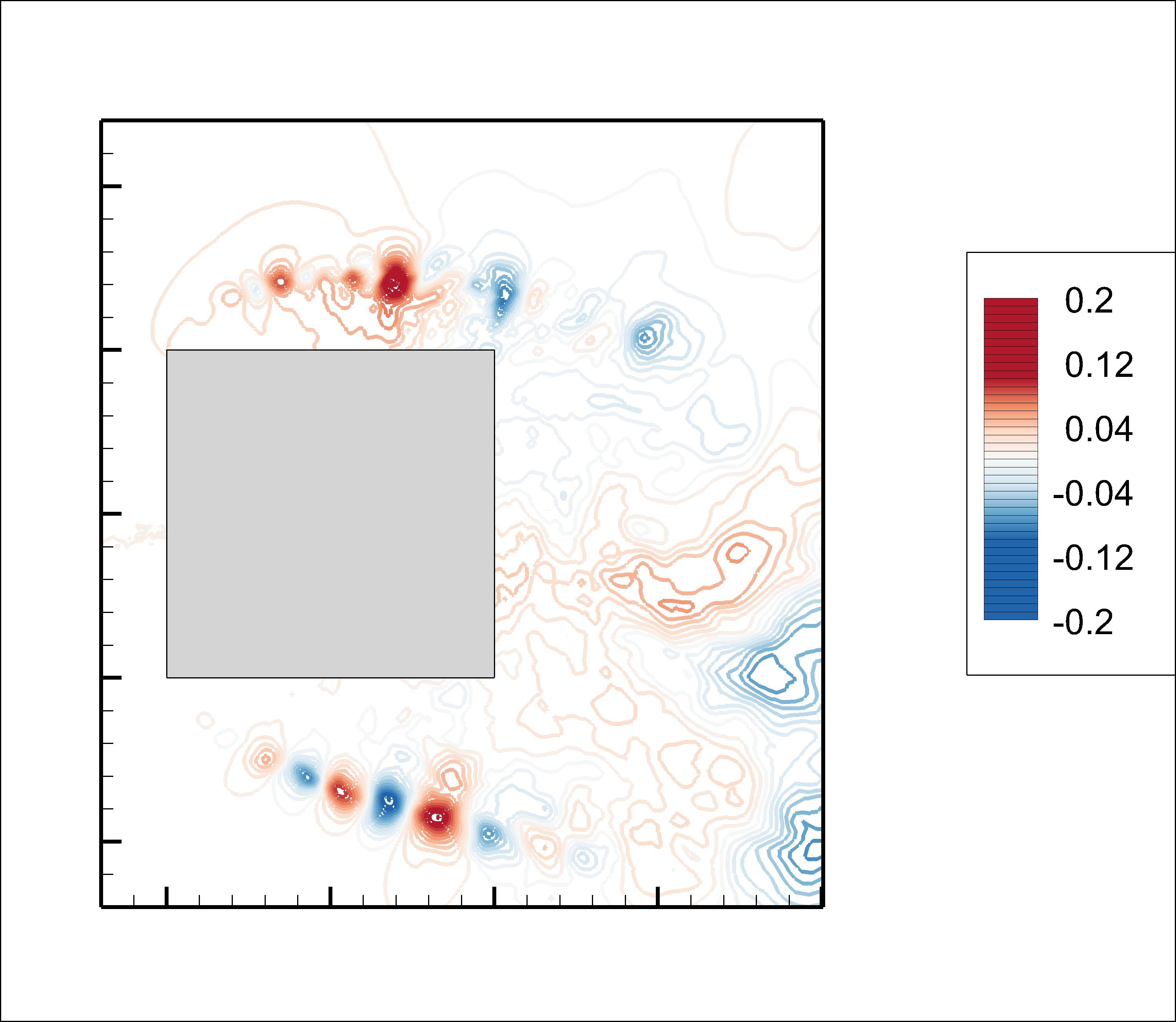}&
    \includegraphics[trim={8cm 10cm 30cm 10cm},clip,height=80px]{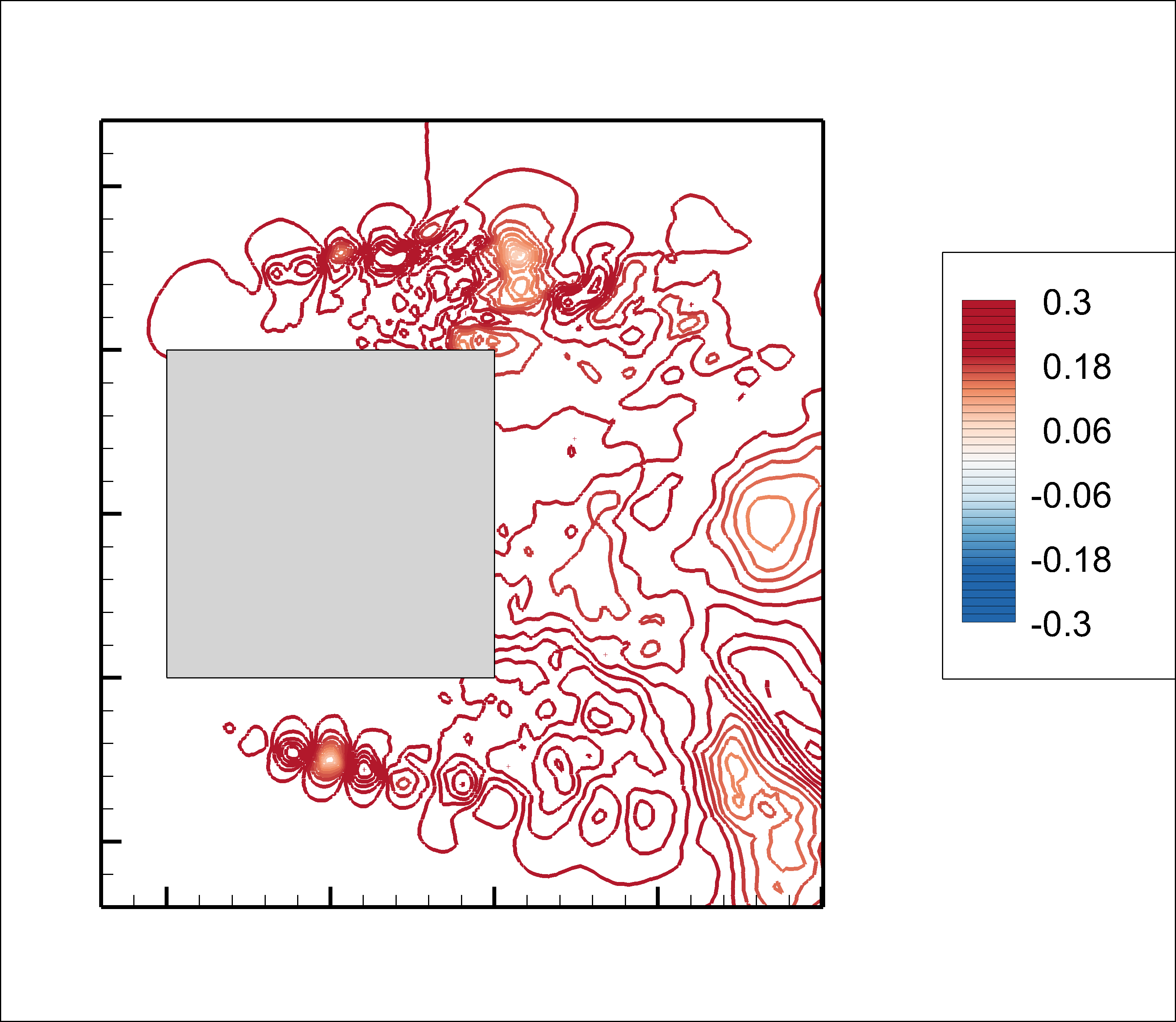}&
    \includegraphics[trim={8cm 10cm 30cm 10cm},clip,height=80px]{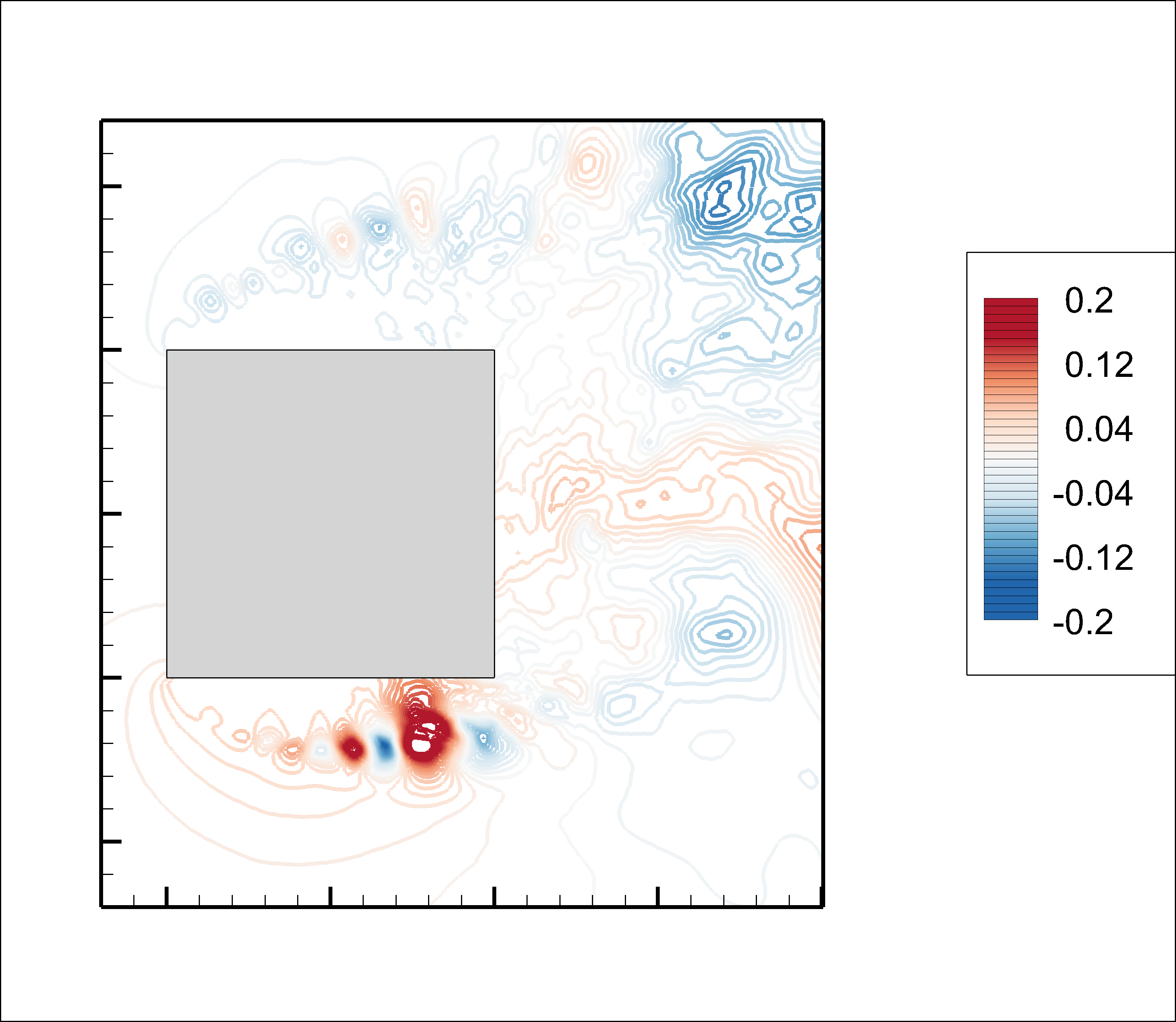}&
    \includegraphics[trim={8cm 10cm 30cm 10cm},clip,height=80px]{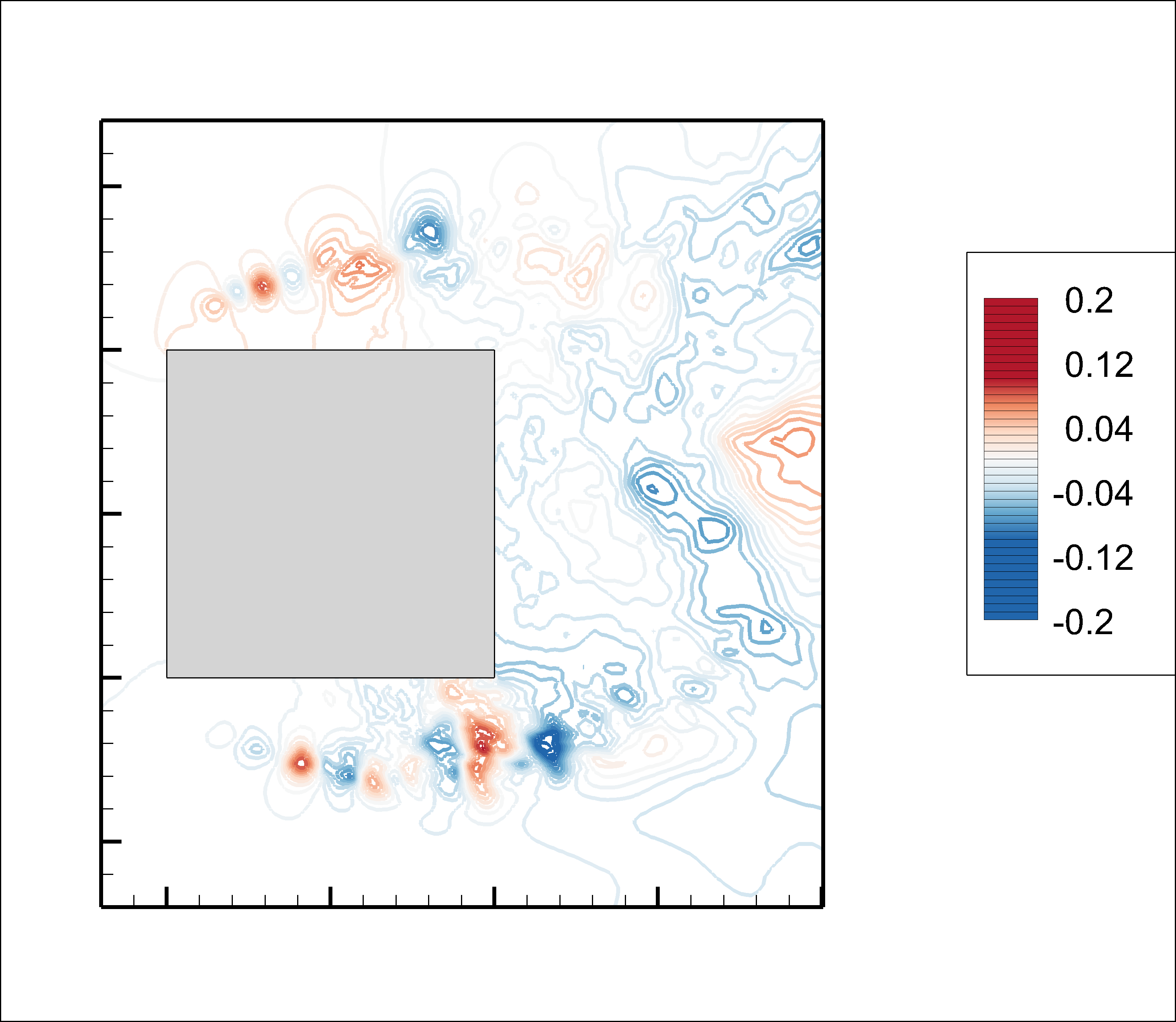} \\
    \end{tabular}
    \caption{Deterministic periodic spanwise-averaged pressure fluctuation field, computed from harmonic-averages from a given bin $\phqq$ (a) and the fluctuation $\qq'$, obtained by subtracting the raw snapshots from the deterministic field (b). Both series of plots are presented for the same phases as in figure \ref{fig:Sketch}(b). 
    }
	\label{fig:fluctuation}
\end{figure}

In this section, we present the PCL-SPOD results. However, before doing so, we briefly discuss the procedure to separate $\qq'$ and $\phqq$ from $\qq$ within a given bin. In figure \ref{fig:fluctuation}, we provide in (a) the deterministic periodic field (the span-wise pressure field), computed based on the mean-flow and first four harmonic averages based on snapshots just taken from the given bin. We can see that the VS motion is properly recovered and that all small scale features seen in the raw data $\qq$ (figure \ref{fig:Sketch}) for the exact same phases have been removed. In figure (b), we provide the fluctuation field $\qq'$, computed by subtracting $\phqq$ (figure (a)) from the raw snapshots $\qq$ (figure \ref{fig:Sketch}(b)). We can see that the large-scale vortices associated to the VS fluctuation field have been removed. The smooth and regular lines above and below the cylinder in (a) together with the large-scale structures highlighted by red circles in figure \ref{fig:Sketch}(a) are now gone. The remaining structures are composed mostly of more complex high-frequency fluctuations, which still clearly display its dependency on the VS motion. We remark, however, that, in some phases (namely the second phase), some uncoherent large-structures may still be present, making, for example, its colorbar to tilt toward the positive range, in red. In what follows, the PCL-SPOD will further filter some of those uncoherent motion, especially at high frequency.


\begin{figure}
	\centering
    \raisebox{1in}{(a)} \includegraphics[trim={0.1cm 0.1cm 0.1cm 0.1cm},clip,height=100px]{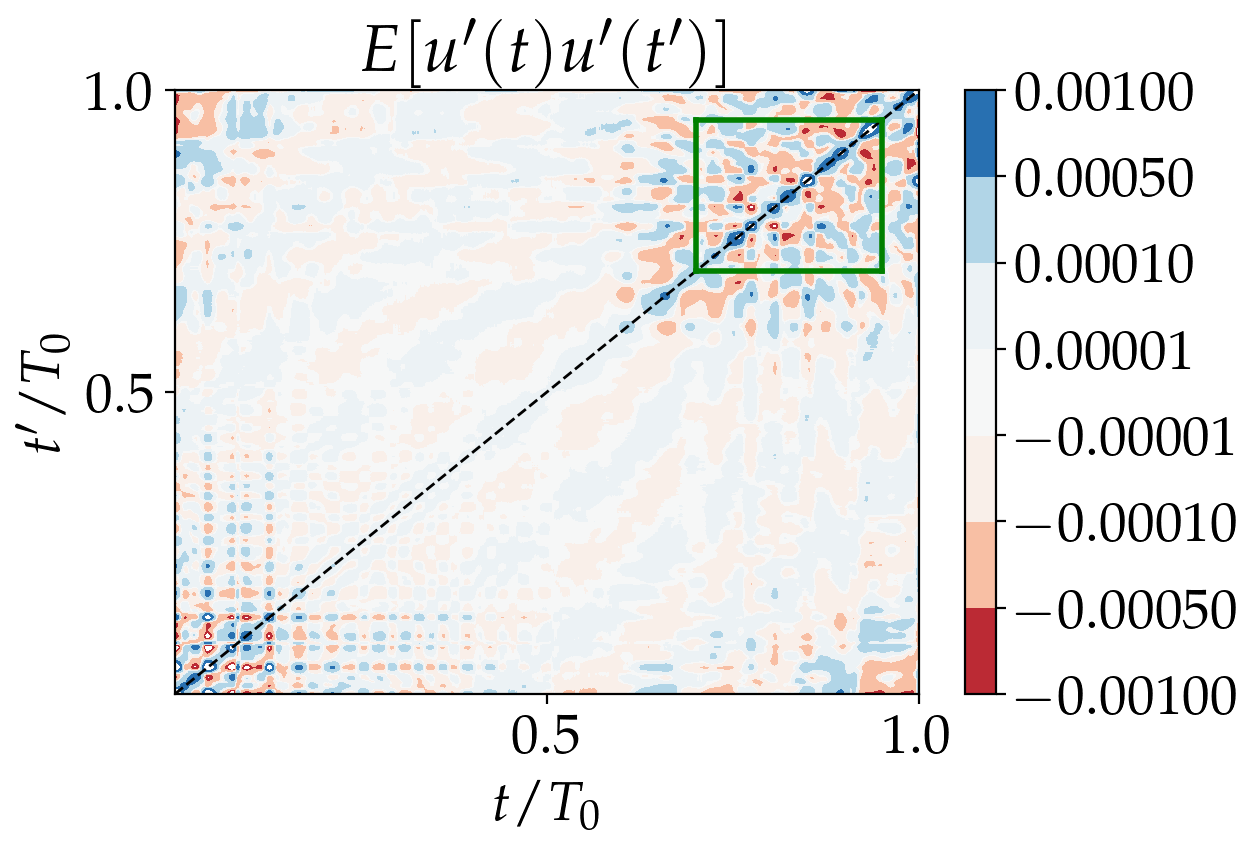} \raisebox{1in}{(b)} \includegraphics[trim={0.1cm 0.1cm 0.1cm 0.1cm},clip,height=100px]{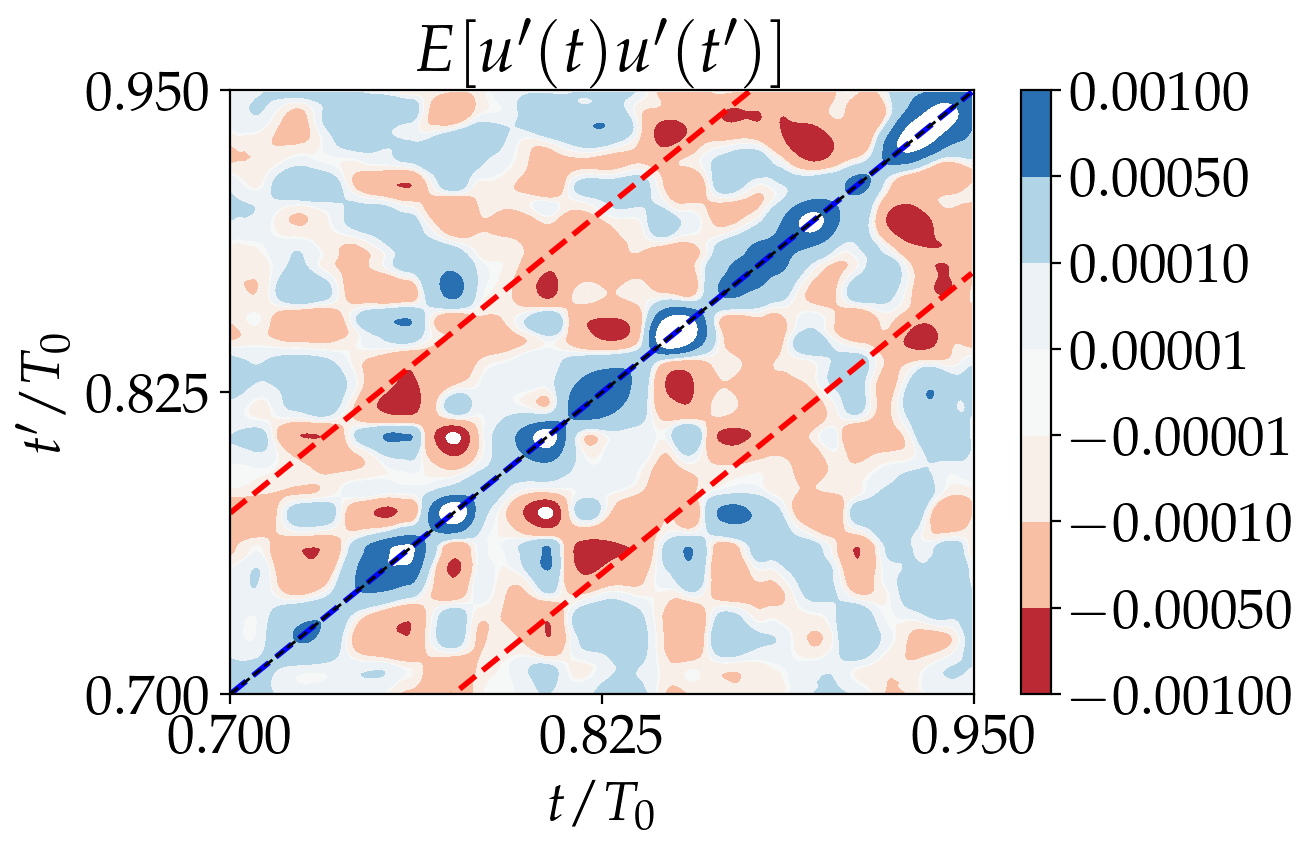}
    
	\caption{Autocorrelation (a) for the signal shown in figure \ref{fig:Sketch}(a) for the fluctuation $\qq'$, motivating the approximation $\mathbf{C}_{\tau}(t,t') \rightarrow \mathbf{C}_{\tau}(t-t')$. Zoom in the green window (b) is also provided.} \label{fig:Correlations}
\end{figure}

The PCL-SPOD analysis relies on the hypothesis that the tensor $\mathbf{C}_{\tau}(t,t')$ may be approximated by $\mathbf{C}_{\tau}(t-t')$ for short time distances $ |t-t'|$. In order to assess this approximation, we  plot in figure \ref{fig:Correlations} the quantity $E[u'(t) u'(t')]$ for the probe whose signal is represented in figure \ref{fig:Sketch}(a) (the expected value refers to averages over 40 non-overlapping bins, each of size $T_0=2\pi/\omega_0$, where $\omega_0$ corresponds to the highest value in the PSD shown in figure \ref{fig:Sketch}(a)). We can in particular see high-frequency structures at some specific phases in the period and very little signal in others. This is coherent with the observation from \citet{brun2008coherent} where, for a given fixed point on the shear-layer, the KH motion was characterized as an intermittent phenomenon, observed mainly at some phases of the period.
The quantity $E[u'_{\tau}(t) u'_{\tau}(t')]$ focuses the analysis on correlations within small windows $(t,t') \in [\tau-T/2,\tau+T/2]\times[\tau-T/2,\tau+T/2]$ (with $T \ll T_0$) around the principal diagonal, an example of which being given by the green window.
Accordingly, we remark that energetic regions on that figure change on a slow time-scale of the order $ T_0 $ and that correlations are approximately constant in windows of size $ T $ (green window) centred along the principal diagonal ($t=t'$, given by a dashed line). Despite the noisiness of the correlations (due to the difficulty of converging second-order statistics), we indeed distinguish blue and red diagonal segments (see in particular green window, zoomed in figure \ref{fig:Correlations}(b)), which indicate that $\mathbf{C}_{\tau}(t,t') $ approximately exhibits constant values along $t-t'=\text{cste}$ so that $\mathbf{C}_{\tau}(t,t')=\mathbf{C}_{\tau}(t-t') $. LF: je change legerement la figure (b), je mets bleu rouge rouge, et pas noire, bleur, rouge 

We now turn our attention to the PCL-SPOD results. They are obtained by considering the dataset split in 40 overlapping bins, each of size $T_0+T$. The short-time (fast) Fourier transform is computed for each bin using a time-window of size $ T = T_0/6$, leading to a fundamental frequency discretization of $\Delta \omega = 2 \pi / T = 6 \omega_0 \approx 5 $, meaning that the short-time Fourier modes are computed for the following frequencies, $j \Delta \omega, j=1,2,3,\cdots \approx 5,10,15, \cdots$. If the size of the window was smaller, the frequency discretization would be much poorer. 
The PCL-SPOD is subsequently performed by solving the eigenvalue problem \eqref{eqn:SPOD} using the integration domain $\Omega=(-0.5 \leq x \leq 0.7) \times (0.5 \leq y \leq 1.2) \cup (-0.5 \leq x \leq 0.7) \times (-1.2 \leq y \leq -0.5)$, shown with green boxes in figure \ref{fig:Harmo}.

\begin{figure}
	\centering
	\begin{tabular}{cccc}
        & $\max_{\tau}\lambda_{\tau,0}^2$ & & $\sum_{k=1}^{N_b} \frac{1}{T_0} \int_0^{T_0} \lambda_{\tau,k}^2 d\tau $ \\ 
        \raisebox{1in}{(a)} & \includegraphics[trim={1cm 1cm 10cm 22cm},clip,height=90px]{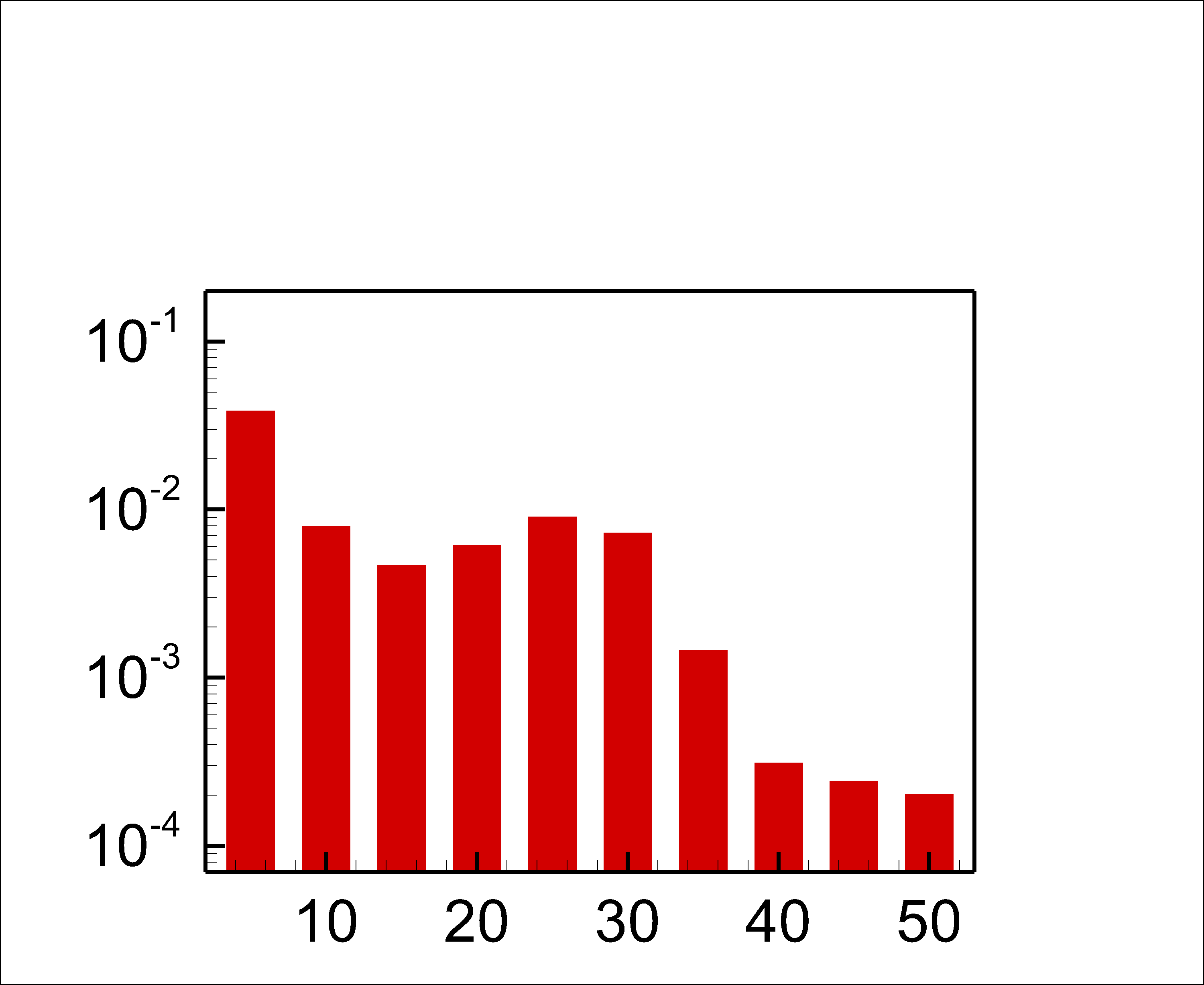} & \raisebox{1in}{(b)} & \includegraphics[trim={1cm 1cm 10cm 22cm},clip,height=90px]{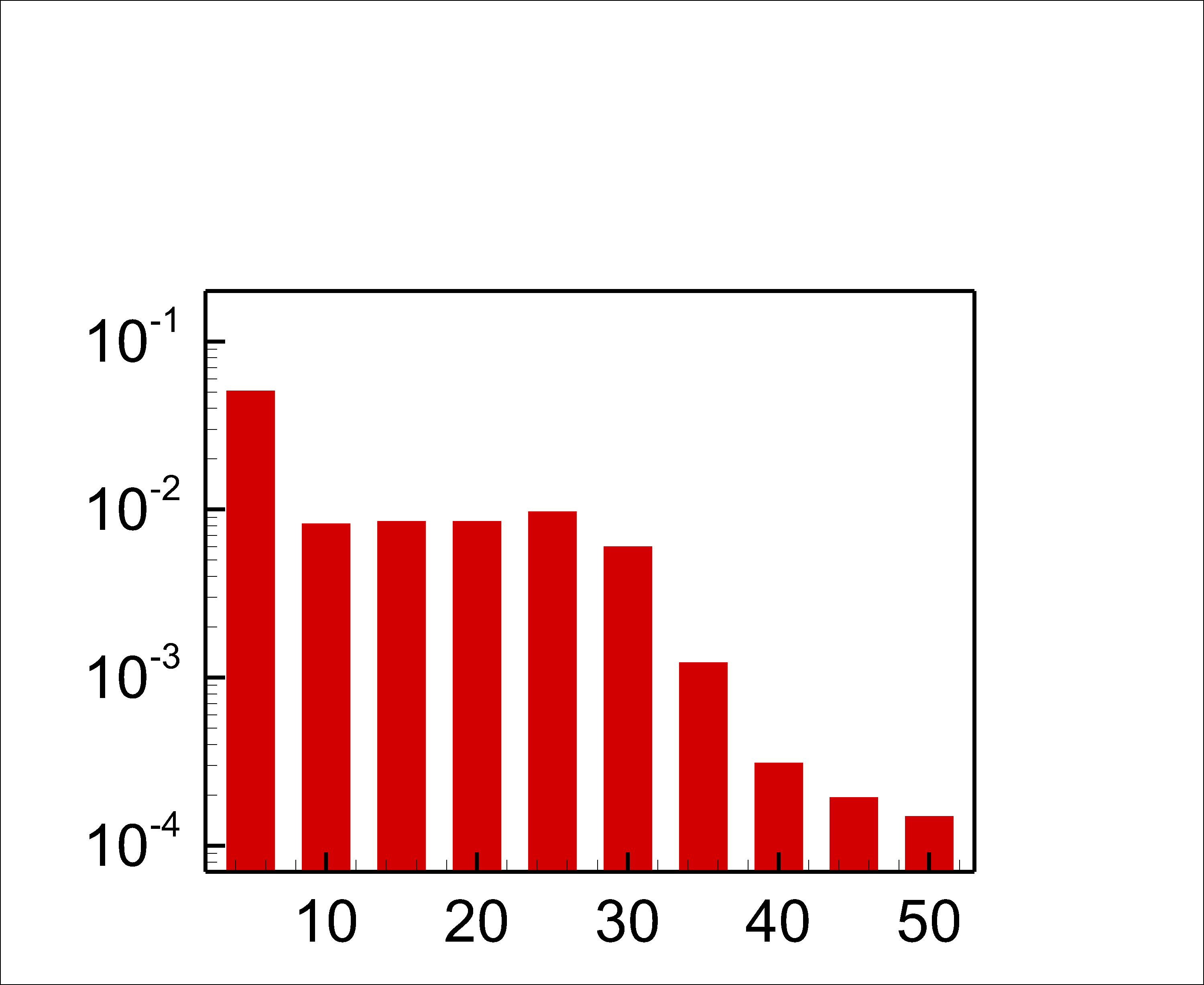} \raisebox{0.1in}{$\omega$}
    \end{tabular}
    \caption{PCL-SPOD: (a) maximum value of optimal energy gain $\lambda_{\tau}^2$ over $0 \leq \tau < T_0$ as a function of frequency and (b) sum of all energy gains averaged over the same time interval as a function of frequency.}\label{fig:SPODgains}
\end{figure}

In figure \ref{fig:SPODgains}, we present some characteristics of the energy distribution of the PCL-SPOD modes as a function of frequency $\omega$. Figure \ref{fig:SPODgains}(a) shows the maximal value of the dominant energy gain over $(0,T_0)$, i.e.  $\max_{\tau} \lambda_{\tau,0}^2$. A bump is clearly observed around frequencies $\omega=20,25$ and $30$, indicating  that the dominant optimal PCL-SPOD mode, which is the most coherent among all of them, exhibits its strongest features within this frequency band. We have also computed the sum of the eigenvalues $\lambda_{\tau,k}^2$ over all modes and averaged along the phase of the period. The distribution of this (full-) energy over frequency is shown in figure \ref{fig:SPODgains}(b). In contrast, this plot exhibits a "plateau" over $10 \leq \omega \leq 30$, suggesting that part of the energy content at frequencies $\omega=15,20$ stems from sub-optimal branches. This behavior will be discussed further in the following, especially for $\omega=20$ and $30$ for which a detailed analysis will be provided. The case $\omega=25$, the dominant one, will not be presented due to its similarities with the two others. We also remark that, for higher frequencies ($\omega \geq 35$), there is a clear cut-off in the energy content, while for the lowest frequency ($\omega=5,10$), we can see an energy increase of the dominant mode. However, we believe this large amount of energy is due to low-frequency dynamics around $\omega_0$ that cascades nonlinearly up to $\omega\approx 5,10$ and thus does not necessarily represent fluctuations arising from linear mechanisms triggered at high frequencies.

\begin{figure}
	\centering
	\raisebox{0.8in}{(a)}\raisebox{0.6in}{$\lambda_{\tau}^2$}\includegraphics[trim={1cm 1cm 7cm 4cm},clip,height=80px]{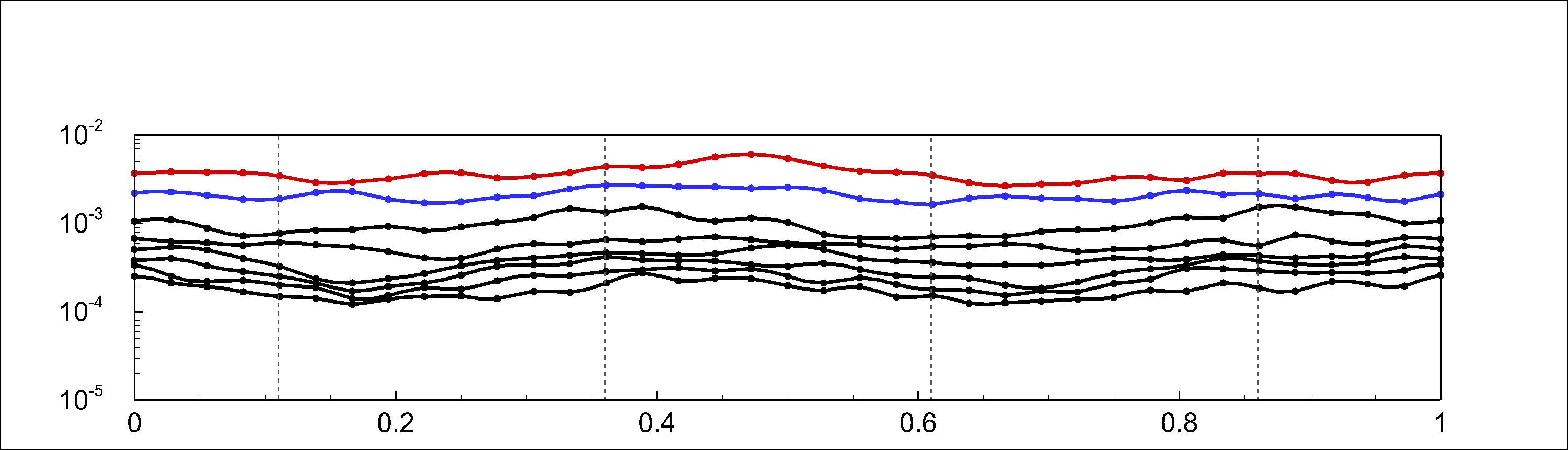}\raisebox{0.1in}{$\;\;\;\; \tau/T_0$}
	\linebreak
	\\
	\raisebox{0.5in}{(b)}\includegraphics[trim={1cm 10cm 30cm 10cm},clip,height=80px]{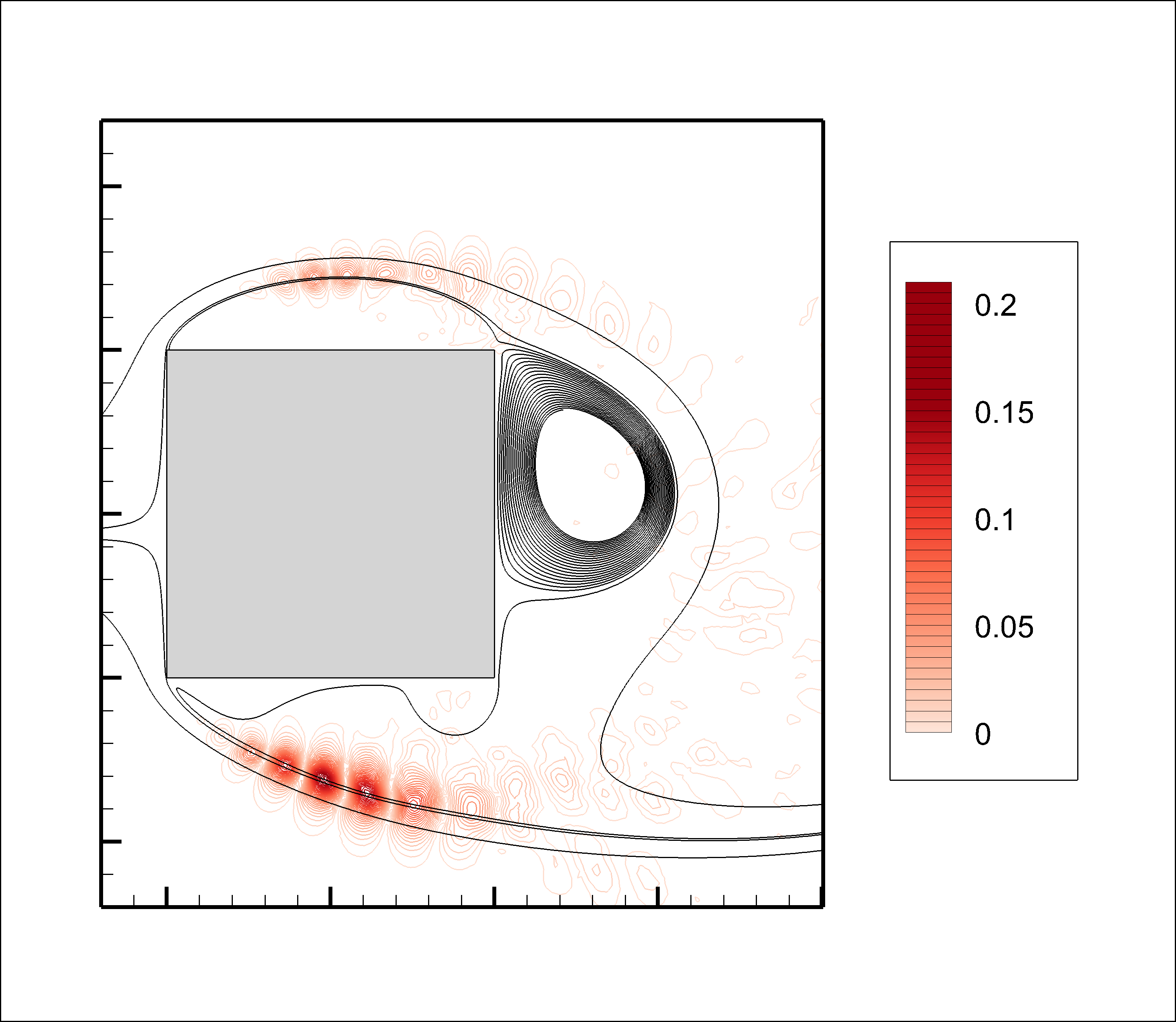}
	\includegraphics[trim={8cm 10cm 30cm 10cm},clip,height=80px]{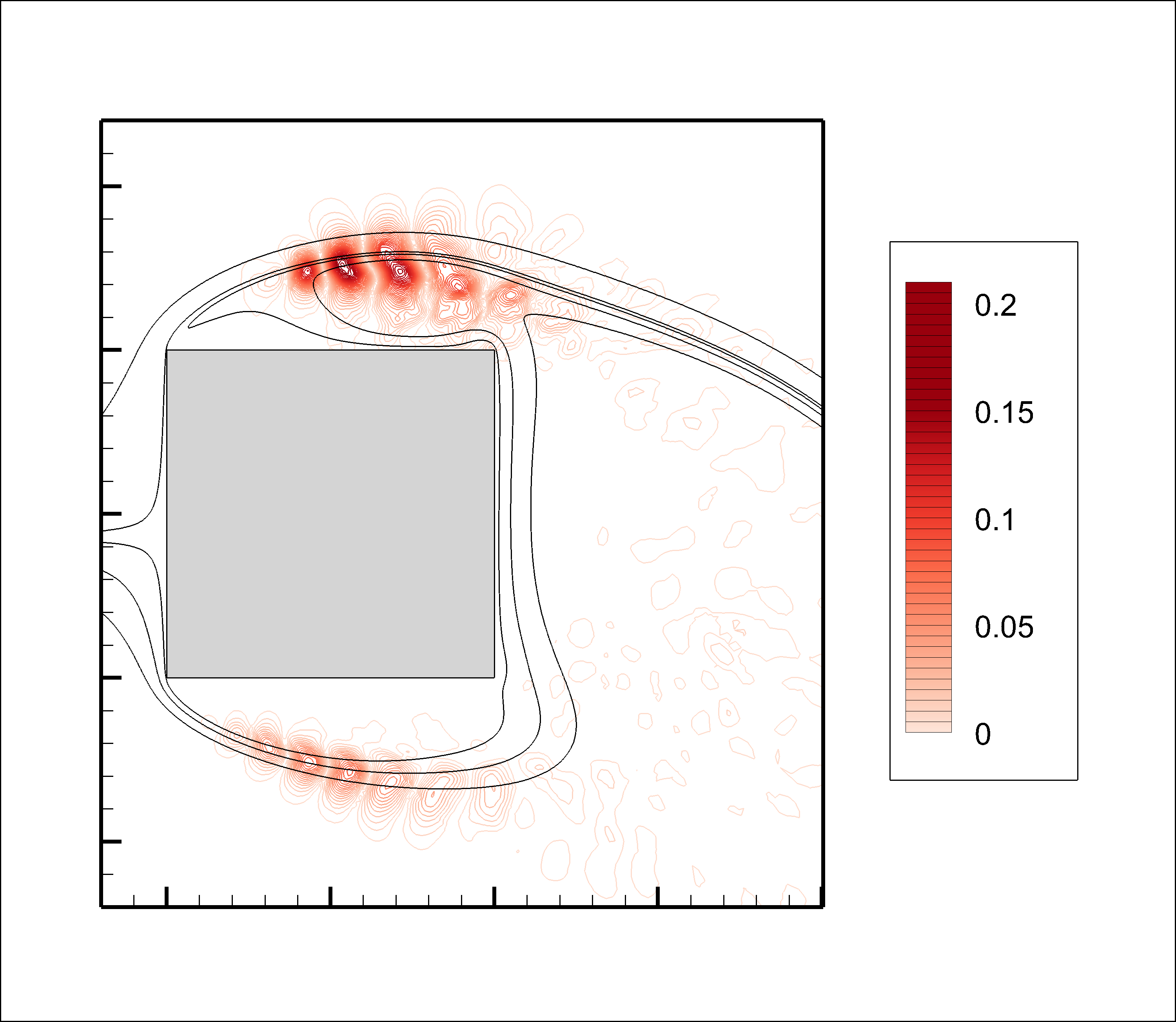}
	\includegraphics[trim={8cm 10cm 30cm 10cm},clip,height=80px]{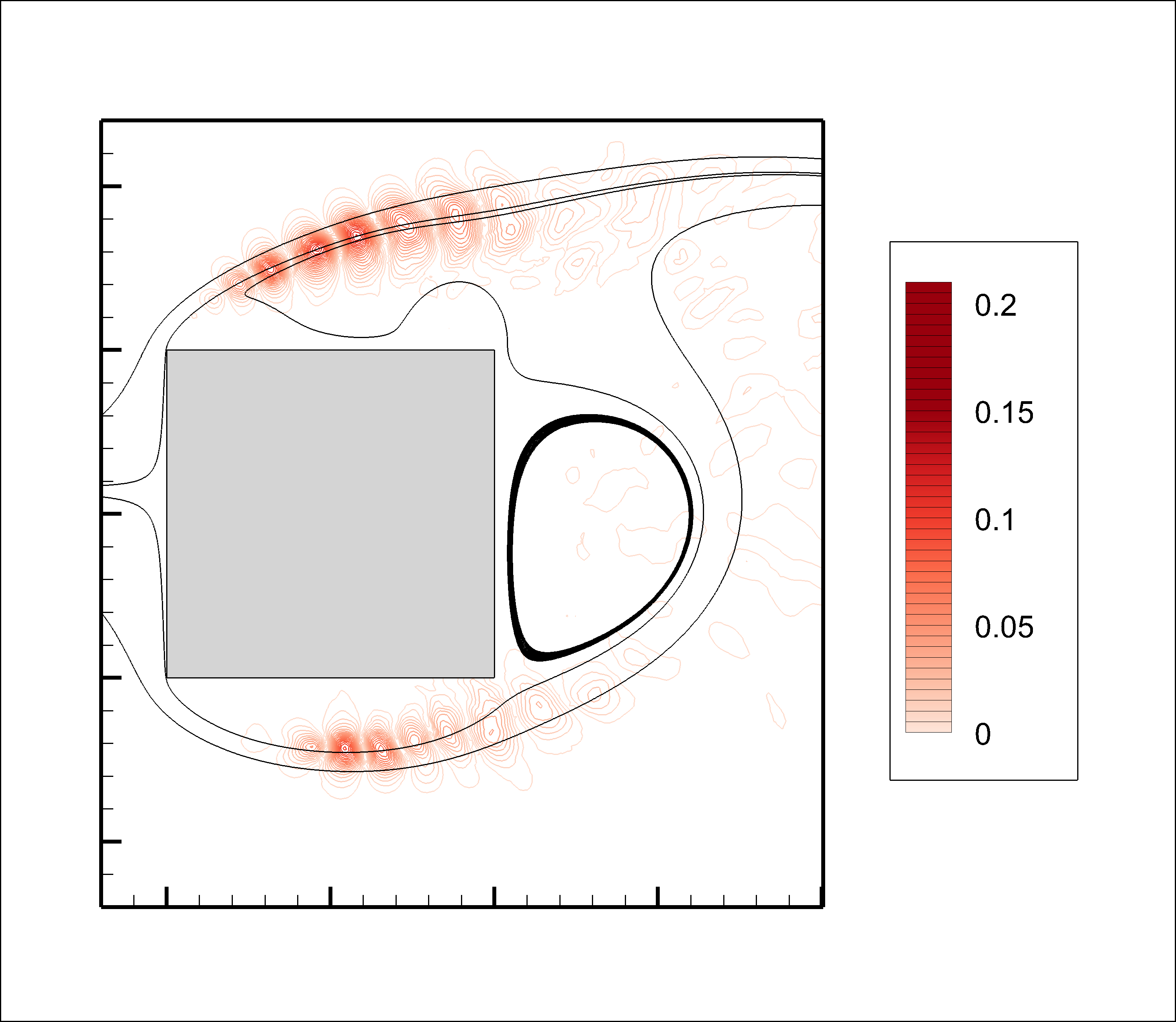} 
	\includegraphics[trim={8cm 10cm 5cm 10cm},clip,height=80px]{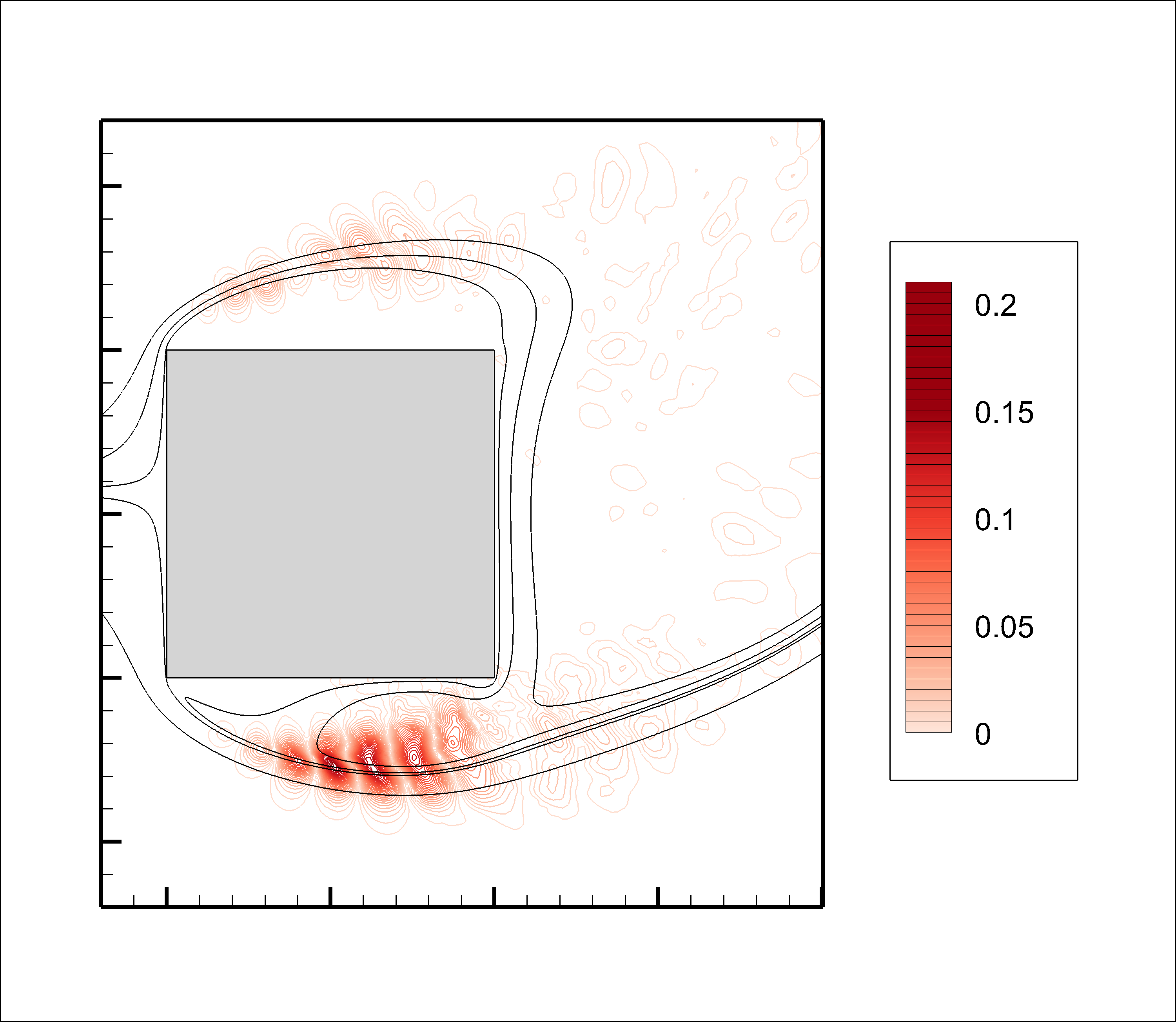}
	\\
	\raisebox{0.5in}{(c)}\includegraphics[trim={1cm 10cm 30cm 10cm},clip,height=80px]{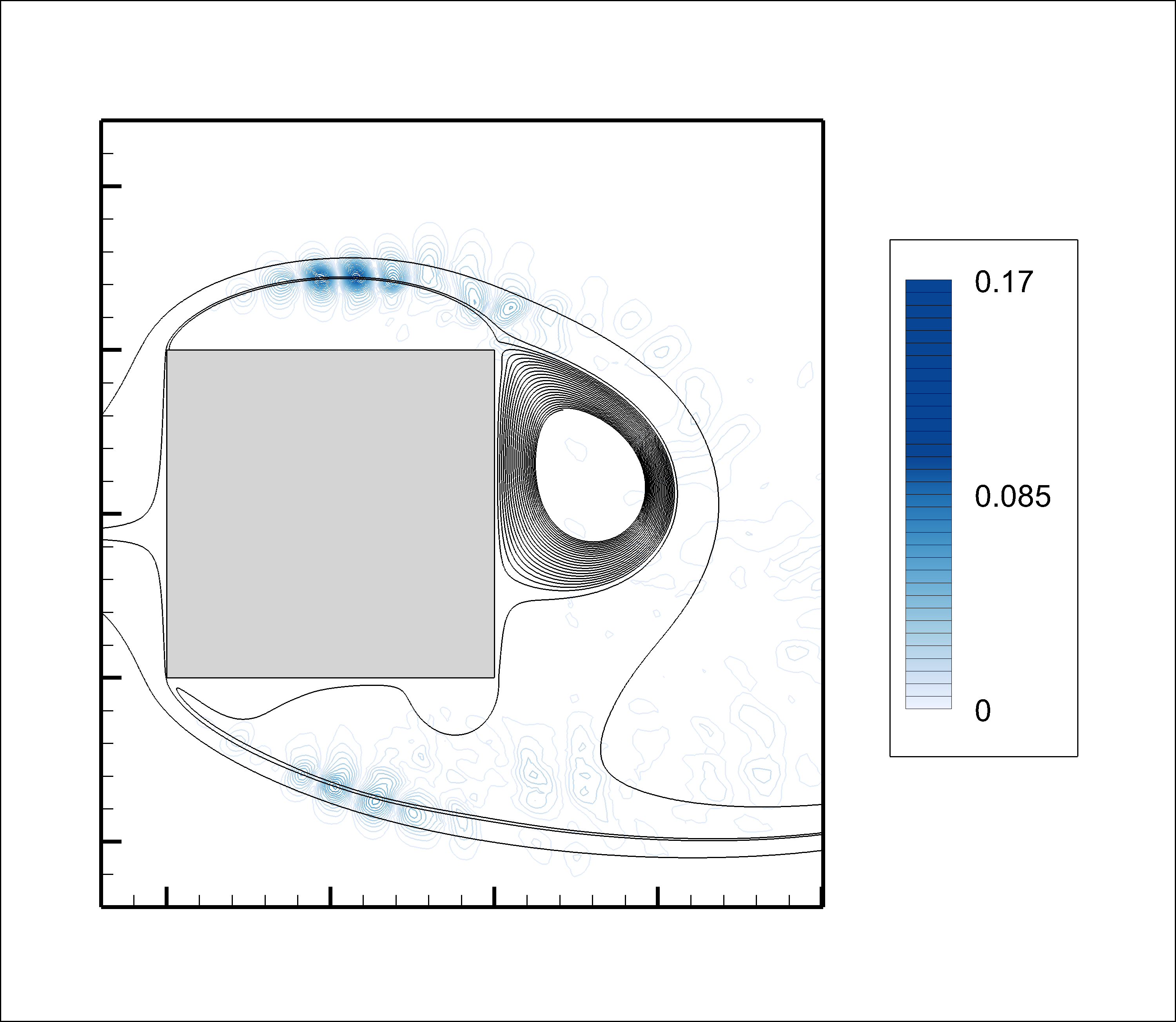}
	\includegraphics[trim={8cm 10cm 30cm 10cm},clip,height=80px]{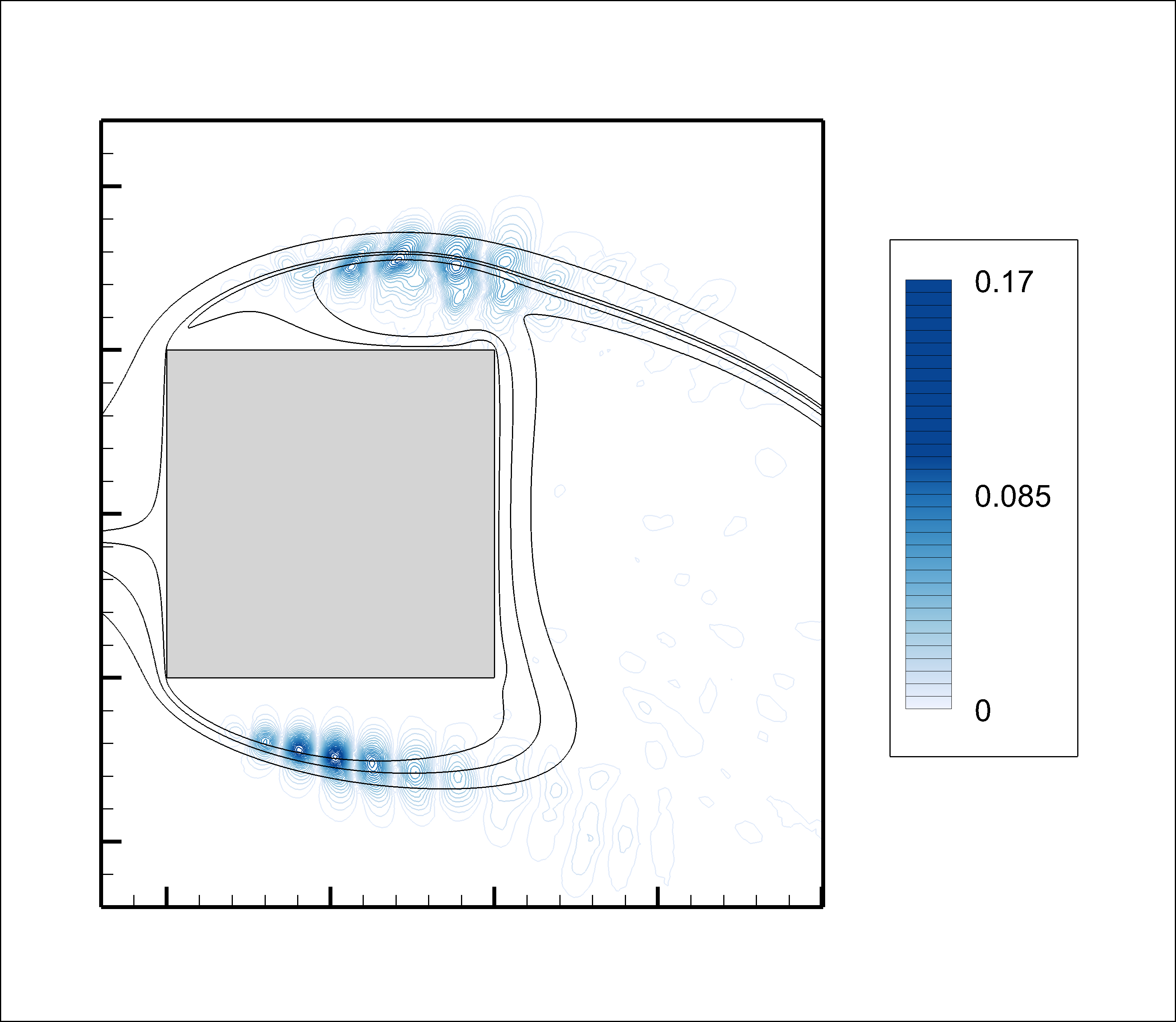}
	\includegraphics[trim={8cm 10cm 30cm 10cm},clip,height=80px]{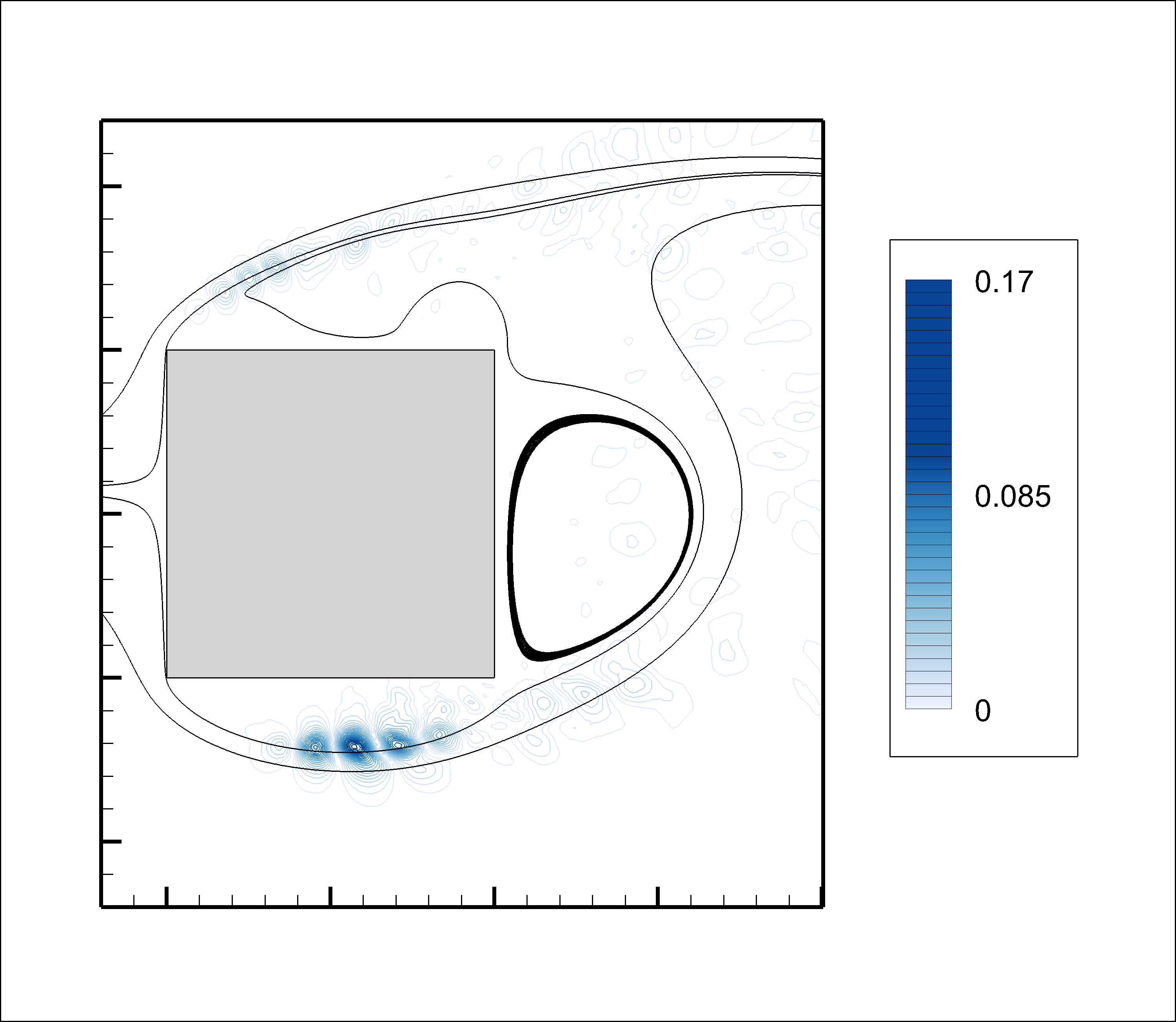} 
	\includegraphics[trim={8cm 10cm 5cm 10cm},clip,height=80px]{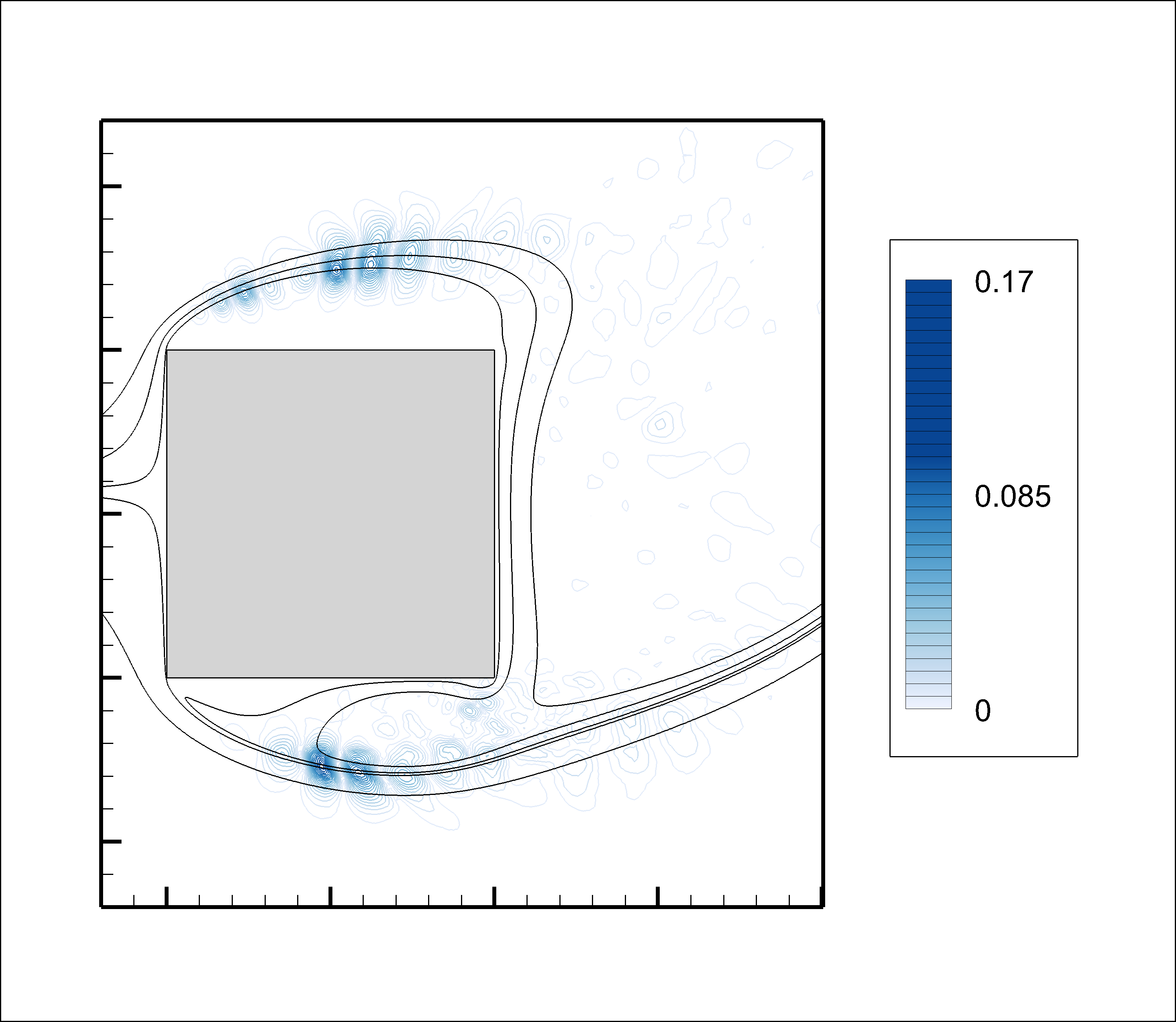}
	\caption{PCL-SPOD modes for $\omega=20$: eigenvalues (a) $\lambda_{\tau,1,\cdots,8}^2$ as a function of the phase $ \tau/T-0$, absolute value of pressure fluctuations for the optimal (b) and suboptimal (c) modes $\lambda_{\tau,1,2} \hat{\phi}_{\tau,1,2}$ at four different (and equidistant) phases (indicated by vertical lines in (a) and corresponding to the same phases as in figure \ref{fig:Sketch}(b)). The red (resp. blue) colour refers to the dominant (resp. sub-dominant) mode in all figures.}\label{fig:SPOD20}
\end{figure}

In figure \ref{fig:SPOD20}, we now represent the results of the PCL-SPOD analysis as a function of the phase $\tau/T_0 \in (0,1)$ for $\omega=20$. In figure \ref{fig:SPOD20}(a), we plot the eight strongest branches $\lambda_{\tau,k}^2 (\tau),k=1,2,\cdots,8 $. The two dominant branches are highlighted with red and blue colors. We can see that those branches do not clearly display any preferential phase within $(0,T_0)$ and present similar energies. The latter point is in accordance with the contrast between the bump present in figure \ref{fig:SPODgains}(a) and the plateau in \ref{fig:SPODgains}(b). The scaled SPOD modes, $ \lambda_\tau \hat{\phi}_\tau $, corresponding to the two dominant branches, are respectively represented in figures \ref{fig:SPOD20}(b) and (c), at the four phases of the fundamental period marked by vertical lines in figure \ref{fig:SPOD20}(a). There are the same phases as shown in figures \ref{fig:Sketch}(b) and \ref{fig:fluctuation}. We can clearly recognize the KH structures (colors) that evolve according to the VS motion (contours). 

\begin{figure}
	\centering
	\raisebox{0.8in}{(a)}\raisebox{0.6in}{$\lambda_{\tau}^2$}\includegraphics[trim={1cm 1cm 7cm 4cm},clip,height=80px]{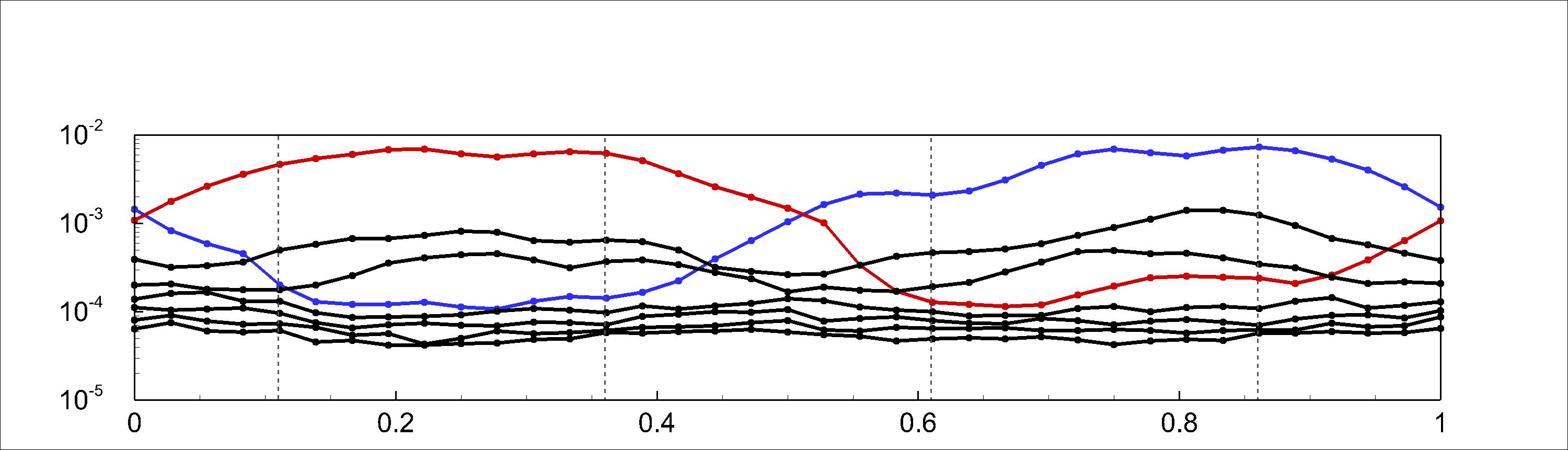}\raisebox{0.1in}{$\;\;\;\; \tau/T_0$}
    \linebreak
	\\
	\raisebox{0.5in}{(b)}\includegraphics[trim={1cm 10cm 30cm 10cm},clip,height=80px]{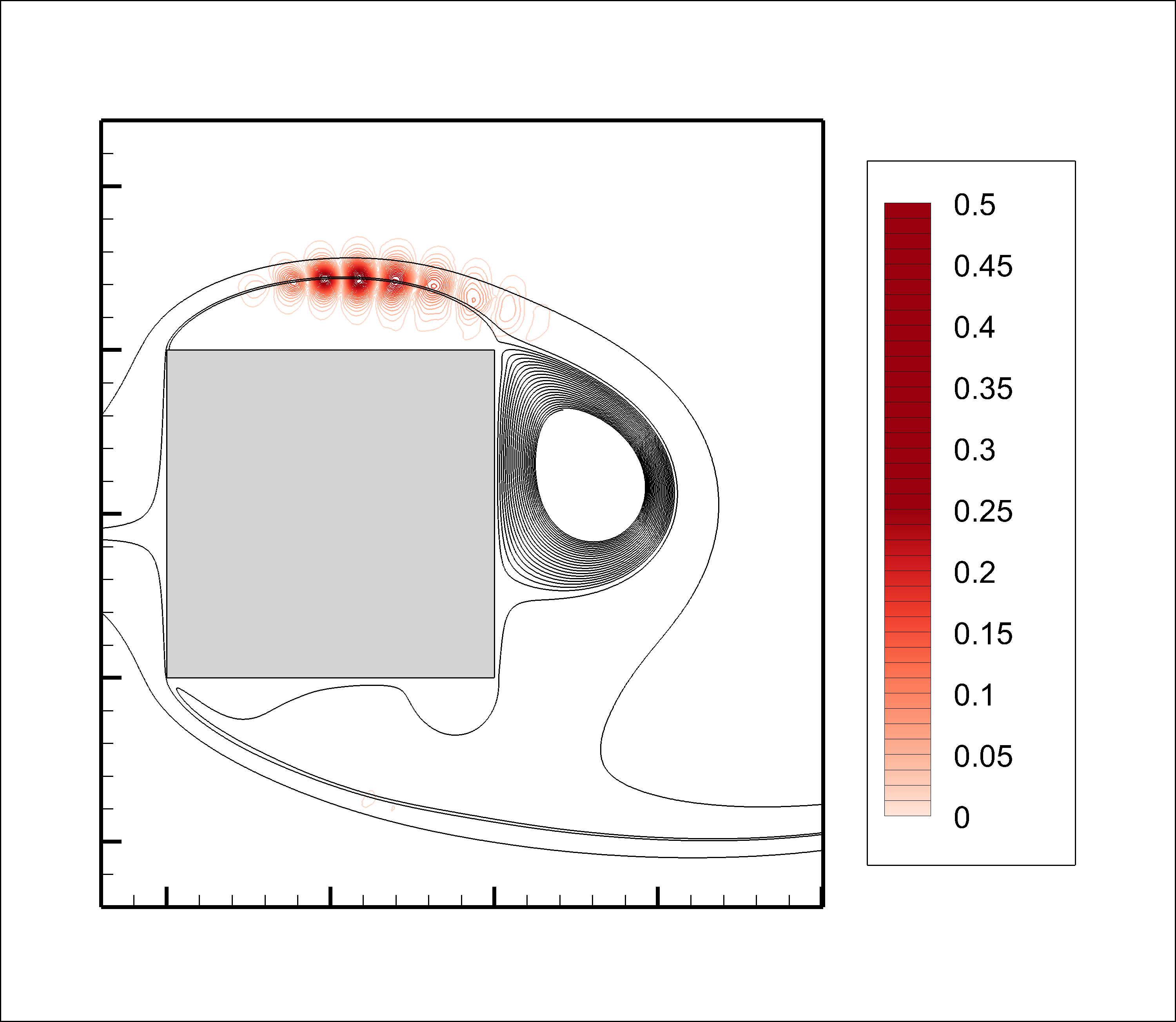}
	\includegraphics[trim={8cm 10cm 30cm 10cm},clip,height=80px]{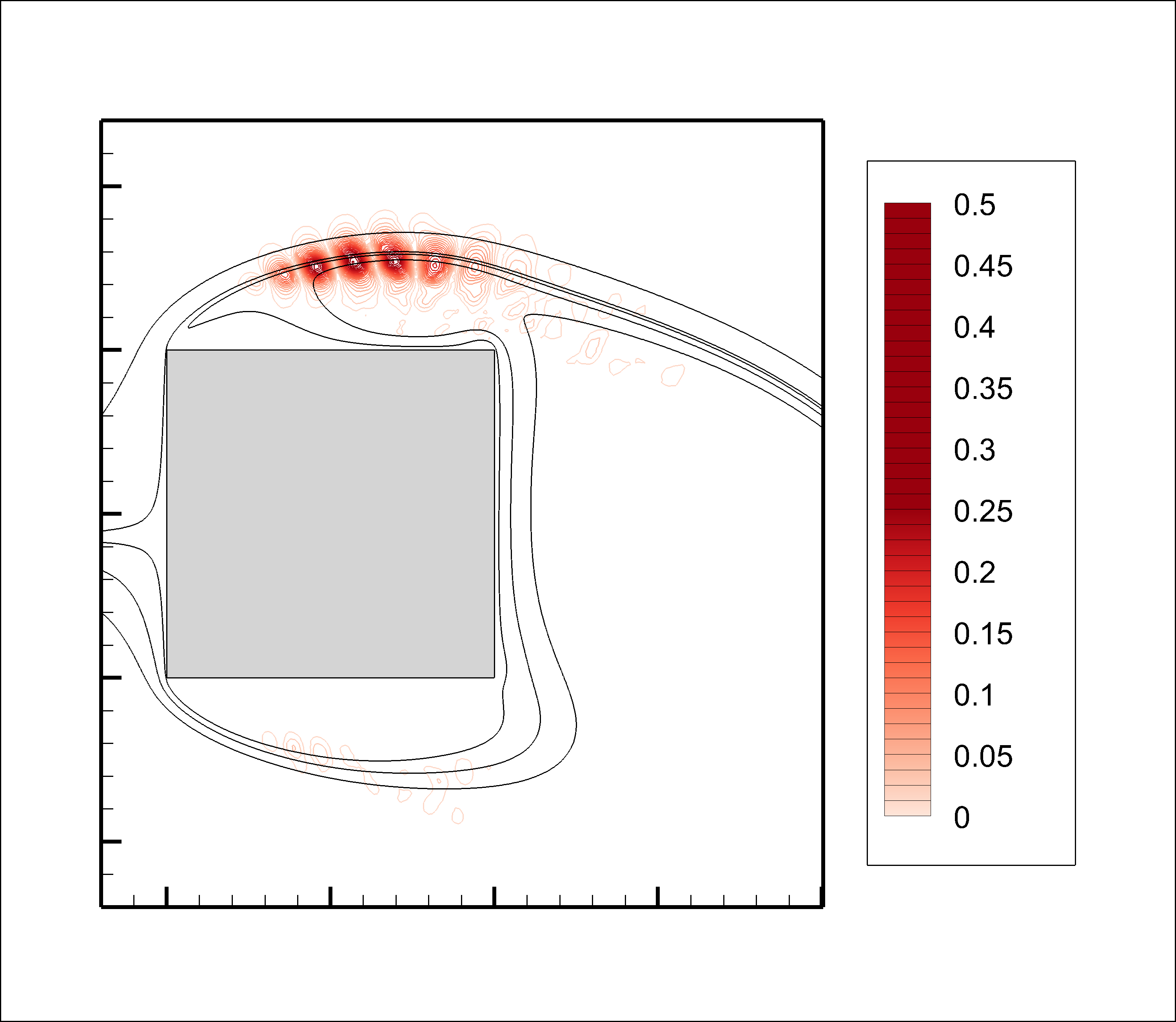}
	\includegraphics[trim={8cm 10cm 30cm 10cm},clip,height=80px]{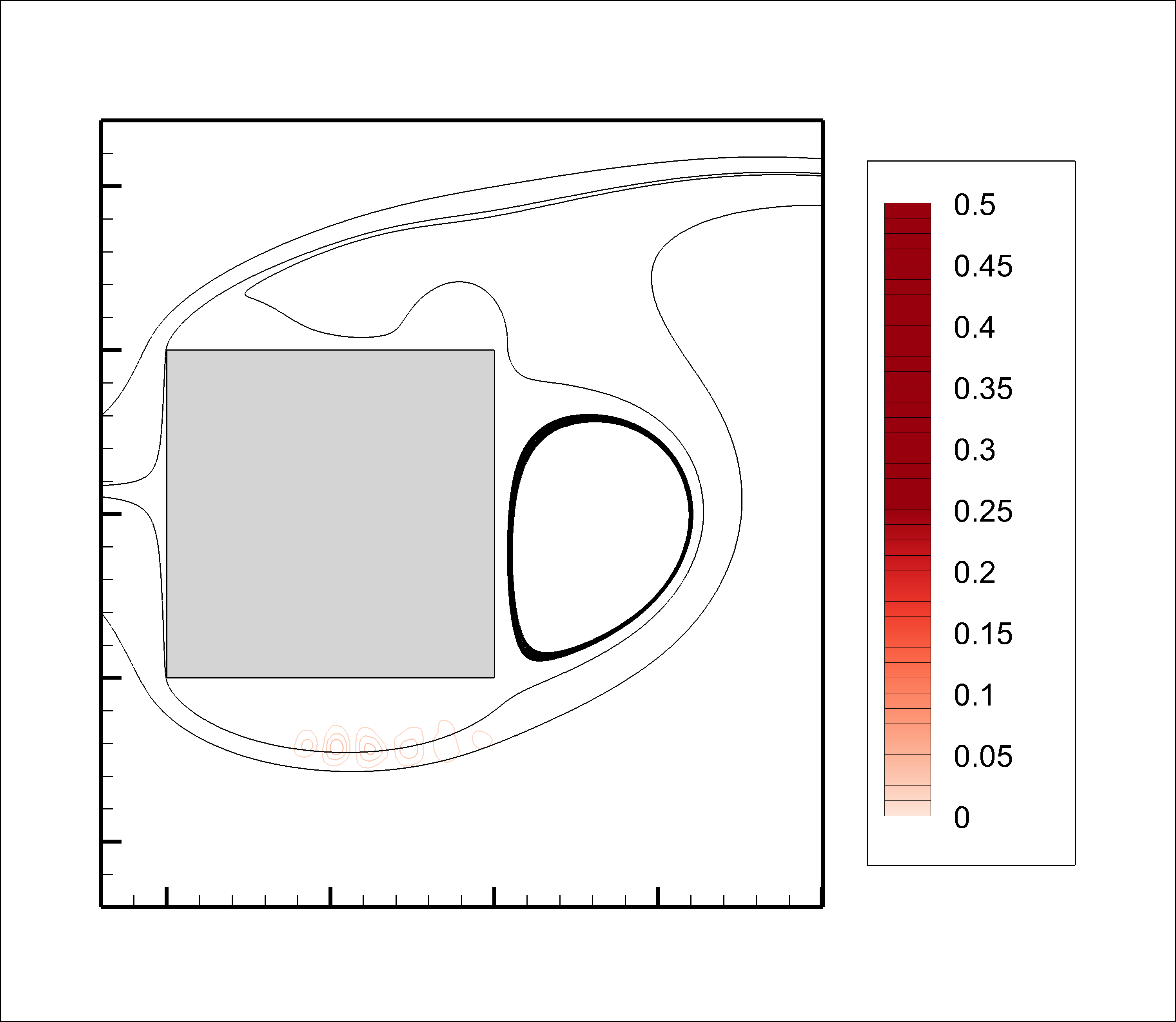} 
	\includegraphics[trim={8cm 10cm 5cm 10cm},clip,height=80px]{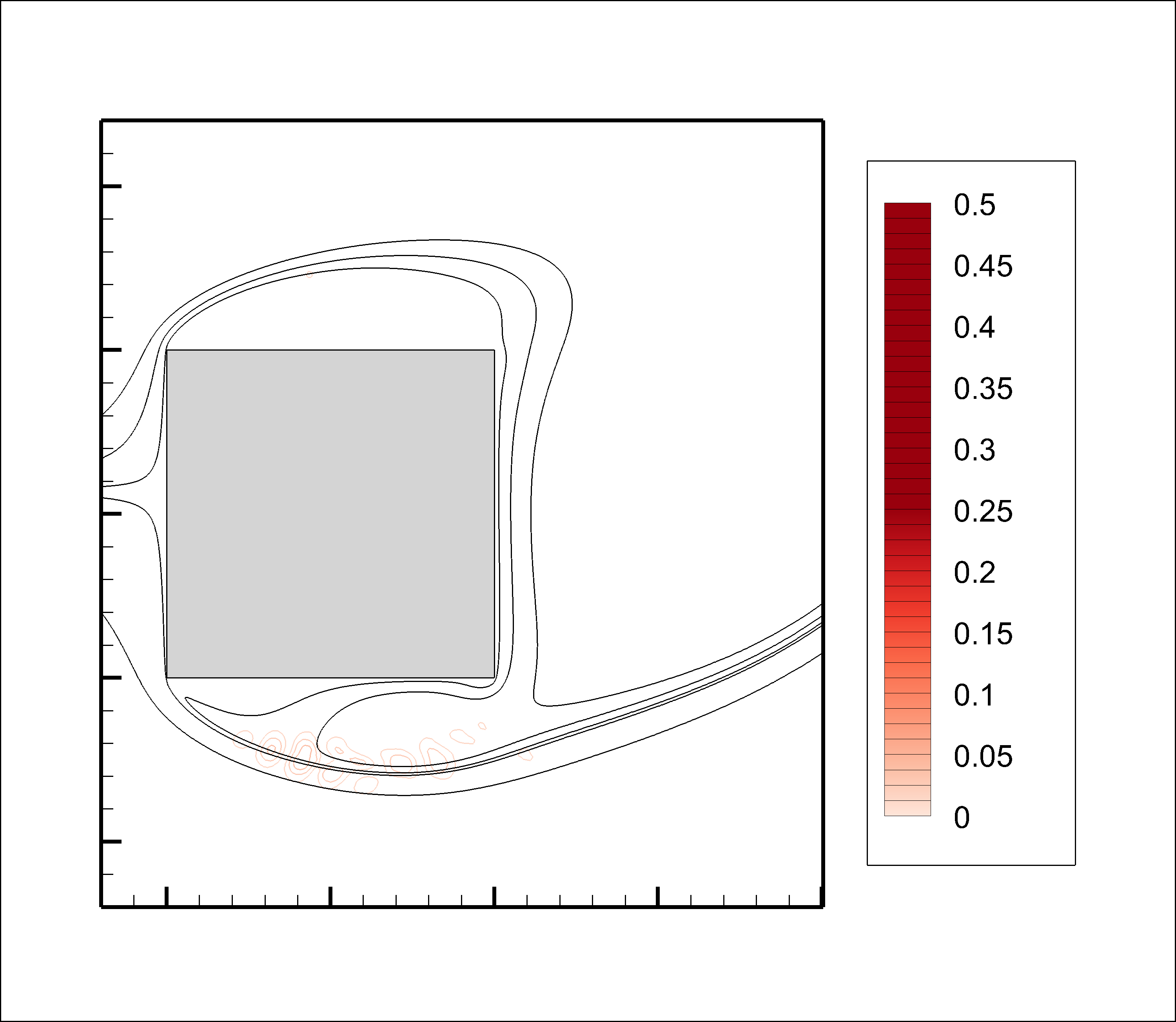}
	\\
	\raisebox{0.5in}{(c)}\includegraphics[trim={1cm 10cm 30cm 10cm},clip,height=80px]{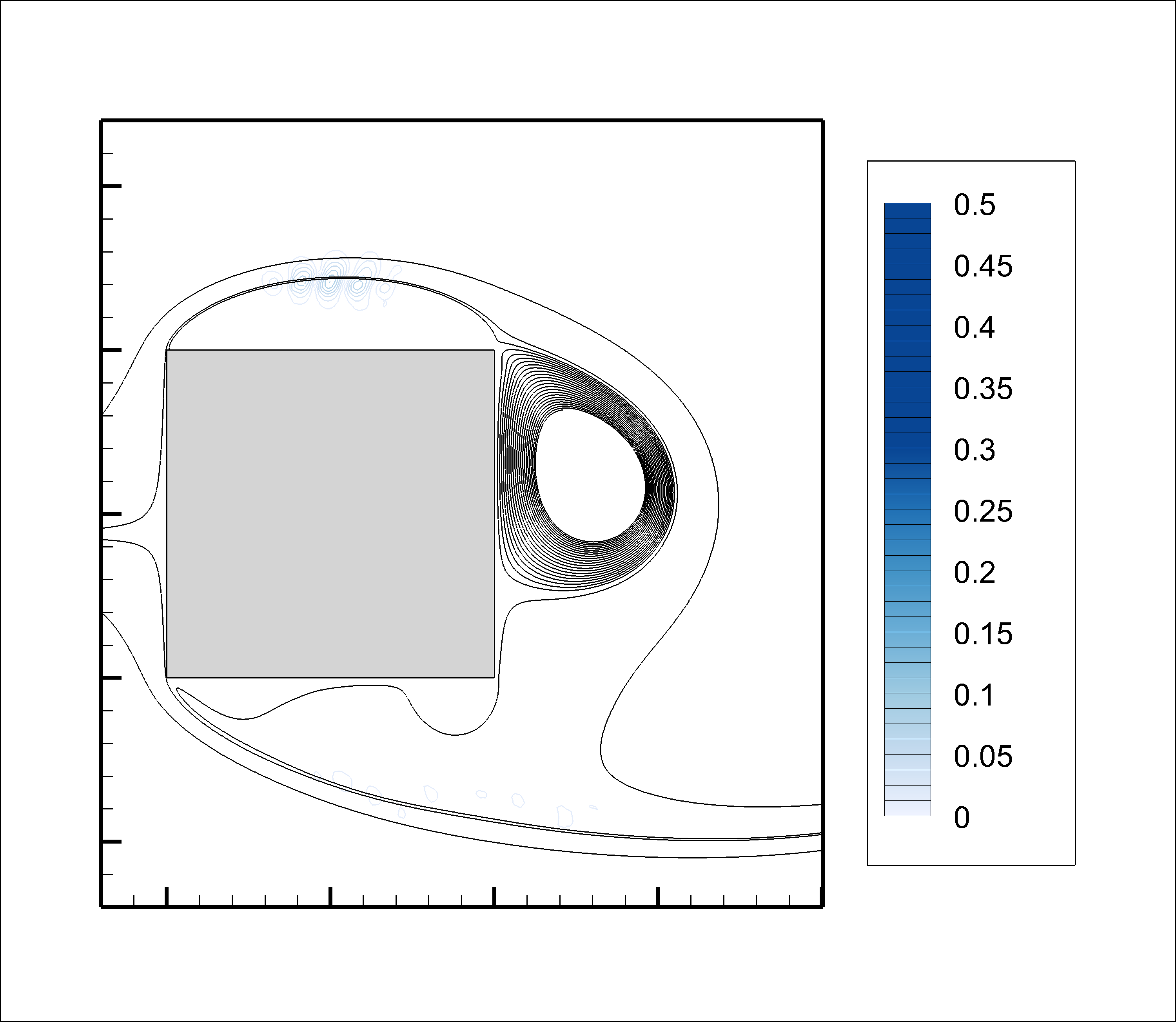}
	\includegraphics[trim={8cm 10cm 30cm 10cm},clip,height=80px]{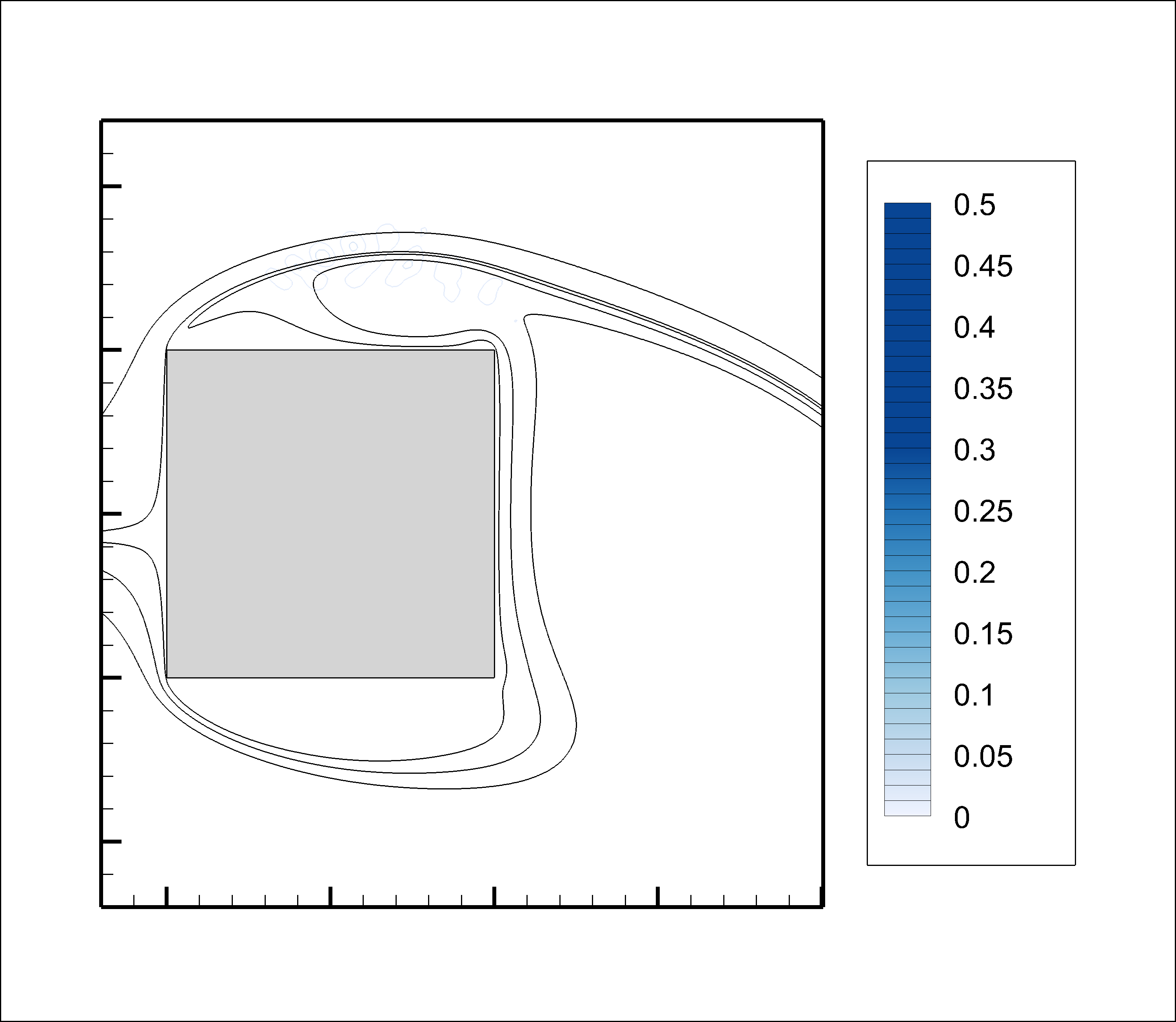}
	\includegraphics[trim={8cm 10cm 30cm 10cm},clip,height=80px]{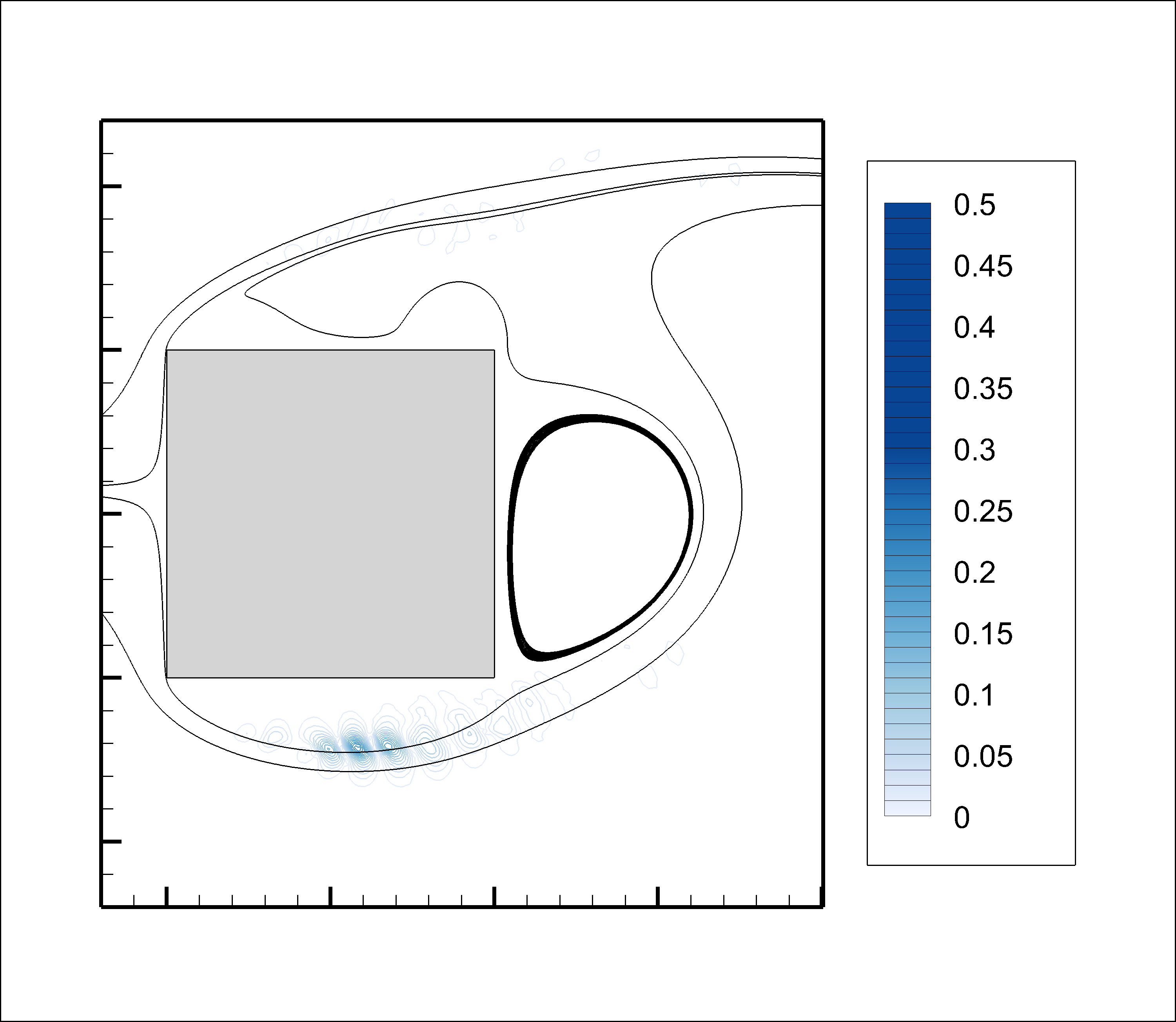} 
	\includegraphics[trim={8cm 10cm 5cm 10cm},clip,height=80px]{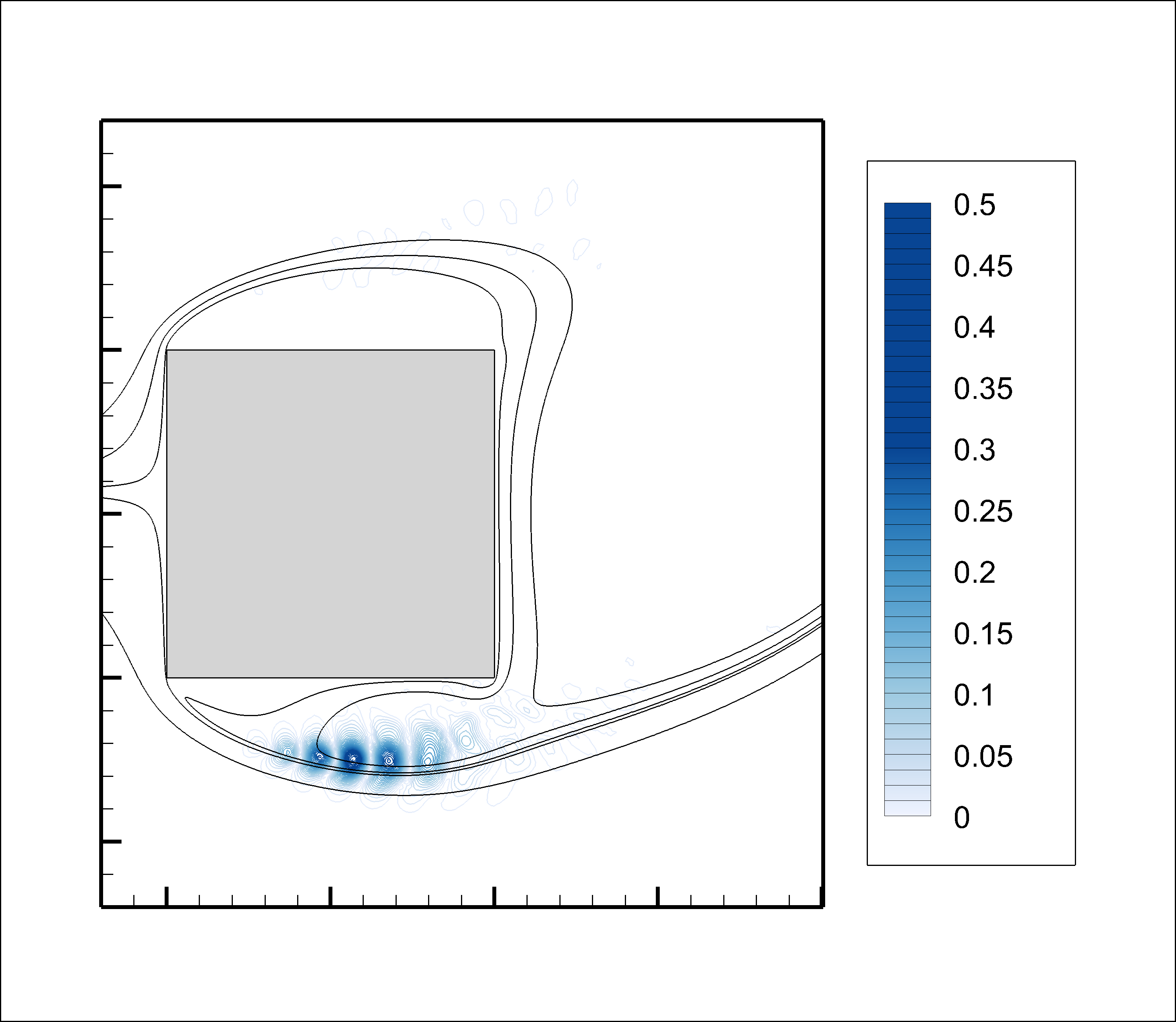}
	\caption{PCL-SPOD modes for $\omega=30$: eigenvalues (a) $\lambda_{\tau,1,\cdots,8}^2$ and absolute value of pressure fluctuations for the two dominant modes (b,c) $\lambda_{\tau,1,2} \hat{\phi}_{\tau,1,2}$ corresponding to the red/blue curves in (a) at four different phases (same as in \ref{fig:SPOD20}).}\label{fig:SPOD30}
\end{figure}

Figure \ref{fig:SPOD30} displays similar results but for $\omega=30$. By continuation in phase, we distinguish several branches that clearly exhibit an oscillating behavior within the fundamental period. For example, the two most energetic ones (red and blue) oscillate in an anti-phase manner. The red curve displays a bump for $\tau/T_0 \in (0.1,0.6)$ and a valley for $\tau/T_0 \in (0,0.1) \cup (0.6,1)$. A similar behavior is observed for the blue curve but the "bump" occurs at phases where the red curve presents a valley and vice-versa. Moreover, since only one branch (either red or blue) is dominant for a given phase, it holds most of the fluctuation energy, which is in accordance with previous comments on figure \ref{fig:SPODgains}. We discuss now the two dominant PCL-SPOD modes (shown in (b,c), scaled with $\lambda_{\tau,k}$). We can clearly see that the bump in the red (resp. blue) curve is associated to KH structures developing only on the upper (resp. lower) shear-layer. Interestingly, those largest energetic amplification occur when the shear layers are closest to the walls, that is when the gradients of $\phqq$ are strongest. This observation agrees with figure \ref{fig:fluctuation} where more fluctuations can be spotted at those phases and locations as well. The symmetry observed is in accordance with the symmetry of the VS motion where the (almost) mirror image of the field $\phqq(t)$ is observed at $t+T_0/2$. Also, it is seen that all the modes at this frequency have their support only on the upper or the lower shear layer. 
This is due to a statistical decorrelation property between the dynamics at the top and bottom of the cylinder, which stems from a separation of the spatial supports of the modes. Indeed, as the number of bins increases, the spectral correlation matrix $\hat{\QQ}_{\tau} \hat{\QQ}_{\tau}^*  =
(1/N_b) \sum_{m=1}^{N_b} \left[ \hat{\qq}_{\tau_m} \hat{\qq}_{\tau_m}^{*} \right]
\equiv \mathbf{A}$ can be split in an upper left block $\mathbf{A}_{uu}$ and a lower right block $\mathbf{A}_{ll}$, where crossing terms, $\mathbf{A}_{ul}$ and $\mathbf{A}_{lu}$, tend to zero as $N_b \rightarrow +\infty$ due to the separation of the supports and therefore the decorrelation of the top and bottom dynamics.
This separation trend was already slightly present for $ \omega=20 $ (see for example figure \ref{fig:SPOD20} (a) in first, second and fourth phases and (b) first and third phases) and is enforced here due to a smaller and more compact spatial support of the modes on the top/bottom shear-layers.
This decorrelation property is further confirmed using upper/lower localized inner-products in Appendix \ref{apx:UpperLower}, showing that the two leading red/blue modes exhibit dynamics that are fully decorrelated from each other.

In the next section, we will present the results of the PCL-Resolvent analysis and discuss its links with the SPOD counterpart. 

\subsection{PCL-Resolvent results}\label{sec:Resolventres}

We present now the results of the PCL-Resolvent analysis. Those results were produced by discretizing the linear Navier-Stokes equation \eqref{eqn:BroadbandNS} with the Finite-Element Method (FEM) in the open source software FreeFEM++ (see \citet{MR3043640}). The used mesh had a spatial discretization similar to the DNS one. Moreover, since the focus of the present work is on the spanwise averaged fields, we only looked at spanwise invariant modes. To deal with high-Reynolds number flows, we employ a second-order Streamline-Upwind Petrov-Galerkin (SUPG) method (see \citet{brooks1982streamline,franceschini2020mean} for more details). The Resolvent modes are obtained by solving the eigenvalue problem \eqref{eqn:SVDResolvent} using ARPACK, interfaced with FreeFEM++. We recall that the inner product used for the gain definition ($\mu_{\tau}$) is the same as the one used for the SPOD and corresponds to the integration of the velocity fields over $\Omega$ (see green windows in fig. \ref{fig:Harmo}(b)). Note however that, contrary to the PCL-SPOD analysis, the results of the Resolvent analysis turned out to be quite unsensitive to the precise choice of $ \Omega $ (some tests were done where $\Omega$ was much larger and no significant changes in the results were observed).

\begin{figure}
	\centering
	\begin{tabular}{cccc}
        & $\max_{\tau}\mu_{\tau,0}^2$ & & $\sum_{k=1}^{N_b} \frac{1}{T_0} \int_0^{T_0} \mu_{\tau,k}^2 d\tau $ \\ 
        \raisebox{1in}{(a)} & \includegraphics[trim={1cm 1cm 10cm 22cm},clip,height=90px]{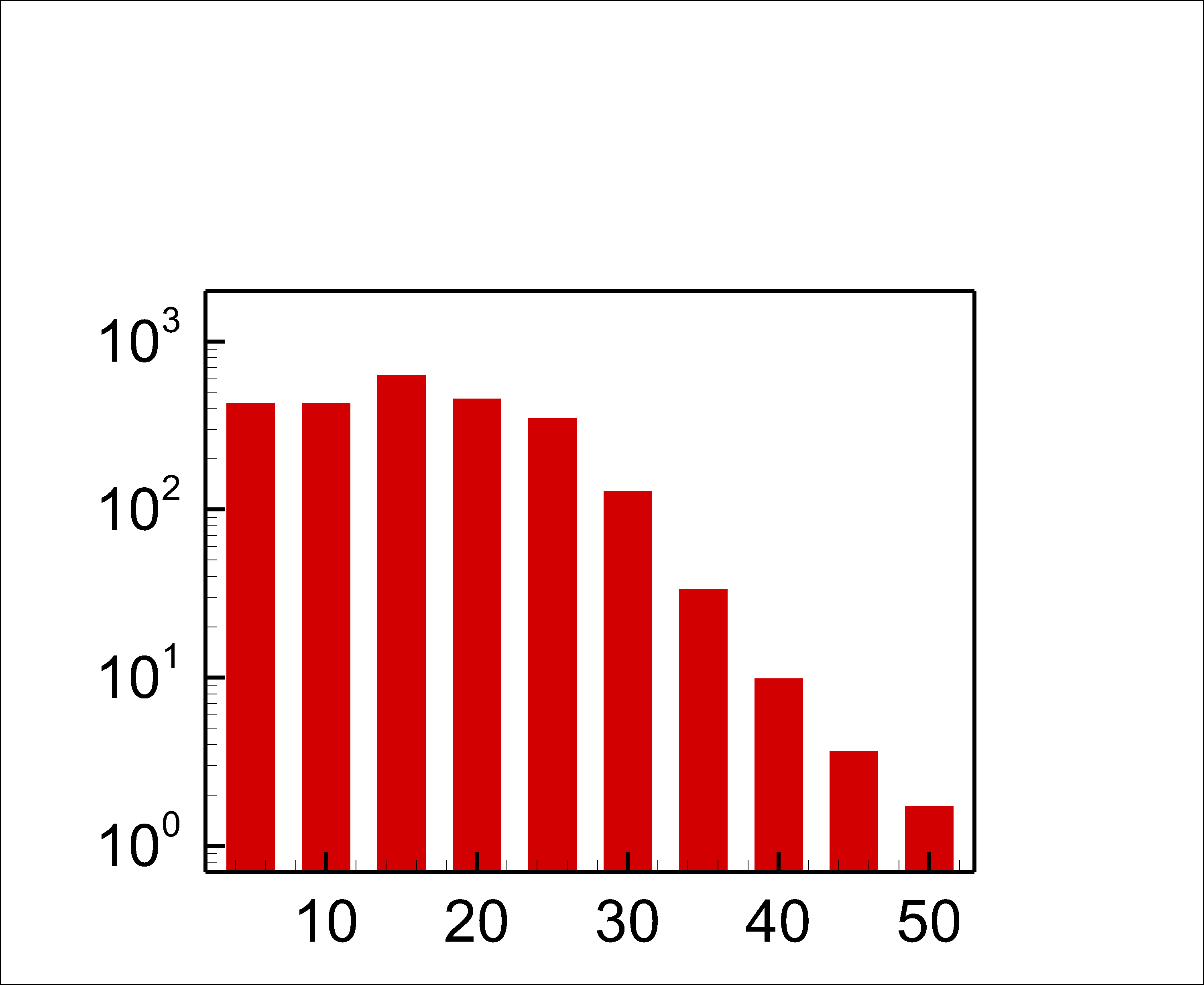} & \raisebox{1in}{(b)} & \includegraphics[trim={1cm 1cm 10cm 22cm},clip,height=90px]{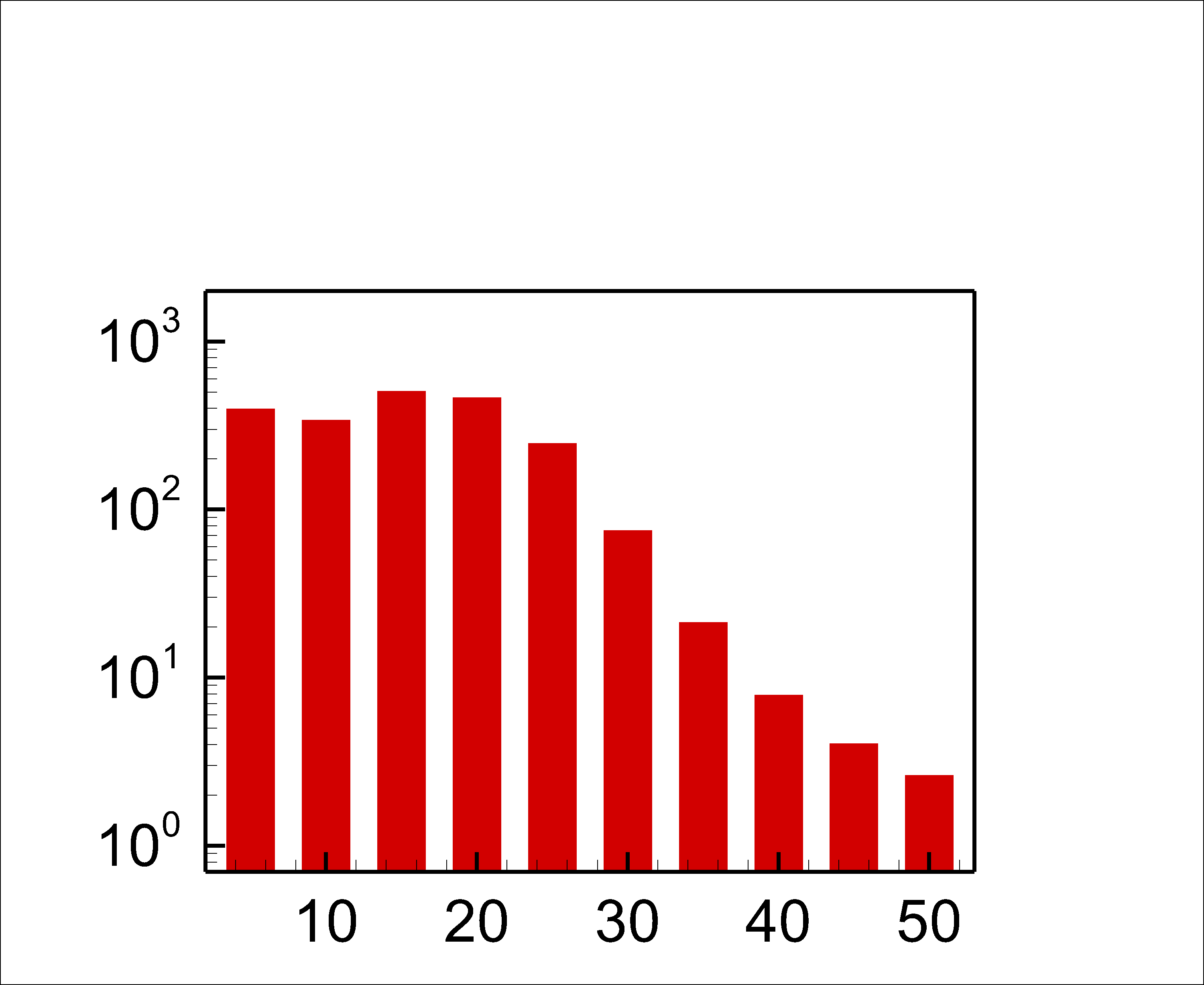} \raisebox{0.1in}{$\omega$}
    \end{tabular}
    \caption{PCL Resolvent modes: (a) maximum value of dominant energy gain $\mu_{\tau,1}^2$ over $ 0 \leq \tau < T_0$ as a function of frequency and (b) average energy gain over $ 0 \leq \tau < T_0$, summed over all eigenvalues, as a function of freqeuncy.}\label{fig:ResolventGains}
\end{figure}

First, similarly to what was done for the SPOD, we plot overall characteristics of the gains as a function of frequency in figure \ref{fig:ResolventGains}. In (a) we show the maximal value of $\mu_{\tau,1}^2$ for $\tau \in (0,T_0)$ and in (b) we plot their sum, integrated over $\tau \in (0,T_0)$. We can see that both present a maximal value for $\omega=15$ and the "bump" in frequency extends up to $\omega=25$. This maximal frequency is slightly smaller than the one obtained for the SPOD analysis, but we believe this is a minor difference. We remark that the large values at $\omega=5$ observed for the SPOD analysis is much less pronounced here, reinforcing that those large energies in SPOD modes are due to a cascade initiated at lower frequencies rather than to the extraction of energy from $\phqq$ at this frequency through a linear mechanism, as confirmed by the resolvent analysis.

\begin{figure}
	\centering
	\raisebox{0.8in}{(a)}\raisebox{0.6in}{$\mu_{\tau}^2$}\includegraphics[trim={1cm 1cm 7cm 4cm},clip,height=80px]{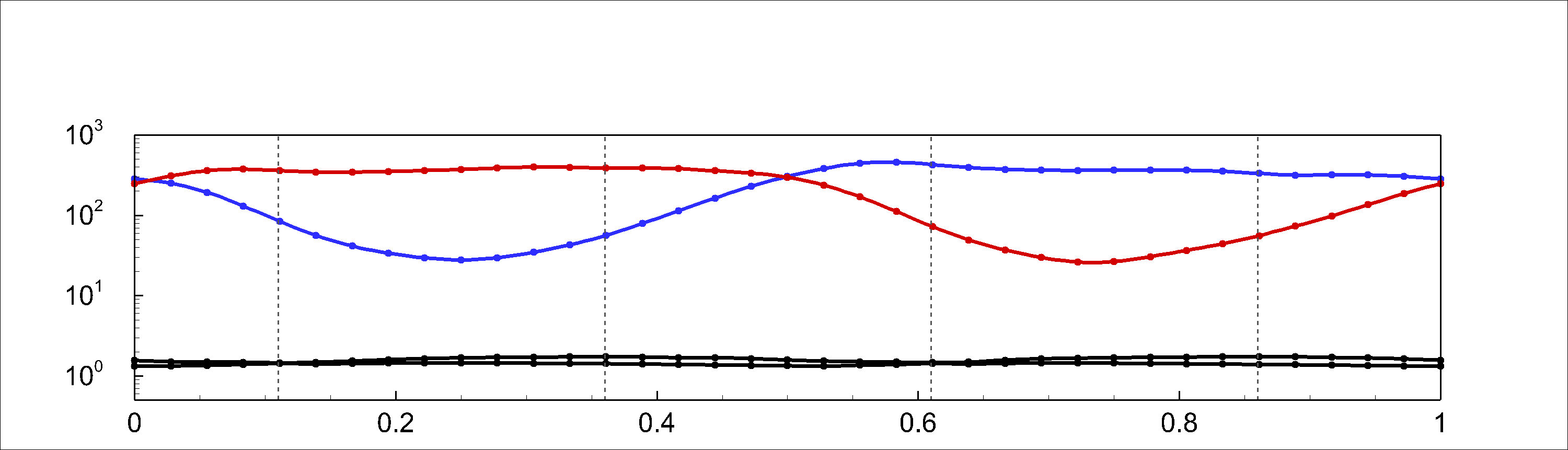}\raisebox{0.1in}{$\;\;\;\; \tau/T_0$}
	\linebreak
	\\
	\raisebox{0.5in}{(b)}\includegraphics[trim={1cm 10cm 30cm 10cm},clip,height=80px]{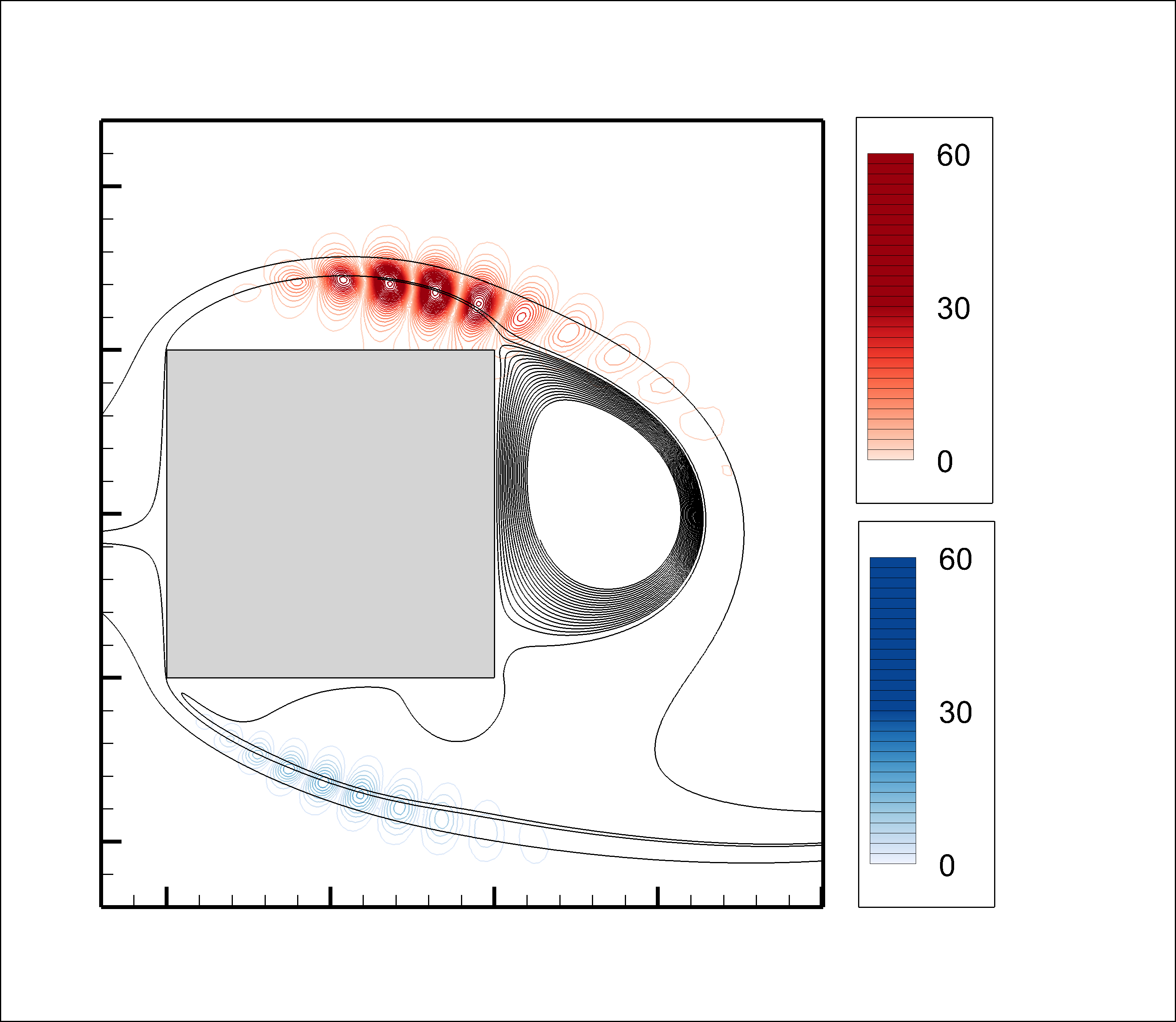}
	\includegraphics[trim={8cm 10cm 30cm 10cm},clip,height=80px]{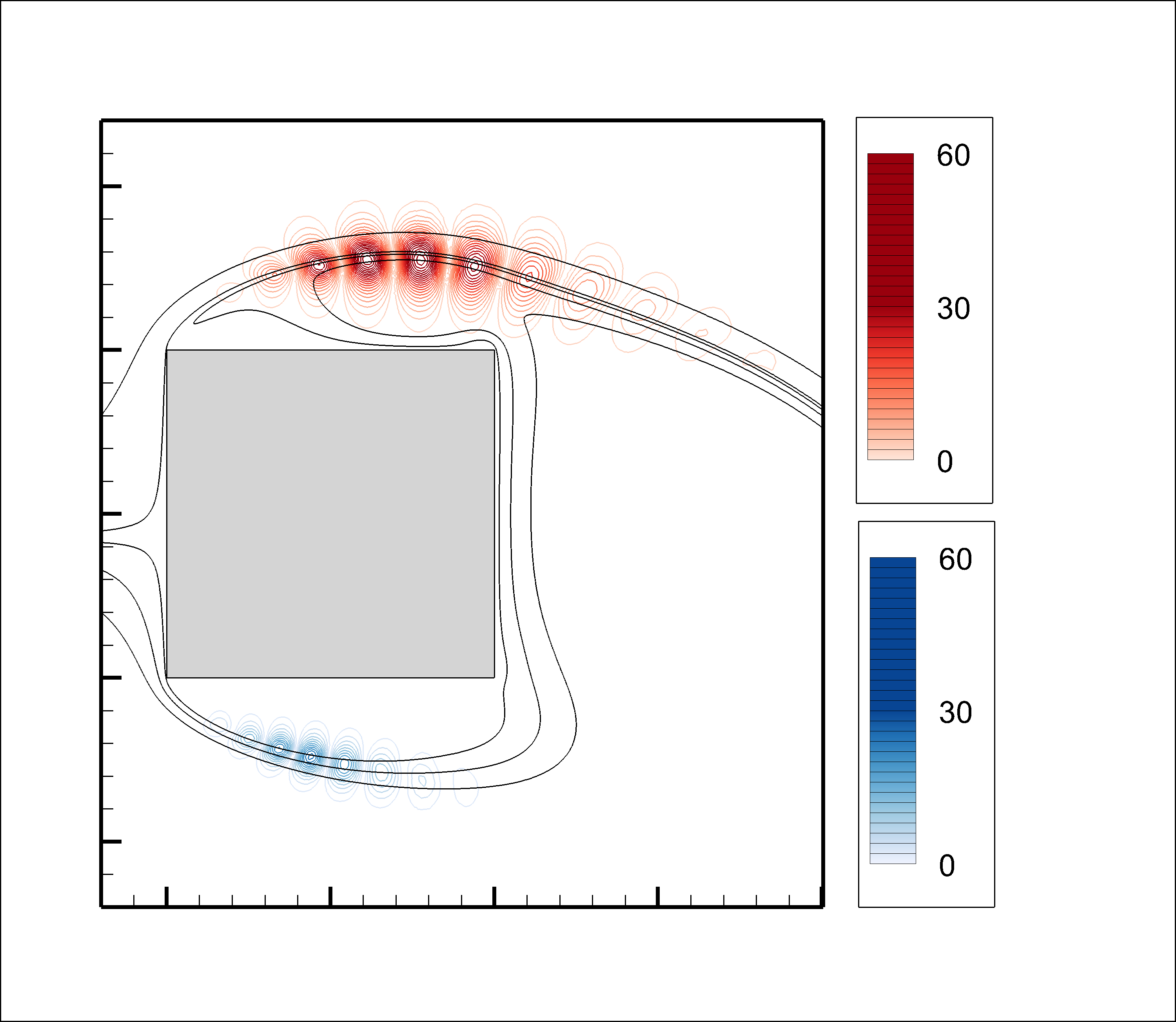}
	\includegraphics[trim={8cm 10cm 30cm 10cm},clip,height=80px]{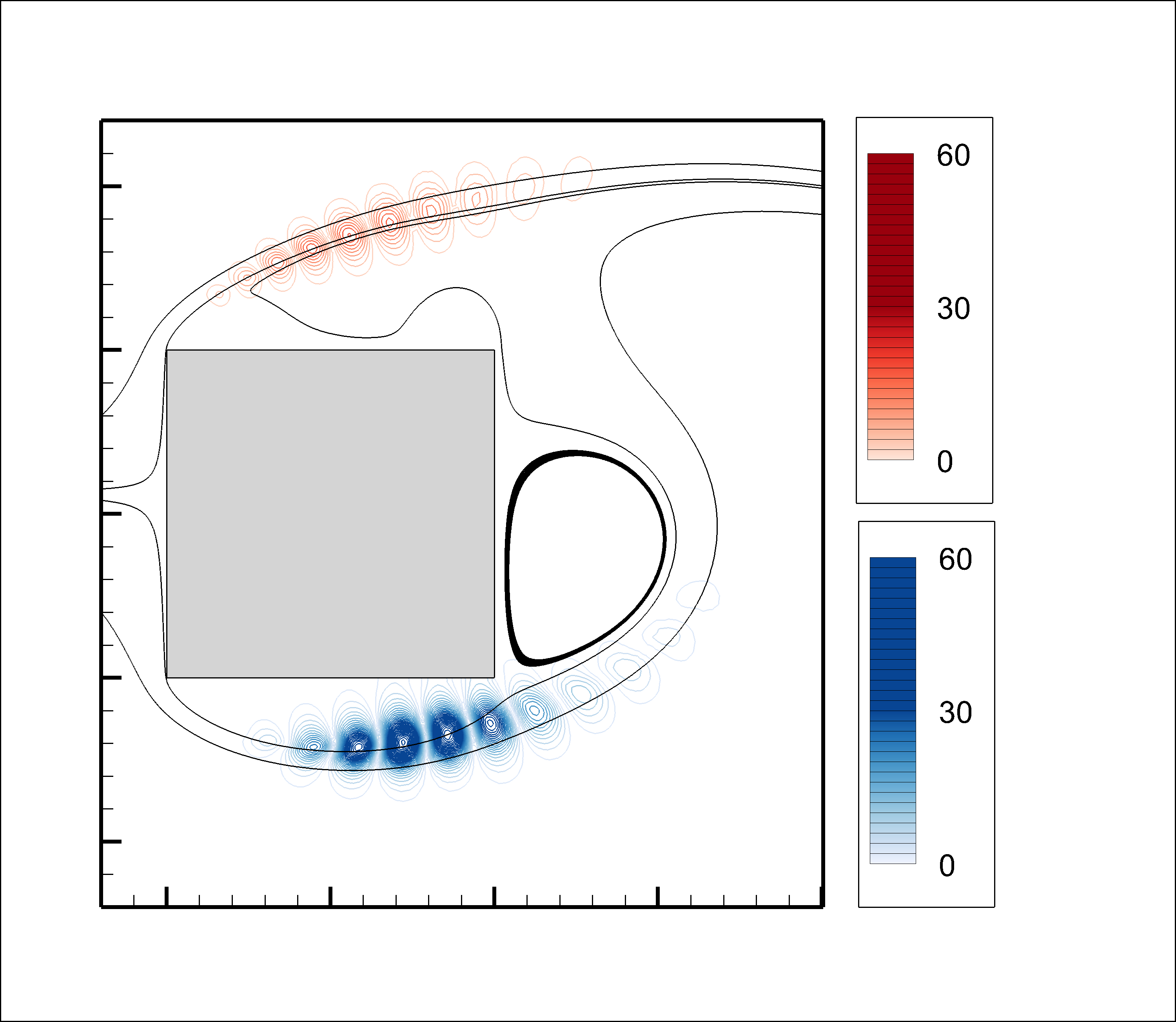} 
	\includegraphics[trim={8cm 10cm 5cm 10cm},clip,height=80px]{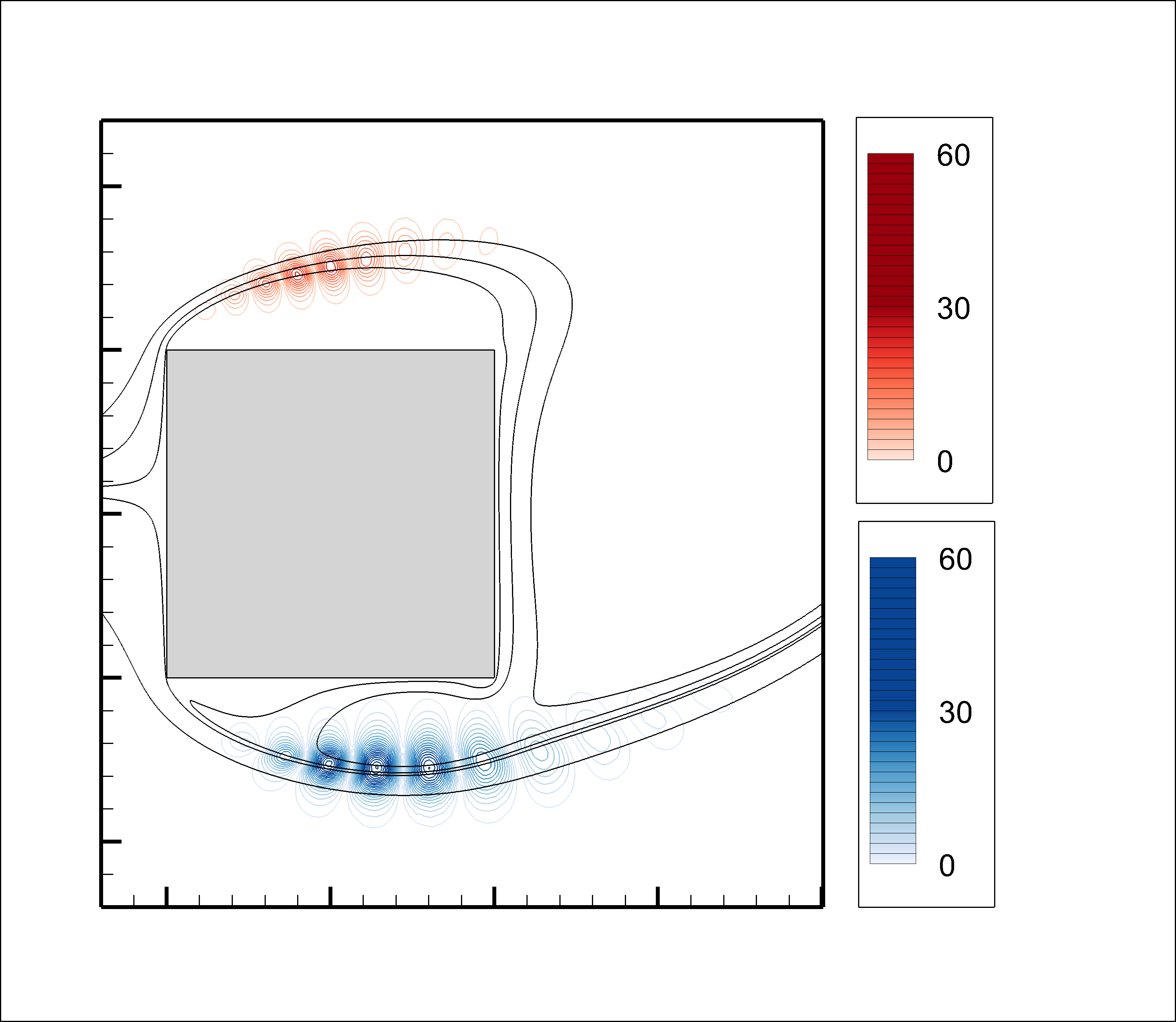}
	\caption{PCL-Resolvent analysis for $\omega=20$: the first four gains $\mu_{\tau,1,\cdots,4}^2$ as a function of time (a) and the absolute value of the pressure fluctuations of the two dominant red/blue modes, scaled by the amplitude, $\mu_{\tau} \hat{\phi}_{\tau}$. The vertical lines in fig. (a) depict the 4 phases represented in fig. (b).}\label{fig:ResolventModes20}
\end{figure}

\begin{figure}
	\centering
	\raisebox{0.8in}{(a)}\raisebox{0.6in}{$\mu_{\tau}^2$}\includegraphics[trim={1cm 1cm 7cm 4cm},clip,height=80px]{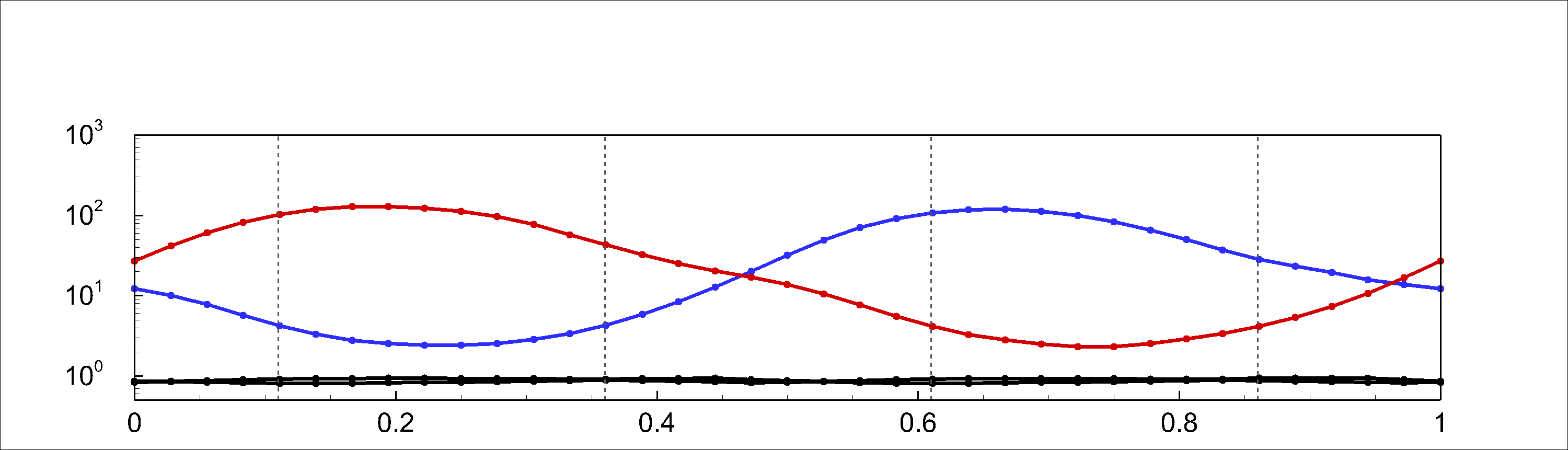}\raisebox{0.1in}{$\;\;\;\; \tau/T_0$}
	\linebreak
	\\
	\raisebox{0.5in}{(b)}\includegraphics[trim={1cm 10cm 30cm 10cm},clip,height=80px]{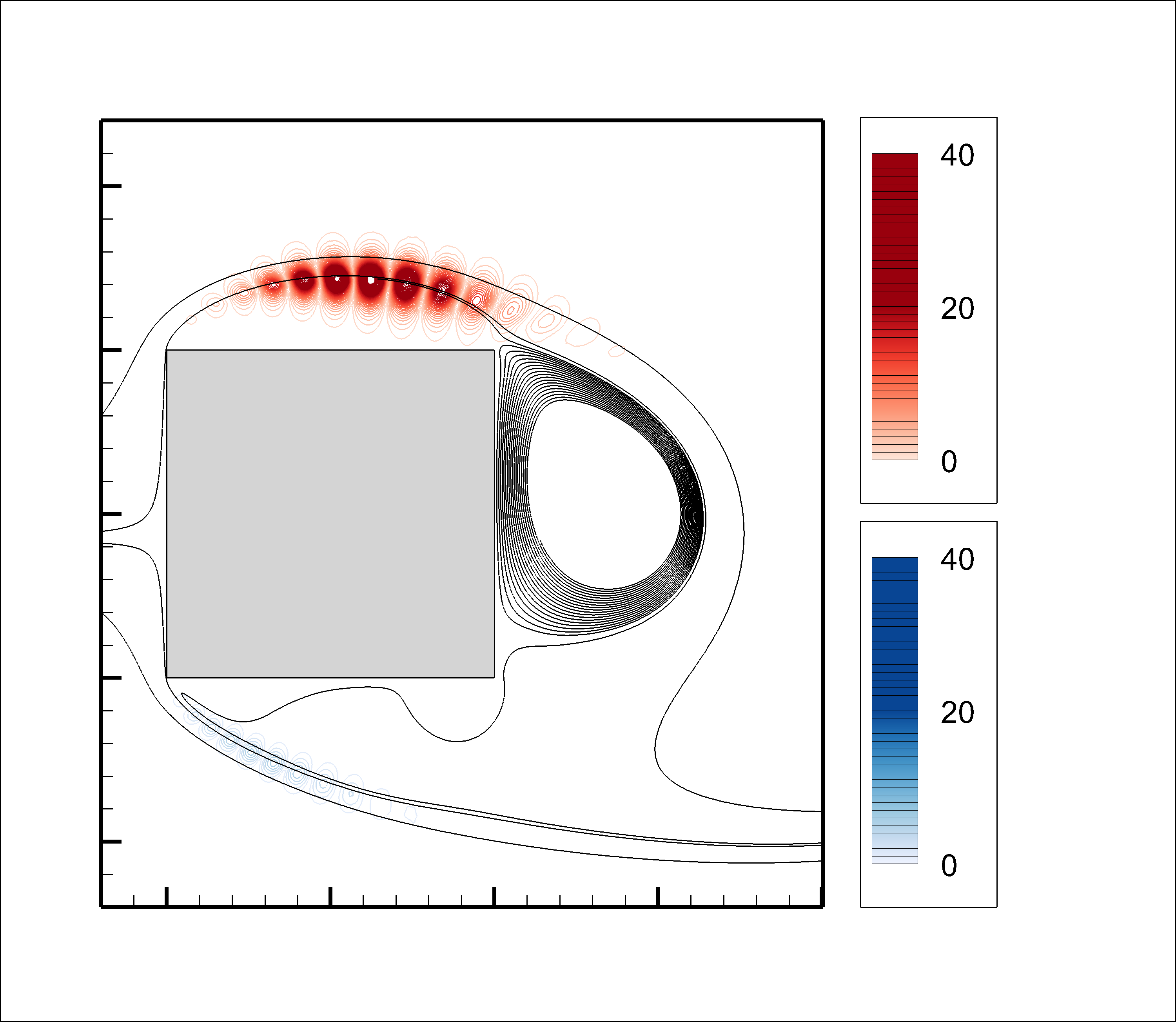}
	\includegraphics[trim={8cm 10cm 30cm 10cm},clip,height=80px]{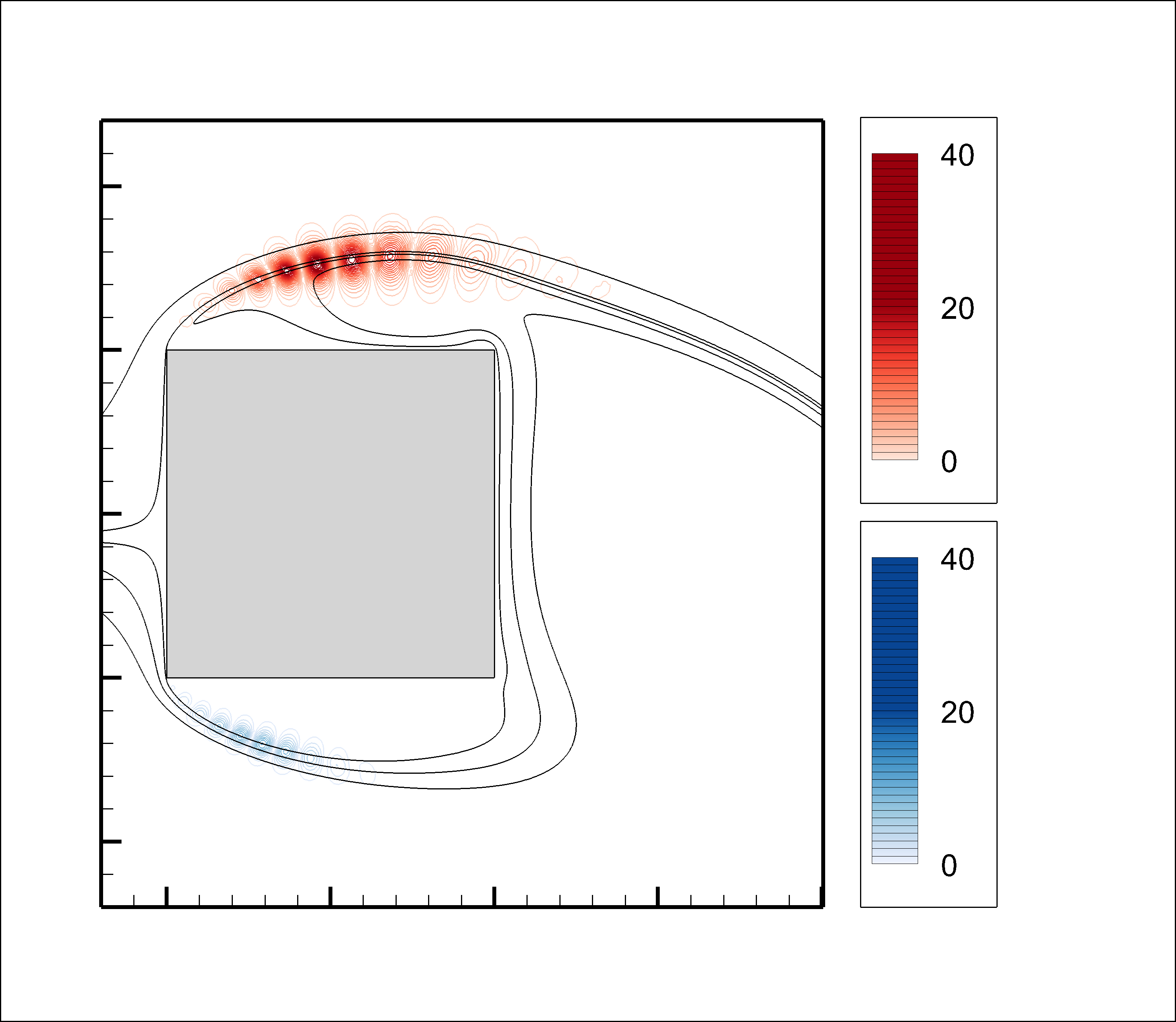}
	\includegraphics[trim={8cm 10cm 30cm 10cm},clip,height=80px]{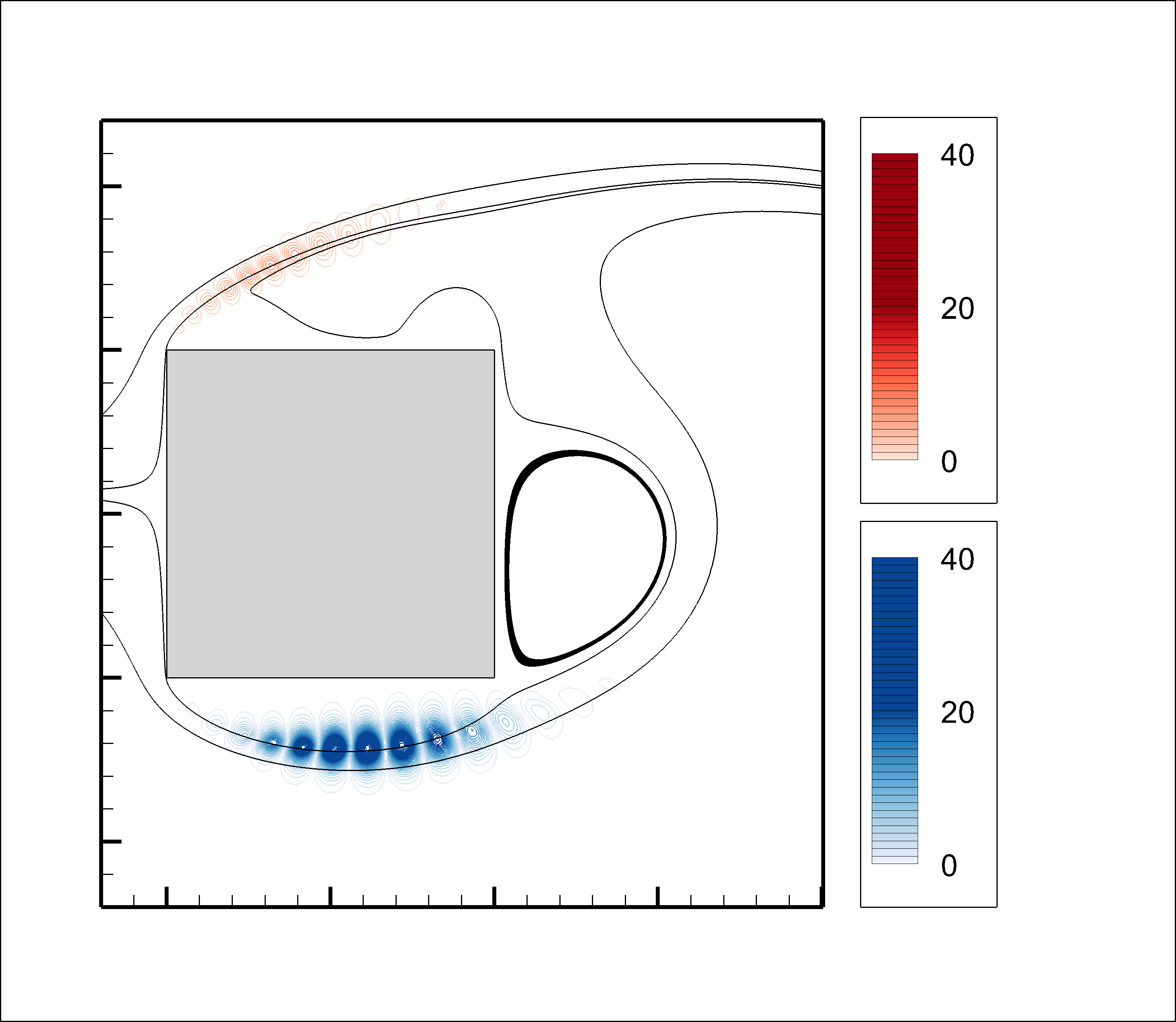} 
	\includegraphics[trim={8cm 10cm 5cm 10cm},clip,height=80px]{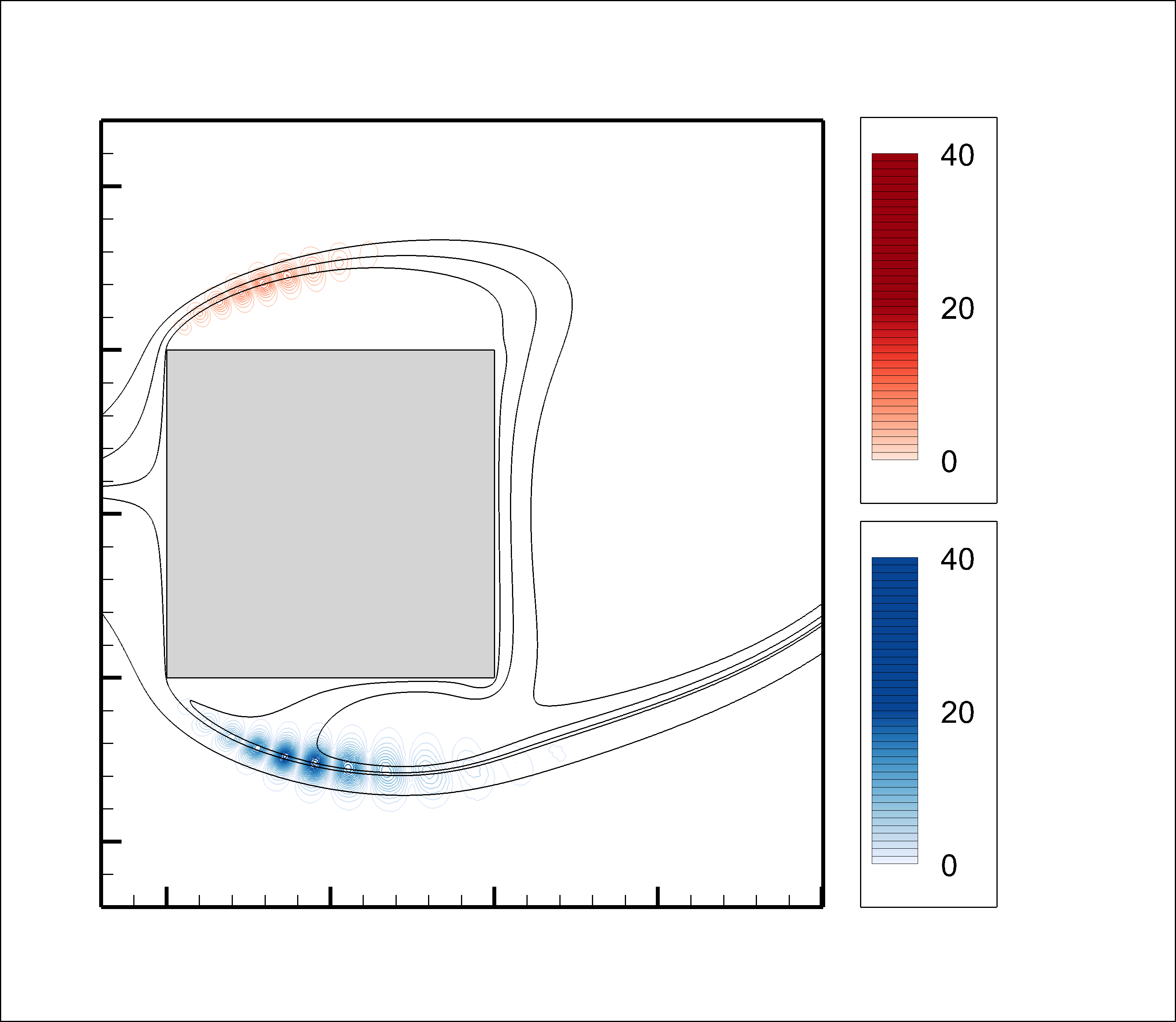}
	\caption{PCL-Resolvent analysis for $\omega=30$: same caption as in figure \ref{fig:ResolventModes20}.}\label{fig:ResolventModes30}
\end{figure}

In figures \ref{fig:ResolventModes20} and \ref{fig:ResolventModes30}, we show the detailed results for the frequencies $\omega=20$ and $30$, respecitvely. We can see, in both cases, two very strong branches (in red and blue) exhibiting the same symmetries in time as those observed for the SPOD modes at $\omega=30$. This clearly stems from the fact that the sole input of the PCL-Resolvent analysis is the $\phuu$ field, which displays the mirror-symmetry discussed before. Also, we can see that the modes are much stronger during the first two phases on the top shear-layer and inversely on the bottom layer during the last two phases, in a similar manner as for the SPOD results. 
Overall, we conclude that the agreement between PCL-SPOD and PCL-Resolvent modes is very good, establishing that high-frequency unsteadiness developing on a low-frequency motion may be well captured by a PCL-Resolvent analysis.

\section{Conclusion}\label{Conclusion}

In this paper, we have proposed a Phase-Conditioned-Localized SPOD and Resolvent analysis for the identification and reconstruction of the broadband high-frequency fluctuation content evolving on top of a periodic low-frequency limit-cycle. The case-study is the flow around a squared-section cylinder at $Re=22000$, exhibiting Kelvin-Helmholtz (KH) structures evolving on top of the Vortex-Shedding (VS) motion \citep{brun2008coherent}. 

The PCL-SPOD analysis is derived from the conditional space-time POD of \citet{schmidt2019conditional}.
We additionally use the short-time Fourier Transform to unveil, at a given phase within the period of the limit-cycle, the existing structures existing at high frequencies. The SPOD analysis corresponds to finding the most coherent and energetic structures at a given phase, for a given frequency. This technique successfully allows us to isolate those structures, together with their dependency on the VS motion. We observed that those modes can be found on the top and bottom shear-layers and tend to de-correlate (especially at higher frequencies) giving rise to modes that exist either on top or bottom shear-layers.

The PCL-Resolvent analysis consists in the classical Resolvent operator analysis with the linearization of the Navier-Stokes equations done around the periodic VS motion at given phases. This leads to an analysis where, for a given phase and frequency, we can identify different physical amplification mechanisms. This approach could successfully identify two energetic branches, whose modes exhibit properties that are similar to those of the PCL-SPOD analysis, namely they exist either on the top or at the bottom of the cylinder.

We believe those techniques could be applied to other flow configurations exhibiting high-frequency fluctuations evolving on free low-frequency deterministic motion, for instance to describe turbulence developing around the periodic shock motion in buffet (\citet{sartor2014stability,sartor2015unsteadiness,bonne2019analysis})
or the low-frequency oscillation around an airfoil in stalling condition  (see \citet{almutairi2013large})
, or to analyse limit cycle oscillations of spring-mounted wings in transitional Reynolds number flows \citet{yuan2013simulations}. It could also be applied to flow configurations with forced low-frequency deterministic motion, 
as wind turbines (\citet{lignarolo2015tip}) and turbomachines (\citet{tucker2011computation}).


\textbf{Declaration of interests}. The authors report no conflict of interest.

\appendix
\section{PCL-SPOD for upper/lower inner product}\label{apx:UpperLower}

\begin{figure}
	\centering
	\raisebox{0.8in}{(a)}\raisebox{0.6in}{$\lambda_{\tau}^2$}\includegraphics[trim={1cm 1cm 7cm 4cm},clip,height=80px]{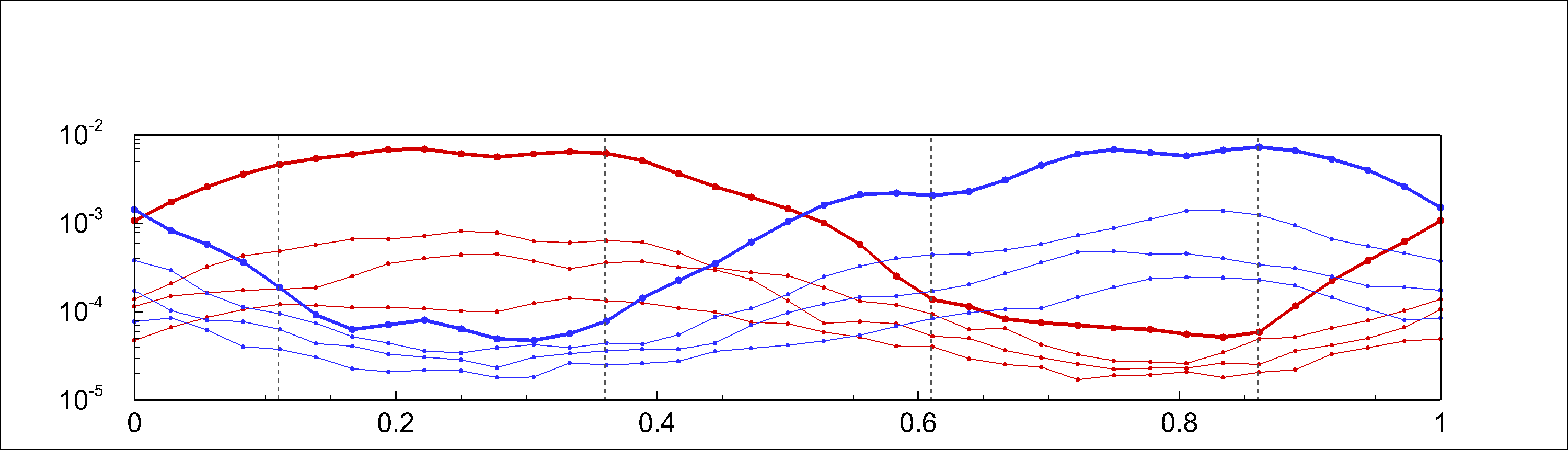}\raisebox{0.1in}{$\;\;\;\; \tau/T_0$}
	\linebreak
	\\
	\raisebox{0.5in}{(b)}\includegraphics[trim={1cm 10cm 30cm 10cm},clip,height=80px]{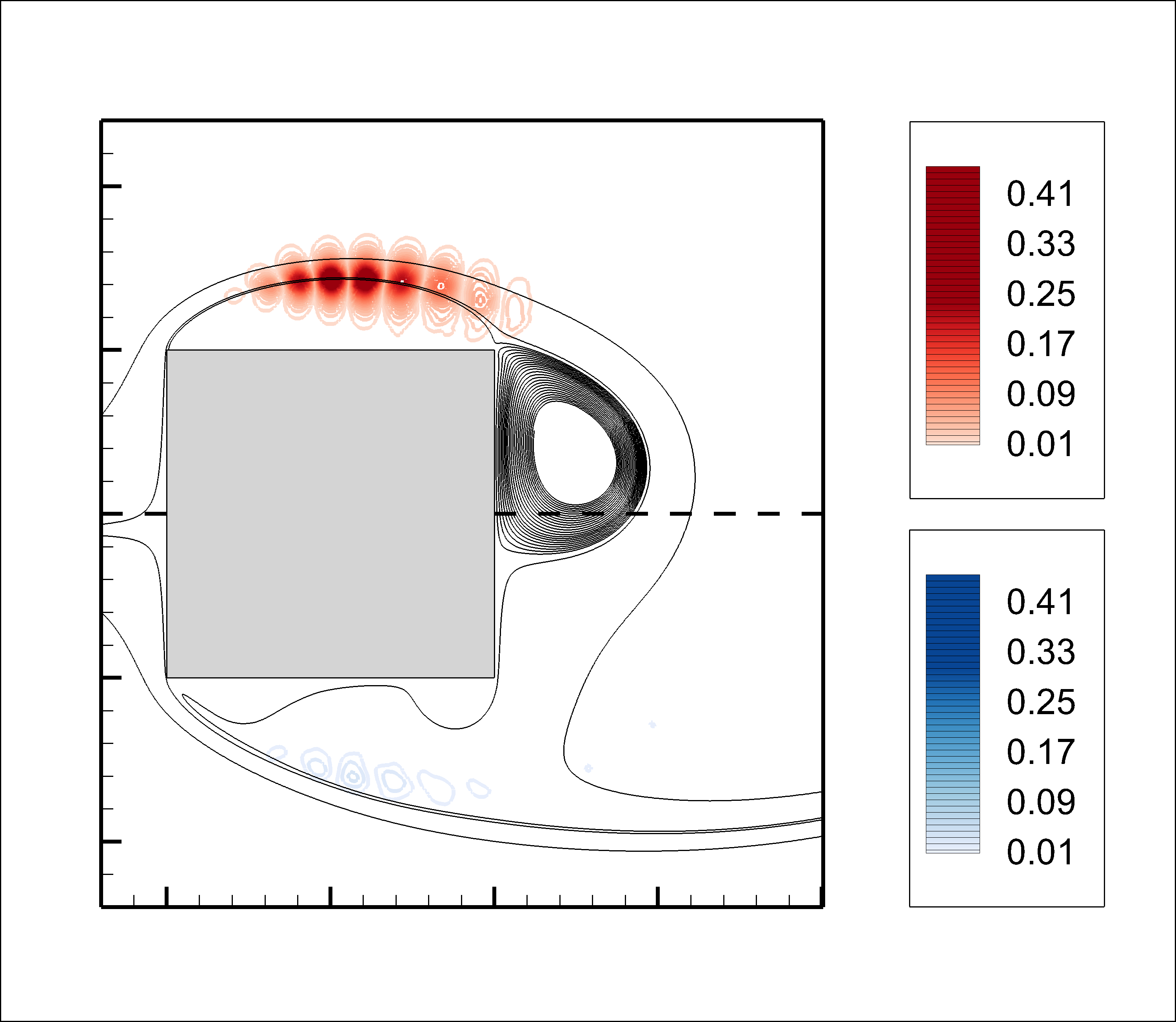}
	\includegraphics[trim={8cm 10cm 30cm 10cm},clip,height=80px]{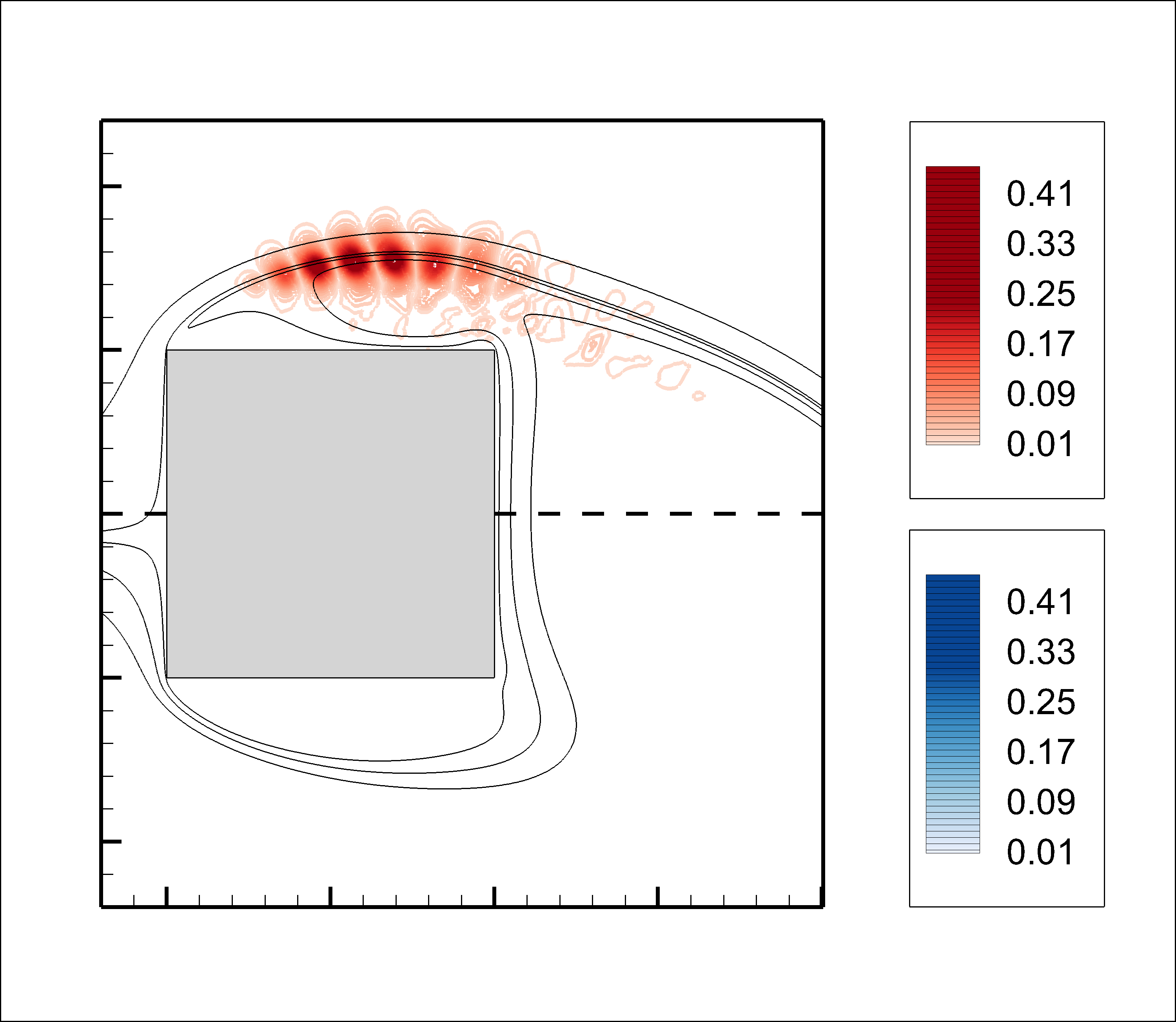}
	\includegraphics[trim={8cm 10cm 30cm 10cm},clip,height=80px]{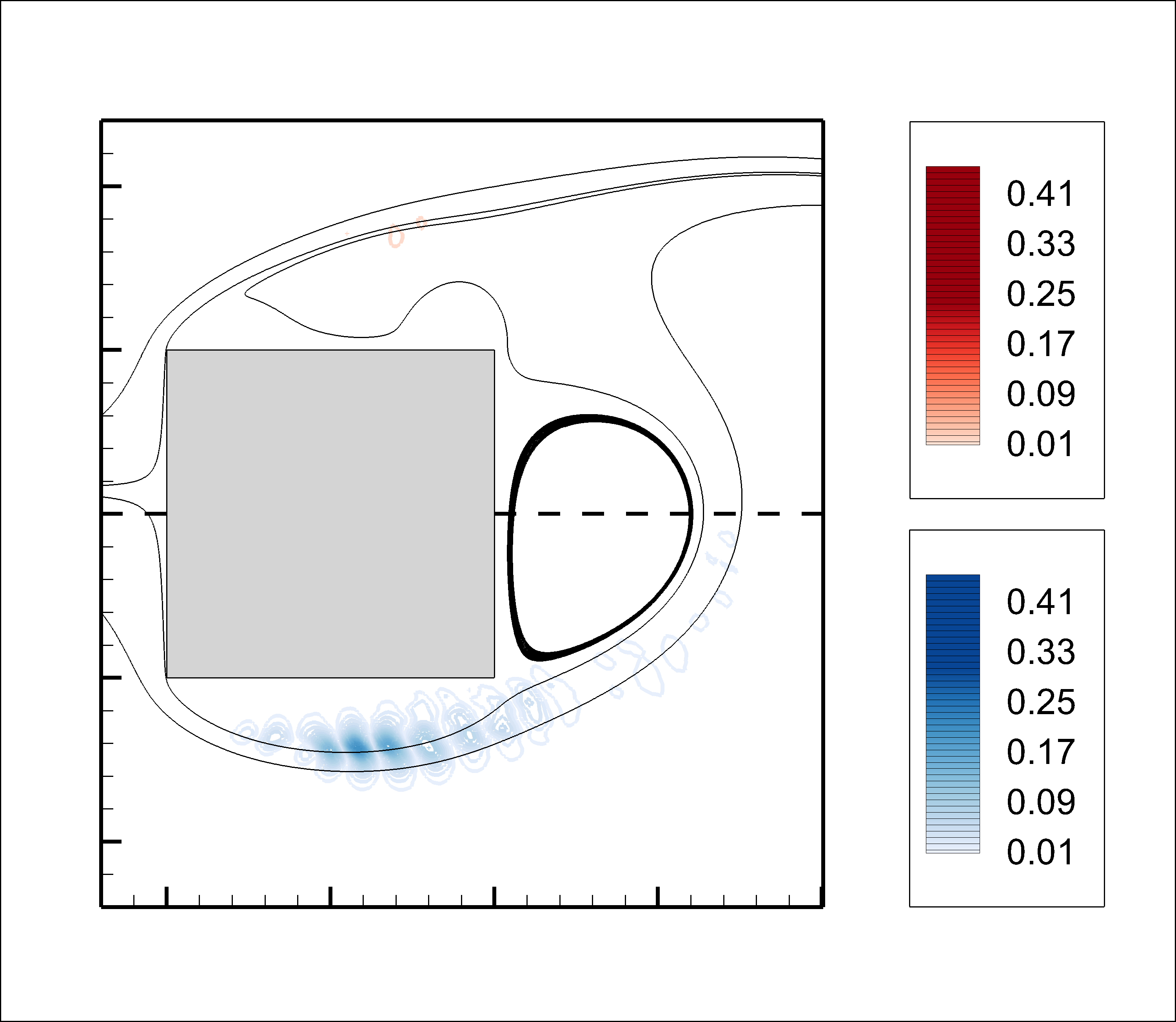} 
	\includegraphics[trim={8cm 10cm 5cm 10cm},clip,height=80px]{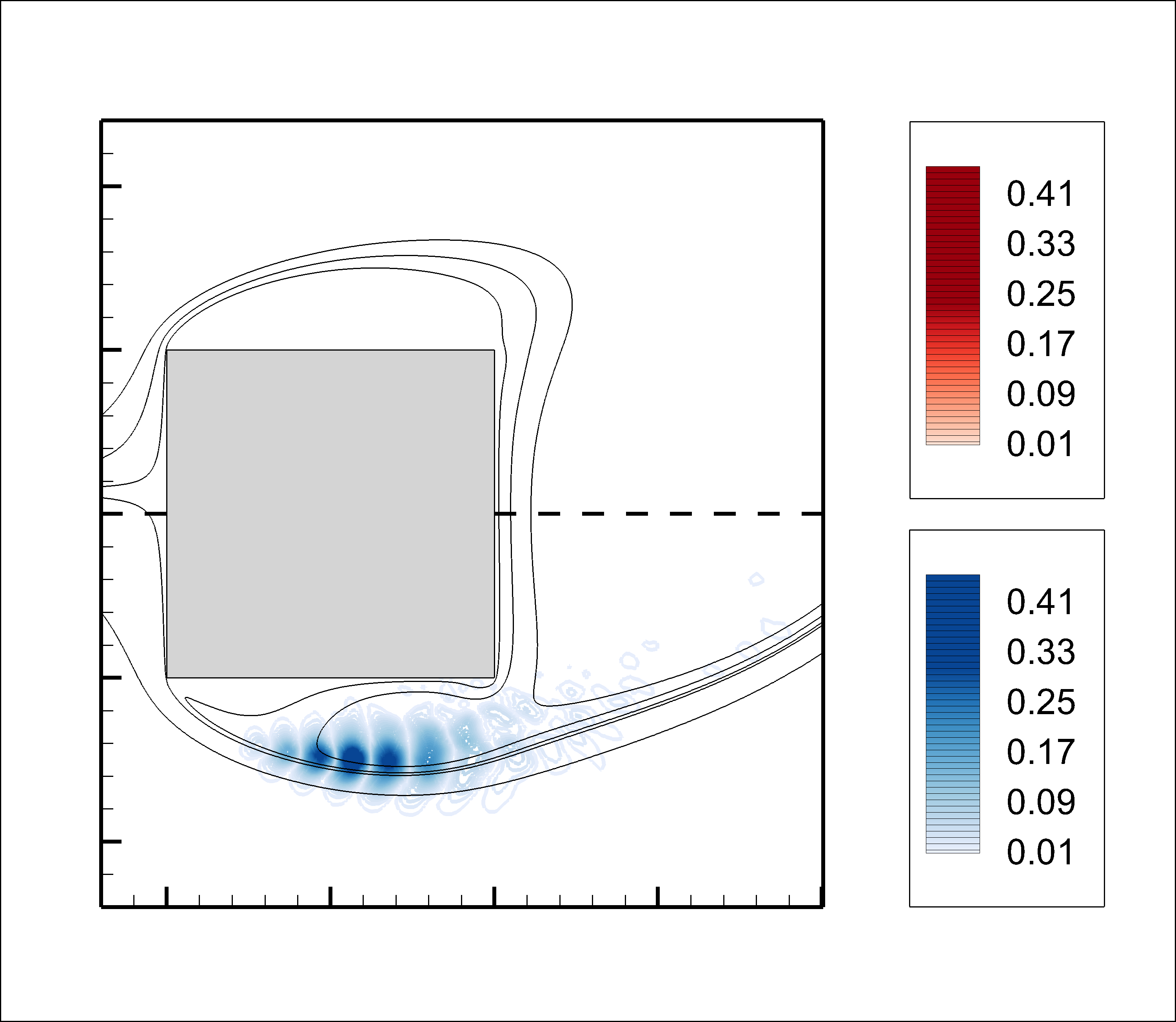}
    \caption{PCL-SPOD analysis for $\omega=30$ and upper/lower localized inner product: (a) gains $\lambda_{\tau}$ given in red/blue for upper/lower regions and (b) the modes $\lambda_{\tau} \hat{\phi}_{\tau}$ are given according to the same color code. Those modes are given only in the half plane corresponding to their inner product region.}\label{fig:ResolventModes30LU}
\end{figure}

In this appendix we provide a few results on the PCL-SPOD when focusing the analysis either on the upper ($y\geq 0$) or on the lower part ($y\leq 0$) of the domain $\Omega$. In figure \ref{fig:ResolventModes30LU} we report those results for $\omega=30$ where the curves in red in (a) correspond to the gains by restricting the data to the upper domain and the blue ones to the lower one. We can see that the optimal mode 
s, either red or blue, present almost exactly the same energy gains as the ones presented in the body of the paper, obtained using the full domain $\Omega$. Moreover, the spatial structure of the modes are also almost unchanged. Those modes are obtained by finding the eigenvalues of the matrix $\mathbf{A}_{uu} = (1/N_b) \sum_{m=1}^{N_b} \left[ \hat{\qq}_{\tau_m}^{\Omega_{u}} \hat{\qq}_{\tau_m}^{\Omega_{u},*} \right]$ and $\mathbf{A}_{ll} = (1/N_b) \sum_{m=1}^{N_b} \left[ \hat{\qq}_{\tau_m}^{\Omega_{l}} \hat{\qq}_{\tau_m}^{\Omega_{l},*} \right]$, where 
$\hat{\qq}_{\tau_m}^{\Omega_{u}}$ and $\hat{\qq}_{\tau_m}^{\Omega_{l}}$ are the fields restricted to the upper and lower domains.
We can see that, indeed the matrix $\mathbf{A}_{lu}$ is indeed negligible.

\bibliographystyle{jfm}
\bibliography{Unsteady-Data-Assimilation-Turbulent}

\end{document}